\documentstyle[emlines2]{article}

\def\hybrid{\topmargin 0pt      \oddsidemargin 0pt
        \headheight 0pt \headsep 0pt
        \textwidth 6.25in       
        \textheight 9.5in       
        \marginparwidth 0.0in
        \parskip 5pt plus 1pt   \jot = 1.5ex}
\catcode`\@=11
\def\marginnote#1{}

\newcount\hour
\newcount\minute
\newtoks\amorpm
\hour=\time\divide\hour by60
\minute=\time{\multiply\hour by60 \global\advance\minute by-\hour}
\edef\standardtime{{\ifnum\hour<12 \global\amorpm={am}%
        \else\global\amorpm={pm}\advance\hour by-12 \fi
        \ifnum\hour=0 \hour=12 \fi
        \number\hour:\ifnum\minute<10 0\fi\number\minute\the\amorpm}}
\edef\militarytime{\number\hour:\ifnum\minute<10 0\fi\number\minute}

\def\draftlabel#1{{\@bsphack\if@filesw {\let\thepage\relax
   \xdef\@gtempa{\write\@auxout{\string
      \newlabel{#1}{{\@currentlabel}{\thepage}}}}}\@gtempa
   \if@nobreak \ifvmode\nobreak\fi\fi\fi\@esphack}
        \gdef\@eqnlabel{#1}}
\def\@eqnlabel{}
\def\@vacuum{}
\def\draftmarginnote#1{\marginpar{\raggedright\scriptsize\tt#1}}

\def\draftlabel#1{{\@bsphack\if@filesw {\let\thepage\relax
   \xdef\@gtempa{\write\@auxout{\string
      \newlabel{#1}{{\@currentlabel}{\thepage}}}}}\@gtempa
   \if@nobreak \ifvmode\nobreak\fi\fi\fi\@esphack}
        \gdef\@eqnlabel{#1}}
\def\@eqnlabel{}
\def\@vacuum{}
\def\draftmarginnote#1{\marginpar{\raggedright\scriptsize\tt#1}}

\def\draft{\oddsidemargin -.5truein
        \def\@oddfoot{\sl preliminary draft \hfil
        \rm\thepage\hfil\sl\today\quad\militarytime}
        \let\@evenfoot\@oddfoot \overfullrule 3pt
        \let\label=\draftlabel
        \let\marginnote=\draftmarginnote
   \def\@eqnnum{(\theequation)\rlap{\kern\marginparsep\tt\@eqnlabel}%
\global\let\@eqnlabel\@vacuum}  }

\def\numberbysection{\@addtoreset{equation}{section}
        \def\theequation{\thesection.\arabic{equation}}}

\def\underline#1{\relax\ifmmode\@@underline#1\else
        $\@@underline{\hbox{#1}}$\relax\fi}

\def\titlepage{\@restonecolfalse\if@twocolumn\@restonecoltrue\onecolumn
     \else \newpage \fi \thispagestyle{empty}\c@page\z@
        \def\thefootnote{\fnsymbol{footnote}} }

\def\endtitlepage{\if@restonecol\twocolumn \else  \fi
        \def\thefootnote{\arabic{footnote}}
        \setcounter{footnote}{0}}  
\relax

\numberbysection
\hybrid

\def\beq{\begin{equation}}
\def\eeq{\end{equation}}
\def\p{\partial}

\def\l{\lambda}
\def\a{\alpha}
\def\b{\beta}

\newtheorem{th}{Theorem}[section]
\newtheorem{prop}{Proposition}[section]

\begin{document}

\begin{titlepage}

\title{A survey of Hirota's difference equations}

\author{A. Zabrodin
\thanks{Joint Institute of Chemical Physics, Kosygina str. 4, 117334,
Moscow, Russia and ITEP, 117259, Moscow, Russia}}

\maketitle

\begin{abstract}

A review of selected topics in Hirota's bilinear
difference equation (HBDE) is given. This
famous 3-dimensional difference equation
is known to provide a canonical integrable discretization
for most important types of soliton equations.
Similarly to the continuous theory, HBDE is a member of
an infinite hierarchy. The central point of our exposition
is a discrete version of the zero curvature condition
explicitly written in the form of discrete Zakharov-Shabat
equations for $M$-operators realized as
difference or pseudo-difference operators.
A unified approach to various
types of $M$-operators and zero curvature representations
is suggested. Different reductions of HBDE to 2-dimensional
equations are considered. Among them discrete counterparts of
the KdV, sine-Gordon, Toda chain, relativistic
Toda chain and other typical examples are discussed in detail.

\end{abstract}

\vfill

\end{titlepage}

\section{Introduction}

In 1981 R.Hirota published the paper \cite{Hirota1} which summarized
his earlier studies of discretizing nonlinear integrable equations
\cite{HirotaKdV}-\cite{Hirota5}. The main result is a compact
bilinear equation which can be viewed as
an integrable discrete analogue of
the 2-dimensional Toda lattice. In Hirota's original
notation it has the form
\beq
\big [ Z_1 \exp (D_1 )+Z_2 \exp (D_2 )+Z_3 \exp (D_3 )\big ]
\tau \cdot \tau =0\,,
\label{HBDE1}
\eeq
where $Z_i$ are arbitrary constants, $\tau =\tau (x_1 , x_2 ,x_3 )$,
$D_i \equiv D_{x_i}$ and Hirota's $D$-operator is defined for a
linear differential operator $F(\p _{x})$ by
\beq
\left.\phantom{{a\over b}}
F(D_x )f(x)\cdot g(x)=F(\p _y )f(x+y)g(x-y)
\right|_{y=0 _{\phantom{ff}}}\,.
\label{D}
\eeq
In the more explicit notation eq.\,(\ref{HBDE1}) looks as follows:
\begin{eqnarray}
&&Z_1 \tau (x_1 +1,x_2 ,x_3 )
\tau (x_1 -1, x_2, x_3 ) \nonumber \\
&+&Z_2 \tau (x_1 ,x_2 +1, x_3 )
\tau (x_1 , x_2 -1, x_3 ) \nonumber \\
&+&Z_3 \tau (x_1 ,x_2 , x_3 +1)
\tau (x_1 , x_2 , x_3 -1)=0\,.
\label{HBDE2}
\end{eqnarray}
This equation is often called Hirota's bilinear difference equation
(HBDE). Its simplicity is
surprizing and delusive at the same time:
each detail is controlled by integrability and hides
meaningful mathematical structures whereas
some even simpler looking equations turn out to be untractable
by analytical methods.

One of the most impressive outcomes of
Hirota's works is that HBDE is shown to unify
many if not all known soliton equations. More precisely,
it contains them in an
encoded form. Performing a scaling continuum limit for
appropriate combinations of parameters
and variables, one is able to obtain
the Korteweg - de Vries (KdV) equation,
Kadomtsev-Petviashvili (KP) equation, modified KdV (MKdV)
and modified KP (MKP) equations,
two-dimensional Toda lattice (2DTL) equation, sine-Gordon (SG)
equation, Benjamin-Ono equation, etc.
Their discrete analogues are produced from HBDE by choosing
suitable dependent and independent variables.
Furthermore, eq.\,(\ref{HBDE1}) was shown to
possess soliton solutions and B\"acklund transformations for generic
values of parameters. These facts suggest to consider HBDE as a
{\it fundamental} classical soliton equation,
from which the typical examples can be obtained as particular cases.

Recently, bilinear equations of this form emerged
\cite{KLWZ},\,\cite{Z}
in the context of {\it quantum}
integrable systems as the model-independent functional relations
\cite{KP},\,\cite{Kuniba1}
for eigenvalues of quantum transfer matrices. This was our motivation
for revisiting classical nonlinear difference equations.

These notes aim at reviewing selected topics in HBDE and further
clarifying basic elements of the theory. In our exposition, we
deal solely with equations themselves saying almost no word about
their solutions\footnote{Just because of this we do not draw any
distinction between {\it discrete} and {\it difference} equations.
Usually, in the latter case it is implied that solutions are functions
of a continuous variable with certain analytical properties.}.
Likewise their continuous counterparts, completely discretized
nonlinear integrable equations are known to possess soliton and
finite-gap solutions.
However, a systematic treatment of these
and other particular classes of
solutions could be a separate
enterprise which requires much more
space. We shall confine ourselves to elaborating discrete versions
of commutation representations and auxiliary linear problems on a
formal algebraic level. At the same time some important elements
of our approach are essentially motivated by the finite-gap
theory.

The difference soliton equations are intimately connected with
the differential ones. We already mentioned that the latter are
obtained from the former by a scaling limit.
Better to say, HBDE was
just designed to enjoy this property.
The fact that such an equation does exist is by no means trivial.
A link in the opposite direction was established
by T.Miwa \cite{Miwa1} who noticed that discrete Hirota's
equations can be obtained from the continuous KP hierarchy by choosing
the time flows to be certain infinite combinations of the standard flows
of the hierarchy. This idea was further developed in the papers
\cite{Miwa2},\,\cite{Miwa3} as a method to produce discrete
soliton equations from continuous ones.
The interrelation between discrete
and continuous integrable hierarchies
looks like a kind of Fourier duality:
they provide complementary descriptions of the same object, namely,
of the infinite dimensional grassmannian
\cite{Sato}-\cite{SW}.

In this survey we do not give a systematic treatment of the
connection between discrete and continuous hierarchies.
The problem to describe the limiting
procedure that would be compatible with
the whole hierarchy is technically involved.
However, it is impossible not to refer to continuous hierarchies
at all. We agree to a compromise restricting ourselves by
a few typical examples.

It is assumed that the reader is
familiar with the basic
notions of the continuous theory such as
Lax and Zakharov-Shabat equations, zero curvature conditions
in scalar and matrix forms, commuting flows,
infinite hierarchies, $\tau$-function etc.

Let us outline the contents of the paper. (More detailed descriptions
are given in short introductions to each section.)

Sect.\,2 can be considered as
a part of the Introduction.
We tried to collect here different forms of 3-dimensional HBDE
known in the literature. All of them are equivalent.
Simple transformations between them are listed.

Sections 3, 4 and 5 form the main body of the paper.
To figure out the key principles underlying the variety
of integrable difference equations, we need a number of
definitions and axioms. They are given in Sect.\,3.
All the notions explained in this section are essentially
used in the sequel.
In Sect.\,4, the discrete version of
the zero curvature representation is presented.
Filling in some gaps in the existing literature, we give explicit forms
of the $M$-operators (realized as difference operators) for
discrete flows.
Sect.\,5 is devoted to various types
of associated auxiliary linear problems. They provide a "linearization"
of the original nonlinear equation.
The related notion of B\"acklund transformation
is discussed. The Baker-Akhiezer functions
are introduced as special formal solutions to the linear
problems.

Sections 6 and 7 are more technical and might
be interesting mostly for experts. They can be
skipped without loss of understanding.
In Sect.\,6 it is explained how to extend
the $M$-operator approach to {\it arbitrary} discrete flows
defined in Sect.\,3. In the general case the $M$-operators
contain negative powers of first order difference operators.
In Sect.\,7 we dwell upon hierarchies of bilinear discrete equations
and suggest the notion of "higher" discrete flows with the
corresponding zero curvature representation.
We conjecture that all "higher" ($N$-term)
discrete Hirota's equations known in the literature are consequences
of the 3-term ones. This assertion is proved for the first nontrivial
example of the 4-term equations in 4 variables.

Sect.\,8 deals with 2-dimensional reductions of HBDE, the corresponding
$L$-$M$ pairs and auxiliary linear problems. The list
of reductions includes discrete
analogues of the KdV equation, the 1D Toda chain, the AKNS system,
the relativistic Toda chain,
the sine-Gordon equation, the Liouville equation and some others.

In the Appendix, we present main elements of a different approach
to HBDE based on Miwa's transformation.
This is the method for generating discrete soliton equations
suggested in ref.\,\cite{Miwa2}. We show how it works for the
simplest examples and comment on the continuum limit which is in a sense
an "inverse" Miwa transformation.

\section{Equivalent forms of the bilinear equation}

Hirota's difference equation exists in several forms.
Historically, they emerged as integrable discretizations
of particular continuous hierarchies (e.g. KP, 2DTL).
In this section we give a list of most popular forms
of Hirota's difference equation
and explicitly demonstrate that they are equivalent.
However, it is useful to bear in mind all of
them since one or another
may be more convenient in a particular problem.

\begin{itemize}
\item a) {\it Hirota's original form} (\ref{HBDE2}):
\beq
Z_1 \tau (x_1 +1)\tau (x_1 -1)+
Z_2 \tau (x_2 +1)\tau (x_2 -1)+
Z_3 \tau (x_3 +1)\tau (x_3 -1)=0
\label{HBDE2a}
\eeq
\end{itemize}
(here and below we often skip the variables that do not
undergo shifts).
Note that the 3 variables enter in a symmetric fashion and the
equation is invariant under their permutations and a simultaneous
permutation of $Z_i$'s. The equation is also invariant under
changing the sign of any one of the variables and under the
transformation
\beq
\tau (x_1 , x_2 , x_3 )\rightarrow
\chi _{0}(x_1 + x_2 + x_3 )
\chi _{1}(x_2 + x_3 - x_1 )
\chi _{2}(x_1 + x_3 - x_2 )
\chi _{3}(x_1 + x_2 - x_3 )
\tau (x_1 , x_2 , x_3 )\,,
\label{inv1}
\eeq
where $\chi _{i}$ are arbitrary functions.
The transformation
\beq
\tau (x_1 , x_2 , x_3 )\rightarrow
Z_{1}^{-x_{1}^{2}/2}Z_{2}^{-x_{2}^{2}/2}Z_{3}^{-x_{3}^{2}/2}
\tau (x_1 , x_2 , x_3 )
\label{NE1}
\eeq
converts eq.\,(\ref{HBDE2}) into the {\it canonical form},
\beq
\tau (x_1 +1)\tau (x_1 -1)+
\tau (x_2 +1)\tau (x_2 -1)+
\tau (x_3 +1)\tau (x_3 -1)=0\,.
\label{HBDE3}
\eeq
This equation does not
contain any free parameters.

\begin{itemize}
\item a$'$) {\it "Gauge invariant" form}:
\beq
Y(x_1 , x_2 +1, x_3 )
Y(x_1 , x_2 -1, x_3 )=\frac{
(1+Y(x_1 ,x_2 , x_3 +1))
(1+Y(x_1 ,x_2 , x_3 -1))}
{(1+Y^{-1}(x_1 +1, x_2 , x_3 ))
(1+Y^{-1}(x_1 -1 , x_2 , x_3 ))},
\label{Ysys}
\eeq
\end{itemize}
where
\beq
Y(x_1 , x_2 , x_3 )\equiv \frac{
\tau (x_1 , x_2 , x_3 +1)\tau (x_1 , x_2 , x_3 -1 )}
{\tau (x_1 +1, x_2 , x_3 )\tau (x_1 -1, x_2 , x_3 )}
\label{Y}
\eeq
is a gauge invariant quantity: the "gauge" transformation (\ref{inv1})
does not change it. This form is a discrete counterpart of nonlinear
integrable equations written in terms of potentials and fields
rather than $\tau$-functions. Some particular cases of this equation
emerge naturally in thermodynamic Bethe ansatz \cite{Zam},\,\cite{Tateo}.

\begin{itemize}
\item b) {\it KP-like form}:
\begin{eqnarray}
&&(z_{2}-z_{3})\tau ^{p_1 +1, p_2 , p_3 }
\tau ^{p_1 , p_{2}+1 , p_{3} +1 }+ \nonumber \\
&+&(z_{3}-z_{1})\tau ^{p_1 , p_2 +1, p_3 }
\tau ^{p_1 +1, p_{2} , p_{3} +1 }+ \nonumber \\
&+&(z_{1}-z_{2})\tau ^{p_1 , p_2 , p_3 +1}
\tau ^{p_1 +1, p_{2}+1 , p_{3}  }=0\,,
\label{HBDE4}
\end{eqnarray}
\end{itemize}
where $\tau ^{p_1 ,p_2 , p_3 }$ is a function of the 3 variables
$p_i$'s and $z_i$ are arbitrary constants. This equation is
invariant under cyclic permutations of the variables and simultaneous
permutations of $z_i$'s. Changing signes of all the variables
also leaves it invariant. Besides, the transformation (\ref{inv1}) can
be translated to eq.\,(\ref{HBDE4}): if $\tau ^{p_1 , p_2 , p_3 }$ is
a solution, then
\beq
\chi _{0}(2p_1 + 2p_2 + 2p_3 )
\chi _{1}(2p_1 )
\chi _{2}(2p_2 )
\chi _{3}(2p_3 )
\tau ^{p_1 , p_2 , p_3 }
\label{inv1a}
\eeq
is a solution too. Again, the coefficients in (\ref{HBDE4}) can be
made equal to 1 by means of the transformation
\beq
\tau ^{p_1 , p_2 , p_3 }\rightarrow
\left (\frac{z_{1}-z_{3}}
{z_{1}-z_{2}}\right )^{p_1 p_2 }
\left (\frac{z_{1}-z_{3}}
{z_{2}-z_{3}}\right )^{p_2 p_3 }
\tau ^{p_1 , p_2 , p_3 }\,,
\label{NE2}
\eeq
bringing eq.\,(\ref{HBDE4}) into its canonical form.

\begin{itemize}
\item b$'$) {\it MKP-like form}:
\begin{eqnarray}
&&(z_{0}-z_{1})(z_{2}-z_{3})
\tau ^{p_1 +1, p_2 , p_3 }_{p_0 }
\tau ^{p_1 , p_{2}+1 , p_{3} +1 }_{p_0 +1}+ \nonumber \\
&+&(z_{0}-z_{2})(z_{3}-z_{1})
\tau ^{p_1 , p_2 +1, p_3 }_{p_0 }
\tau ^{p_1 +1, p_{2} , p_{3} +1 }_{p_0 +1}+ \nonumber \\
&+&(z_{0}-z_{3})(z_{1}-z_{2})
\tau ^{p_1 , p_2 , p_3 +1}_{p_0 }
\tau ^{p_1 +1, p_{2}+1 , p_{3}  }_{p_0 +1}=0\,.
\label{HBDE4a}
\end{eqnarray}
\end{itemize}
Note that the combination of the arguments
$p_1 +p_2 +p_3 -p_0 $
is the same for all $\tau$-functions in this equation. In orther words,
the hyperplane
$p_1 +p_2 +p_3 -p_0 =\mbox{const}$ is invariant. Therefore, this
equation actually depends on three variables rather than four: say,
$p_1 , p_2 , p_3 $. Since sum of the coefficients in
eq.\,(\ref{HBDE4a}) is zero, like in eq.\,(\ref{HBDE4}), they
differ by a reparametrization of $z_{i}$'s only.

\begin{itemize}
\item c) {\it 2DTL-like form}:
\beq
\nu \tau _{n}^{l,\bar l +1}\tau _{n}^{l+1, \bar l}+
(\mu - \nu )
\tau _{n}^{l,\bar l }\tau _{n}^{l+1, \bar l +1}=
{\mu }\tau _{n+1}^{l,\bar l +1}\tau _{n-1}^{l+1, \bar l}\,,
\label{HBDE6}
\eeq
\end{itemize}
where $\tau _{n}^{l, \bar l}$ is a function of the 3 variables and
$\mu$, $\nu$ are arbitrary constants. The variables
$l, \bar l$ are called light cone coordinates.
Note that in this
form the permutation symmetry is lost. However, an analogue of
eq.\,(\ref{inv1}) holds true: if $\tau _{n}^{l, \bar l}$ solves
eq.\,(\ref{HBDE6}), then
$
\chi _{0}(2n+2l)
\chi _{1}(2l)
\chi _{2}(2\bar l )
\chi _{3}(2n-2\bar l )
\tau _{n}^{l,\bar l}
$
is a solution, too. The transformation
\beq
\tau _{n}^{l,\bar l} \rightarrow
\left (\frac{{\mu}}{\nu} -1\right )^{-l\bar l}
\left (-\frac{\mu}{\nu}\right )^{-n^{2}/2}
\tau _{n}^{l,\bar l}
\label{NE3}
\eeq
allows one to hide the coefficients in eq.\,(\ref{HBDE6}):
\beq
\tau _{n}^{l,\bar l +1}\tau _{n}^{l+1, \bar l}+
\tau _{n}^{l,\bar l }\tau _{n}^{l+1, \bar l +1}+
\tau _{n+1}^{l,\bar l +1}\tau _{n-1}^{l+1, \bar l}=0\,,
\label{HBDE7}
\eeq
which is refered to as its canonical form.

For reader's convenience we present here the linear substitutions
making the canonical forms of equations a), b), c) equivalent.

a)$\leftrightarrow$ b):
$\tau (x_1 , x_2 , x_3 )=
\tau ^{p_1 , p_2 , p_3 }$,
\beq
p_1 =\textstyle{ \frac{1}{2}}(-x_1 +x_2 +x_3 )\,,
\;\;\;\;
p_2 =\textstyle{ \frac{1}{2}}(x_1 -x_2 +x_3 )\,,
\;\;\;\;
p_3 =\textstyle{ \frac{1}{2}}(x_1 +x_2 -x_3 )\,,
\label{lin1}
\eeq
\beq
x_1 =p_2 +p_3 \,,
\;\;\;\;\;\;
x_2 =p_1 +p_3 \,,
\;\;\;\;\;\;
x_3 =p_1 +p_2 \,,
\label{lin1a}
\eeq

b)$\leftrightarrow$ c):
$\tau ^{p_1 , p_2 , p_3 }=
\tau ^{l, \bar l}_{n}$,
\beq
n=p_2 +p_3 \,, \;\;\;\;\;\;
l=p_1 \,,\;\;\;\;\;\;
\bar l=p_2 \,,
\label{lin2}
\eeq
\beq
p_1 =l \,, \;\;\;\;\;\;
p_2 =n -\bar l \,,\;\;\;\;\;\;
p_3 =\bar l \,,
\label{lin2a}
\eeq

a)$\leftrightarrow$ c):
$\tau (x_1 , x_2 , x_3 )=
\tau ^{l, \bar l}_{x}$,
\beq
n =x_1\,,
\;\;\;\;\;
l =\textstyle{\frac{1}{2}}(-x_1 +x_2 +x_3 )\,,
\;\;\;\;\;
\bar l =\textstyle{\frac{1}{2}}(x_1 -x_2 +x_3 )\,,
\label{lin3}
\eeq
\beq
x_1 =n \,, \;\;\;\;\;\;
x_2 =n +l -\bar l \,, \;\;\;\;\;\;
x_3 =l+ \bar l \,.
\label{lin3a}
\eeq

Clearly, these linear substitutions are not unique. All other
possibilities can be obtained from the given one by applying a
transformation of the form
$(x_1 , x_2 , x_3 )\rightarrow
(\pm x_{P(1)},\pm x_{P(2)},\pm x_{P(3)})$, where $P$ is a permutation.
Using formulas (\ref{lin1})-(\ref{lin3a}) one can easily obtain gauge
invariant forms of equations b) and c).

\section{Definitions: the nomenclature of flows}

Here we introduce a practical set of
definitions and axioms which will help
us to develop a systematic viewpoint to the zoo of non-linear
integrable equations and their commutation representations.
This viewpoint is in fact more general than we need for
HBDE itself.
Differential as well as "mixed"
differential-difference non-linear equations fit the
scheme, too.
Our approach is motivated by algebro-geometric solutions \cite{kr}
to soliton equations expressed through Riemann theta-functions.
However, since the goal is to clarify formal algebraic
structures, we never refer to the solutions explicitly.

\subsection{The variables and kinematical constraints}

The "unknown function" entering bilinear equations is
always denoted by $\tau$. This function depends on infinite
set of independent variables which are called {\it flows} or
{\it times}. The last two words will be used as synonyms.
For each particular equation only a finite number of
the time variables take non-zero values.

The flows are labeled by points of the complex
plane ${\bf C}$.
Call points $\l
\in {\bf C}$ {\it labels}.
We make distinction between
{\it discrete} and {\it continuous} flows\footnote{These are not
more than conventional names. In general the both time
variables may take complex values.}.
\begin{itemize}
\item {\it Discrete flows:} With each ordered pair of
points $\l ,\mu \in {\bf C}$, $\l \neq \mu$,
a discrete flow $l=l_{\l \mu}$ is associated. To put it
differently, the flows are attached to vectors
$\overrightarrow{\l \mu}$,
i.e., each discrete flow has {\it two} labels.
\item {\it Continuous flows:} With each
point $\l \in {\bf C}$ an infinite sequence
of times
$\{ t_1 , t_2 , t_3 , \ldots \, \}^{(\l )}$ is associated. All the
variables $t_j$ have the common label $\l$.
\end{itemize}

In each particular equation which we are going to consider only
a finite number of labels are involved. To axiomatize this
situation, we say that
for all but a finite number of labels
$\l
\in {\bf C}$ and for all but
a finite number of ordered pairs of labels
the corresponding variables
are implied to be equal to zero. This condition makes
the definition very close to the adelic ideology from the
algebraic number theory. The definition in such an abstract
form may seem to be overcomplicated and too general.
However, this standpoint is useful
since it provides an adequate formalization of
the simple fact that the number of independent
variables in equations of an integrable
hierarchy can be arbitrary but finite.

Sometimes it is convenient to say that those variables
which are non-zero are {\it swithed on} while all others
are {\it swithed off}. According to the above definition,
the set of labels corresponding to the swithed on variables
is always finite.

Having this in mind, it is worthwhile to rephrase the definition
making it a little bit more concrete\footnote{For
each concrete example this unified notation is still not
very convenient to work with.
In the technical part of the paper this will
be changed and simplified.
However, for the sake of clarity and definiteness,
it is better to introduce general notions and definitions using
the unified notation.}.

Let $\{ \l _{\a}\}$ ,
$\a \in I$, be a finite set of marked points in
${\bf C}$. Here, $I$ is just the
finite set of labels corresponding to the variables that
are swithed on. By $l_{\a \b}$ ($\a \neq \b $) denote the discrete
variable associated with $\overrightarrow{\l _{\a}\l _{\b}}$.
By $t_{j}^{(\a )}$, $j=1,2, \ldots $ denote the continuous times
associated with $\l _{\a}$.
The $\tau$-function is a function of these variables:
$$
\tau = \tau \big ( \{ l_{ \a \b } \}; \{ t_{j}^{(\a )} \} \big ).
$$

Let ${\cal G}$ be the graph whose vertices are the marked points
(labels) $\l _{\a}$, $\a \in I$, and whose edges are vectors
$\overrightarrow{ \l _{\a} \l _{\b}}$. The edges have orientation
that is indicated by an arrow looking from $\a$ to $\b$. This
graph will be refered to as the {\it graph of flows}. It encodes
the kinematic structure of the equation.

We stress that the
only essential elements of the graph are vertices and their ordered
pairs. All other graphical elements are introduced for convenience
of the visualization. In particular, the vectors may intersect
on the complex plane but the intersection points
should be considered as not belonging
to the graph. It is also worth emphasizing that the vectors
are just convenient names of flows. They should not be mixed
with "directions" of the flows in any sense of this word.

The introduced variables are not independent. There are certain
"kinematical" constraints imposed on them.

The first group of constraints involves discrete variables only.
The constraints arise when the graph of flows
${\cal G}$ has cycles. It is enough to fix the constraints for
the following two cases:
\begin{itemize}
\item[i)] The 2-cycle:
\beq
\tau \big ( l_{\a \b }+1 ,
l_{\b \a }+1 \big ) =
\tau \big ( l_{\a \b },
l_{\b \a } \big ).
\label{2-cycle}
\eeq

\vspace{0.3cm}

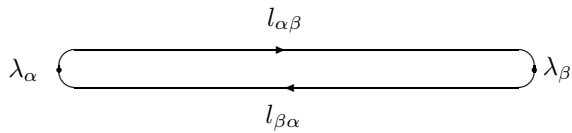
\begin{center}
\special{em:linewidth 0.4pt}
\unitlength 0.8mm
\linethickness{0.4pt}
\begin{picture}(102.83,30.90)
\put(59.50,22.83){\oval(79.00,6.33)[]}
\put(20.00,22.67){\circle*{0.67}}
\put(99.00,22.67){\circle*{0.67}}
\put(57.50,19.61){\vector(-1,0){0.2}}
\emline{58.41}{19.61}{1}{57.50}{19.61}{2}
\put(14.12,22.64){\makebox(0,0)[cc]{$\lambda _{\alpha}$}}
\put(102.83,22.64){\makebox(0,0)[cc]{$\lambda _{\beta}$}}
\put(57.54,30.90){\makebox(0,0)[cc]{$l_{\alpha \beta}$}}
\put(57.27,14.65){\makebox(0,0)[cc]{$l_{\beta \alpha}$}}
\put(57.56,25.98){\vector(1,0){0.2}}
\emline{55.19}{25.98}{3}{57.56}{25.98}{4}
\end{picture}
\end{center}

\noindent
Informally, this means that $l_{\b \a }$ is identified
with $-l_{\a \b }$.
\item[ii)] The 3-cycle:
\beq
\tau \big ( l_{\a \b }+1 ,\,
l_{\b \gamma }+1 ,\, l_{\gamma \a }+1  \big ) =
\tau \big ( l_{\a \b },\,
l_{\b \gamma }, \, l_{\gamma \a } \big ).
\label{3-cycle}
\eeq

\vspace{0.3cm}

\begin{center}
\special{em:linewidth 0.4pt}
\unitlength 1mm
\linethickness{0.4pt}
\begin{picture}(42.04,35.44)(10.0,0.0)
\emline{11.67}{19.33}{1}{34.33}{32.67}{2}
\emline{34.33}{32.67}{3}{39.67}{10.00}{4}
\emline{39.67}{10.00}{5}{12.00}{19.33}{6}
\put(22.55,25.77){\vector(3,2){0.2}}
\emline{22.09}{25.47}{7}{22.55}{25.77}{8}
\put(37.13,20.86){\vector(1,-2){0.2}}
\emline{36.97}{21.17}{9}{37.13}{20.86}{10}
\put(25.62,14.73){\vector(-3,1){0.2}}
\emline{26.08}{14.57}{11}{25.62}{14.73}{12}
\put(7.52,19.18){\makebox(0,0)[cc]{$\lambda _{\alpha}$}}
\put(34.21,35.44){\makebox(0,0)[cc]{$\lambda _{\beta}$}}
\put(42.04,8.28){\makebox(0,0)[cc]{$\lambda _{\gamma}$}}
\end{picture}
\end{center}
\end{itemize}
The corresponding rules for longer cycles follow from these two.
According to these rules, one can subsequently remove all the cycles
and reduce the graph to a tree. The tree graphs correspond to
kinematically independent flows. Formally, it is sufficient to
consider graphs without cycles.
However, introducing cycles sometimes makes the set
of variables more symmetric though non-minimal.

The last constraint describes the
interrelation between
a discrete flow
$\overrightarrow{ \l _{\a} \l _{\b}}$
and the "adjacent" continuous ones
(i.e., corresponding to the endpoints
$\l_{\a}$ and $\l_{\b}$):
\begin{itemize}
\item[iii)] Miwa's rule \cite{Miwa1},\,\cite{Miwa2}:
\beq
\tau \big ( l_{\a \b} +1; \,\, t^{(\a )}, t^{(\b )} \big )
=\tau \big ( l_{\a \b} ; \, t^{(\a )}
-[\l_{\b} -\l_{\a}], \, t^{(\b )} \big ),
\label{miwa1}
\eeq
\beq
\tau \big ( l_{\a \b} -1; \,\, t^{(\a )}, t^{(\b )} \big )
=\tau \big ( l_{\a \b} ; \,
t^{(\a )}, \,
t^{(\b )}
-[\l_{\a} -\l_{\b}]
\big ),
\label{miwa2}
\eeq

\vspace{0.3cm}

\begin{center}
\special{em:linewidth 0.4pt}
\unitlength 0.8mm
\linethickness{0.4pt}
\begin{picture}(96.33,15.33)
\emline{10.33}{9.67}{1}{89.67}{9.67}{2}
\put(11.00,9.67){\circle*{2.00}}
\put(89.67,9.67){\circle*{2.00}}
\put(50.00,9.67){\vector(1,0){0.2}}
\emline{49.67}{9.67}{3}{50.00}{9.67}{4}
\put(4.00,9.67){\makebox(0,0)[cc]{$\lambda _{\alpha}$}}
\put(96.33,9.67){\makebox(0,0)[cc]{$\lambda _{\beta}$}}
\put(50.00,15.33){\makebox(0,0)[cc]{$l_{\alpha \beta}$}}
\put(15.67,15.33){\makebox(0,0)[cc]{$t_{j}^{(\alpha )}$}}
\put(88.33,15.33){\makebox(0,0)[cc]{$t_{j}^{(\beta )}$}}
\end{picture}
\end{center}
\end{itemize}
Here $\tau (t)\equiv \tau (\{ t_{j} \})$ and the short-hand notation
\beq
f\left ( t\pm [z] \right )\equiv
f \left ( t_1 \pm z ,\,
t_2 \pm \textstyle{\frac{1}{2}}z^2 , \,
t_3 \pm \textstyle{\frac{1}{3}}z^3 , \,
\ldots \, \right )
\label{sh}
\eeq
is used for a function $f$ of the infinite sequence
of variables $t=\{ t_1 , t_2 , \ldots \, \}$. The second
relation (\ref{miwa2}) follows from the rule
(i) and the first one.

The relations (\ref{miwa1}), (\ref{miwa2}) should be understood
as formal rules which allow one to translate the infinite sequence
of continuous time shifts into the shift of a single discrete
variable and vice versa. We do not care about convergency of
the infinite substitutions, i.e., the $\tau$-function is
considered to be
a formal series in $\l $'s (in the left hand sides,
$\l $'s are implicitly present in the definition of discrete
flows). In known examples of algebro-geometric solutions,
the $\tau$-function is a {\it true function}, not
merely a formal series. In this case there are
some additional restrictions
to the domains of all variables and labels. They ensure
the convergency of the infinite substitutions. Meanwhile,
for algebro-geometric solutions elements of the graph ${\cal G}$
acquire a transparent interpretation on a Riemann surface as
punctures and cuts. Furthermore, the discrete time variables
discribe discontinuities of the Baker-Akhiezer function on the cuts.

We refer to the discrete flows
$\overrightarrow{ \l _{\a} \l _{\b}}$ as
{\it elementary discrete flows}. One may introduce more
complicated flows which can be thought of as
"superpositions" of the elementary ones.
Specifically, fix several
elementary flows, say, $l_1 , l_2 , \ldots , l_M$
(here $l_i \equiv l_{\a _{i} \b _{i}}$ for some
$\a _{i}, \b _{i}$) and consider
the $\tau$-function as a function of a new variable $y$ as
follows: $\tau [y]
\equiv \tau (l_1 +y ,\, l_2 +y,\, \ldots \, , l_M +y)$.
In the time evolution with respect to the new variable
$y$ the "elementary" variables $l_i$ simultaneously
get shifted by $y$ while the others are constants.
Let us call flows of this type
{\it composite discrete flows}.

To put it differently, let $\p _{i}$ be the vector field
corresponding to the flow $l_i$. Then the vector field
corresponding to the composite flow $y$ is
$\p _{y}:=\sum _{i=1}^{M}\p _{i}$, so
$$
\exp \left (\p _{y}\right )
\tau \big (\{ l_{\a _{i}\b _{i}}\}\big )=
\tau \big (\{ l_{\a _{i}\b _{i}} +1\}\big )
\exp \left (\p _{y}\right ).
$$
However, one should be careful since
due to (\ref{3-cycle}) the simultaneous shift of $l_{\a \b}$
and $l_{\b \gamma }$ is equivalent to an elementary flow.

The precise definition is as follows:
\begin{itemize}
\item {\it Composite discrete flows} are labeled by
finite sets of vectors
$\big \{
\overrightarrow{ \l _{\a _{i}} \l _{\b _{i}}} \big \}$,
$i=1,2, \ldots , M$ such that $\b _{i}\neq \a _{j}$ for any $i,j$.
Let $y$ be the corresponding time variable, then the evolution is
defined by
$$
\tau [y]= \tau \big (\{ l_{\a _{i}\b _{i}}+y \}\big ),
$$
where $l _{\a _{i} \b _{i}}$
and other elementary variables are supposed to be constants.
\end{itemize}

The distinction between elementary and composite flows can be
extended to the continuous flows, too.
For the reason which will be more clear later, it is natural
to consider the continuous
times $t_{1}^{(\a )}$ as elementary flows. At this stage
we motivate this definition by the fact that due to Miwa's rule
(\ref{miwa1})
they can be obtained as the result of a scaling limit
from discrete elementary flows. Similarly, higher continuous times
$t_{j}^{(\a )}$ with $j \geq 2$ are limits of composite
discrete flows. Therefore, we call them composite.

Let us summarize. We introduced several notions and definitions
which are extensively used throughout the paper. First of all,
a partial classification of flows and time variables has been
suggested. We have defined discrete and continuous flows and
distinguished between elementary and composite flows. To any
particular equation one may assigne a graph which explicitly
shows the kinematical structure of the equation and possible
constraints imposed on the flows.
In order to make this clear, we give some examples.

\subsection{Examples}

Here we illustrate the above notions by familiar
examples. To make the graphs of flows more informative, let us
add a new graphical element: fat dots mean that
the corresponding continuous times are
non-zero\footnote{Taking
into account
Miwa's rule (iii),
it would be more
precise to say that the continuous times can not be made equal to
zero
by a transformation of the form
(\ref{miwa1}), (\ref{miwa2})}.

In the KP hierarchy the graph of flows consists of one "fat" point
with corersponding continuous "times" $\{ t_j \}$.
The set of discrete flows is empty. The $\tau$-function is
$\tau (t) \equiv \tau (t_1 , t_2 , \ldots )$.

In the 2DTL hierarchy the graph of flows consists of two "fat"
points with the corresponding times $\{ t_j \}$ and
$\{ \bar t_j \}$. The discrete flow associated to the vector
connecting the two points is the discrete "time" $n$ of the 2DTL
$\tau$-function $\tau _{n} (t; \bar t)$.

\vspace{0.1cm}

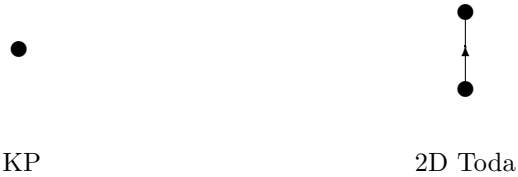
\begin{center}
\special{em:linewidth 0.4pt}
\unitlength 1mm
\linethickness{0.4pt}
\begin{picture}(81.00,26.33)(20.0,5.0)
\put(20.33,20.00){\circle*{2.00}}
\put(20.67,5.00){\makebox(0,0)[cc]{KP}}
\put(79.67,14.67){\circle*{2.00}}
\put(79.67,25.00){\circle*{2.00}}
\emline{79.67}{14.33}{1}{79.67}{25.00}{2}
\put(79.67,20.33){\vector(0,1){0.2}}
\emline{79.67}{20.33}{3}{79.67}{20.33}{4}
\put(79.67,5.00){\makebox(0,0)[cc]{2D Toda}}
\end{picture}
\end{center}

\vspace{0.2cm}

The graphs of flows for the discrete
KP and 2DTL equations are as follows:

\vspace{0.2cm}

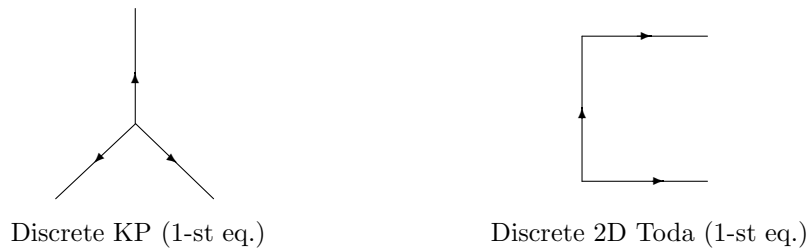
\begin{center}
\special{em:linewidth 0.4pt}
\unitlength 1.00mm
\linethickness{0.4pt}
\begin{picture}(96.33,35.33)
\emline{20.33}{20.00}{1}{9.67}{10.00}{2}
\emline{20.33}{20.00}{3}{30.67}{10.00}{4}
\emline{20.26}{20.00}{5}{20.26}{35.33}{6}
\put(15.00,15.00){\vector(-1,-1){0.2}}
\emline{15.67}{15.67}{7}{15.00}{15.00}{8}
\put(20.26,26.67){\vector(0,1){0.2}}
\emline{20.26}{26.00}{9}{20.26}{26.67}{10}
\put(20.67,4.33){\makebox(0,0)[cb]{Discrete KP (1-st eq.)}}
\emline{79.67}{12.33}{11}{79.67}{31.67}{12}
\emline{79.67}{31.67}{13}{96.33}{31.67}{14}
\emline{79.67}{12.33}{15}{96.33}{12.33}{16}
\put(88.67,31.67){\vector(1,0){0.2}}
\emline{87.00}{31.67}{17}{88.67}{31.67}{18}
\put(79.67,22.00){\vector(0,1){0.2}}
\emline{79.67}{20.67}{19}{79.67}{22.00}{20}
\put(88.67,4.33){\makebox(0,0)[cb]{Discrete 2D Toda (1-st eq.)}}
\put(25.65,14.92){\vector(4,-3){0.2}}
\emline{24.89}{15.52}{21}{25.65}{14.92}{22}
\put(90.34,12.33){\vector(1,0){0.2}}
\emline{89.00}{12.33}{23}{90.34}{12.33}{24}
\end{picture}
\end{center}

\vspace{0.2cm}

\noindent
All continuous times are swithed off.
In both cases only three independent discrete flows are
swithed on.
This agrees with the continuous case where
first non-trivial equations of the KP and 2DTL hierarchies
have three independent variables $t_1 , t_2 , t_3$ and
$t_1$, $\bar t _{1}$, $n$, respectively.
In the continuum limit, all the lines except the vertical one
in the 2DTL figure shrink up to fat points.

The discretized KP and 2DTL {\it hierarchies} are illustrated
by the following graphs:

\begin{center}
\special{em:linewidth 0.4pt}
\unitlength 1.00mm
\linethickness{0.4pt}
\begin{picture}(148.33,70.33)
\emline{40.00}{40.00}{1}{40.00}{70.33}{2}
\emline{40.00}{40.00}{3}{51.67}{67.00}{4}
\emline{40.00}{40.67}{5}{29.00}{66.67}{6}
\emline{40.00}{40.33}{7}{23.33}{25.67}{8}
\emline{39.33}{40.00}{9}{59.33}{25.00}{10}
\emline{40.67}{40.00}{11}{65.00}{33.00}{12}
\emline{40.33}{40.00}{13}{29.00}{19.33}{14}
\emline{40.00}{40.00}{15}{17.00}{33.67}{16}
\emline{40.00}{39.33}{17}{55.00}{17.33}{18}
\put(19.33,56.67){\circle*{0.00}}
\put(17.33,51.00){\circle*{0.00}}
\put(16.33,45.33){\circle*{0.00}}
\put(59.33,57.33){\circle*{0.00}}
\put(63.00,51.33){\circle*{0.00}}
\put(64.67,43.67){\circle*{0.00}}
\put(36.33,16.67){\circle*{0.00}}
\put(41.33,15.00){\circle*{0.00}}
\put(46.33,16.33){\circle*{0.00}}
\put(34.97,30.28){\vector(-3,-4){0.2}}
\emline{35.41}{30.87}{19}{34.97}{30.28}{20}
\put(31.85,33.10){\vector(-3,-4){0.2}}
\emline{32.59}{33.99}{21}{31.85}{33.10}{22}
\put(29.03,36.81){\vector(-2,-1){0.2}}
\emline{30.66}{37.56}{23}{29.03}{36.81}{24}
\put(33.93,54.79){\vector(-2,3){0.2}}
\emline{34.82}{53.45}{25}{33.93}{54.79}{26}
\put(40.02,57.16){\vector(0,1){0.2}}
\emline{40.02}{55.53}{27}{40.02}{57.16}{28}
\put(46.55,55.38){\vector(1,2){0.2}}
\emline{46.11}{54.34}{29}{46.55}{55.38}{30}
\put(51.31,36.96){\vector(4,-1){0.2}}
\emline{50.42}{37.11}{31}{51.31}{36.96}{32}
\put(49.97,32.06){\vector(3,-2){0.2}}
\emline{48.63}{32.95}{33}{49.97}{32.06}{34}
\put(48.04,27.75){\vector(2,-3){0.2}}
\emline{47.15}{29.09}{35}{48.04}{27.75}{36}
\emline{123.89}{32.96}{37}{123.89}{53.33}{38}
\emline{124.26}{52.96}{39}{148.33}{56.30}{40}
\emline{123.89}{53.70}{41}{143.89}{60.74}{42}
\emline{123.89}{53.33}{43}{146.48}{50.74}{44}
\emline{123.52}{53.33}{45}{103.14}{50.37}{46}
\emline{123.52}{53.33}{47}{101.66}{58.89}{48}
\emline{123.14}{53.70}{49}{107.59}{63.70}{50}
\emline{123.52}{32.59}{51}{102.77}{36.30}{52}
\emline{122.77}{32.22}{53}{103.89}{28.89}{54}
\emline{123.14}{32.22}{55}{107.22}{22.96}{56}
\emline{123.14}{32.22}{57}{144.26}{34.07}{58}
\emline{123.14}{32.22}{59}{144.26}{26.30}{60}
\emline{123.55}{32.11}{61}{142.43}{21.37}{62}
\put(118.33,65.56){\circle*{0.00}}
\put(124.63,66.30){\circle*{0.00}}
\put(129.81,65.56){\circle*{0.00}}
\put(116.11,18.15){\circle*{0.00}}
\put(123.89,16.30){\circle*{0.00}}
\put(130.18,17.04){\circle*{0.00}}
\put(133.06,56.95){\vector(3,1){0.2}}
\emline{131.33}{56.33}{63}{133.06}{56.95}{64}
\put(135.25,54.44){\vector(1,0){0.2}}
\emline{133.68}{54.44}{65}{135.25}{54.44}{66}
\put(135.88,51.93){\vector(1,0){0.2}}
\emline{134.00}{52.09}{67}{135.88}{51.93}{68}
\put(114.08,59.62){\vector(-4,3){0.2}}
\emline{115.18}{58.84}{69}{114.08}{59.62}{70}
\put(112.67,56.17){\vector(-4,1){0.2}}
\emline{113.92}{55.86}{71}{112.67}{56.17}{72}
\put(111.26,51.31){\vector(-4,-1){0.2}}
\emline{112.35}{51.62}{73}{111.26}{51.31}{74}
\put(134.00,33.11){\vector(1,0){0.2}}
\emline{132.74}{32.96}{75}{134.00}{33.11}{76}
\put(134.00,29.19){\vector(4,-1){0.2}}
\emline{132.59}{29.51}{77}{134.00}{29.19}{78}
\put(133.29,26.43){\vector(3,-2){0.2}}
\emline{131.56}{27.53}{79}{133.29}{26.43}{80}
\put(114.24,27.15){\vector(-2,-1){0.2}}
\emline{115.33}{27.62}{81}{114.24}{27.15}{82}
\put(111.26,30.13){\vector(-1,0){0.2}}
\emline{112.82}{30.29}{83}{111.26}{30.13}{84}
\put(110.79,34.84){\vector(-4,1){0.2}}
\emline{112.04}{34.52}{85}{110.79}{34.84}{86}
\put(123.96,42.52){\vector(0,1){0.2}}
\emline{123.96}{41.27}{87}{123.96}{42.52}{88}
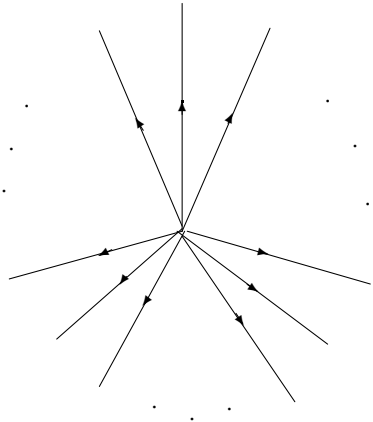
\put(40.74,4.81){\makebox(0,0)[cb]{Discrete KP (the hierarchy)}}
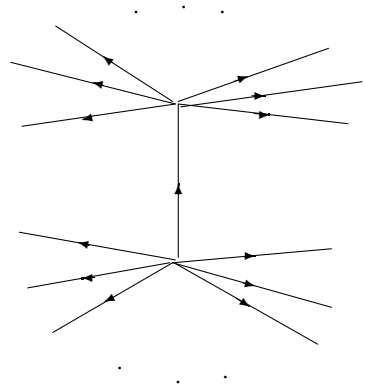
\put(124.63,4.81){\makebox(0,0)[cb]{Discrete 2D Toda (the hierarchy)}}
\end{picture}
\end{center}

\noindent
Higher equations of the hierarchies involve
more than three elementary
flows. Labels of these flows are analogous to the number of
a higher flow in the continuous hierarchies.
The labels are complex numbers. This looks like a kind of
Fourier duality between a parameter marking equations of the
hierarchy and the time variable corresponding to a particular
flow: continuous flows are marked by a discrete "label" whereas
discrete flows are marked by a continuous label.

\section{Discrete Zakharov-Shabat representation of Hirota's
equation}

The reformulation of classical nonlinear integrable equations as
flatness conditions for a two-dimensional connection is the basic
constituent of the theory. The flatness means that subsequent shifts
along any pair of the time flows commute. These conditions are known as
{\it Zakharov-Shabat equations} or {\it zero curvature representation}.
In the paper \cite{Hirota1} R.Hirota gave an example of the
discretized zero
curvature representation for eq.\,(\ref{HBDE1}). In the physical
language, the discrete connection is a lattice gauge field. The
approach emphasizing the relation to gauge field theories on the
lattice was developed by S.Saito and N.Saitoh \cite{SS}.
We present these results in a modified
form which makes the theory
completely parallel to the 2DTL theory \cite{UT}.

The discrete zero curvature condition is equivalent to
commutativity of certain multivariable difference operators.
The existence of such a "commutation representation"
is a hall-mark of itegrability. At the same time, if a commutation
representation exists, it is not unique.
In particular,
there are different (in fact infinitely many) ways to represent
HBDE as a zero curvature condition.

The general scheme is as follows. Choose {\it any} time flow to be the
"reference"
one, i.e., the one in which all the $M$-operators are going to act as
differential or difference operators.
Commutativity of
the flows means that any pair of such $M$-operators obeys a
compatibility condition which is just
one of the Zakharov-Shabat equations.
This fact allows one to relate different hierarchies to
each other. In general
$M$-operators are pseudo-difference\footnote{We use
this shorter name for what is usually called
"quantum pseudo-differential operator".}
or difference operators with matrix
coefficients.
Here we consider only the case of
difference operators. More general examples are given
later in Sect.6.
When the reference flow is taken to be an elementary one,
the coefficients are scalar functions. Sect.\,4.3 contains an example
of the zero curvature condition for
HBDE realized by $2\times2$-matrix
difference operators.

\subsection{Basic $M$-operators}

So far all elementary discrete flows enjoyed equal rights.
No one of them was better than any other one.
We are going to break this equality of rights and distinguish a
{\it reference flow}. It may be {\it any} flow
including composite and continuous ones.
For simplicity, we start with the case when
the reference flow is discrete and elementary. Other cases are discussed
later.

The idea is to assign difference operators to flows.
These operators act to functions of the reference flow
variable. We call them {\it $M$-operators}.
In this section we consider the simplest $M$-operators
which are basic blocks of more general ones.

Let us specify the notation and take the reference
flow to be $\overrightarrow{\l _{0}\l_{1}}$. The double
index notation is inconvenient for practical purposes.
Dealing with a limited number of flows, it is worthwhile to give
them simpler though less systematic names. Unless otherwise
stated, the letter $u$
will be reserved for the reference variable corresponding
to an elementary discrete flow.
So we set $$u=l_{01}.$$

Let $\l _{2}$
be any label different from $\l _{0} , \l _{1}$.
In this situation, the graph of flows is a triangle:

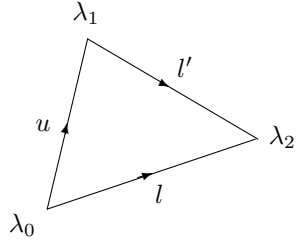
\begin{center}
\special{em:linewidth 0.4pt}
\unitlength 1.00mm
\linethickness{0.4pt}
\begin{picture}(37.60,31.49)
\emline{34.33}{14.66}{1}{11.67}{28.00}{2}
\emline{11.67}{28.00}{3}{6.33}{5.33}{4}
\emline{6.33}{5.33}{5}{34.00}{14.66}{6}
\put(20.15,10.08){\vector(3,1){0.2}}
\emline{18.35}{9.54}{7}{20.15}{10.08}{8}
\put(3.06,3.24){\makebox(0,0)[cc]{$\lambda _{0}$}}
\put(11.51,31.49){\makebox(0,0)[cc]{$\lambda _1$}}
\put(37.60,14.93){\makebox(0,0)[cc]{$\lambda _2$}}
\put(5.76,16.55){\makebox(0,0)[cc]{$u$}}
\put(24.65,24.29){\makebox(0,0)[cc]{$l'$}}
\put(21.23,7.38){\makebox(0,0)[cc]{$l$}}
\put(8.82,16.10){\vector(1,4){0.2}}
\emline{8.71}{15.28}{9}{8.82}{16.10}{10}
\put(22.35,21.74){\vector(2,-1){0.2}}
\emline{21.63}{22.15}{11}{22.35}{21.74}{12}
\end{picture}
\end{center}

\noindent
Its sides
$\overrightarrow{\l _{0}\l_{2}}$
and $\overrightarrow{\l _{1}\l_{2}}$ define flows which we call
adjacent
to the reference flow $u$ in the obvious sense.
In general, a flow
$\overrightarrow{\l _{\a }\l_{\gamma }}$
(resp.,
$\overrightarrow{\l _{\b }\l_{\gamma '}}$) is said to be
{\it left adjacent} (resp., {\it right adjacent}) to the flow
$\overrightarrow{\l _{\a }\l_{\b }}$.

Coming back to the triangle graph of flows, we set
$l_{02}=l$, $l_{12}=l'$.
The $\tau$-function will be
denoted by $\tau _{u}^{l, l'}$.
There are only two independent flows. According to (ii), we have
\beq
\tau _{u+1}^{l,l' +1}= \tau _{u}^{l+1 , l'}.
\label{ul}
\eeq

Coefficients of $M$-operators are expressed through $\tau$.
Let us take $u,l$ as independent variables.
By definition,
the $M$-operator assigned to the left adjacent flow $l$ is
\beq
M^{l}_{u}=e^{\p _{u}}-\lambda _{2}^{01}
\frac{ \tau ^{l}_{u}\tau ^{l+1}_{u+1}}
{ \tau ^{l+1}_{u}\tau ^{l}_{u+1}} ,
\label{Z0}
\eeq
where the coefficient $\l _{2}^{01}$ is expressed through the
three labels:
\beq
\l _{2}^{01}=\frac{1}{\l _{2} -\l _{0}}
-\frac{1}{\l _{1} -\l _{0}}, \;\;\;\;\;\;\;\;\;
\l _{1}^{02}=-
\l _{2}^{01}.
\label{lambda}
\eeq
The shift operator $e^{\p _{u}}$ has standard commutation relations
with functions of $u$: $e^{\pm \p _{u}}f(u)=f(u\pm 1)e^{\pm \p _{u}}$.
Note that $M_{u}^{u}=e^{\p _{u}}$.
It is implied that the $\tau$-function in (\ref{Z0}) might depend
on all the other variables which are swithed off in this particular
case. When they are swithed on, they enter eq.\,(\ref{Z0})
as parameters.
Their values are the same for each of the four $\tau$-functions
in the ratio. As a rule, we do not indicate them
explicitly if this does not lead to a confusion.

Once the $M$-operator for the left adjacent flow is
written, it can be translated into the one for the right
adjacent flow $l'$ by passing to independent variables $u, l'$.
In eq.\,(\ref{Z0}), the (implicit) argument $l'$ is the same
in each $\tau$-function. Using (\ref{ul}), we rewrite them
in such a way that $l$ is the same and implicit.
The rule (ii) tells us that the shift $l' \rightarrow l' +1$
is equivalent the simultaneous
shifts
$u \rightarrow u+1$
and $l \rightarrow l+1$.
Thinking about
$M$-operators as generating shifts of the discrete variables
by 1, it is natural to {\it define} the $M$-operator for the
right adjacent flow $l'$ as follows:
$$
\bar M _{u}^{l'}=e^{-\p _{u}}M_{u}^{l},
$$
or, more explicitly,
\beq
\bar M^{l'}_{u}=1-\l _{2}^{01}
\frac{ \tau ^{l'+1}_{u+1}\tau ^{l'}_{u-1} }
{ \tau ^{l' +1}_{u}\tau ^{l'}_{u}}
e^{-\p _{u}}\,.
\label{Z0a}
\eeq

It is also useful to introduce
\beq
{\cal M}_{u}^{l}=e^{-\p _{l}}M_{u}^{l},
\;\;\;\;\;\;
\bar {\cal M}_{u}^{l'}=e^{-\p _{l'}}\bar M_{u}^{l'}
\label{Z0b}
\eeq
which are difference operators in two variables.
It trivially follows from the
construction that they commute:
\beq [ {\cal M}_{u}^{l},\, \bar {\cal M}_{u}^{l'} ]=0\, .
\label{comm1}
\eeq

We have defined $M$-operators for elementary discrete
flows adjacent to the
reference one. In this case they have especially simple form.
They are first order difference operators in $u$. $M$-operators
corresponding to composite and non-adjacent flows have more
complicated structure.

Let us comment on continuous reference flows.
According to (\ref{miwa1}), the continuum limit in $u$
means $e^{\p _{u}}\rightarrow 1-\l ^{-1}\p _{t_1}+O(\l ^{-2})$,
where $t_1$ is the first continuous flow with labels $\l _{0}$
and $\l =(\l _{1}-\l _{0})^{-1}$. The limiting form of the
$M$-operator (\ref{Z0}) as $\l \rightarrow \infty$ is
$$
M^{(l)}=\p _{t_1} -\p _{t_1}\log \frac{\tau ^{l+1}}{\tau ^{l}}
-(\l _{2}-\l _{0})^{-1}\,.
$$
This is a first order differential operator in the reference
continuous time veriable $t_1$. It generates shifts in the
discrete variable $l$. The fact that it is an operator
of the {\it first} order suggests to call the
continuous flow $t_1$ {\it elementary} (see the end of Sect.\,3.1).

\subsection{Discrete Zakharov-Shabat equations}

Two independent flows are not enough for deriving bilinear
equations. Non-trivial bilinear equations for $\tau$ arise
starting from 3 independent discrete flows, in which case
the graph of flows should contain at least 4 vertices.
So let us fix 4 labels $\l _{\a }$, $\a =0,1,2,3$ and consider
the general graph with 4 vertices:

\vspace{0.3cm}

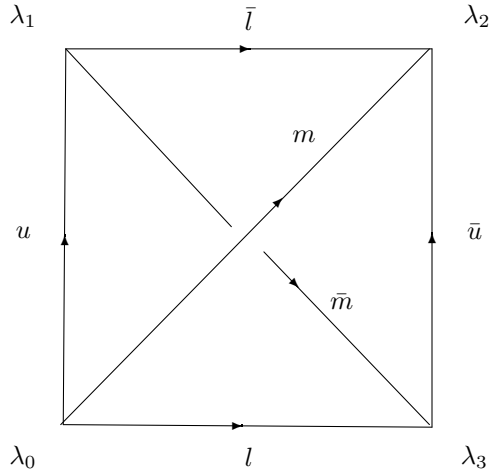
\begin{center}
\special{em:linewidth 0.4pt}
\unitlength 1.00mm
\linethickness{0.4pt}
\begin{picture}(65.67,64.33)
\emline{10.67}{10.00}{1}{59.67}{9.67}{2}
\emline{59.67}{9.67}{3}{59.67}{60.00}{4}
\emline{59.67}{60.00}{5}{11.00}{60.00}{6}
\emline{11.00}{60.00}{7}{10.67}{10.00}{8}
\emline{10.33}{10.00}{9}{59.33}{60.00}{10}
\emline{11.00}{60.00}{11}{33.00}{36.33}{12}
\emline{59.33}{10.00}{13}{37.33}{33.00}{14}
\put(35.33,60.00){\vector(1,0){0.2}}
\emline{35.00}{60.00}{15}{35.33}{60.00}{16}
\put(59.67,35.33){\vector(0,1){0.2}}
\emline{59.67}{35.00}{17}{59.67}{35.33}{18}
\put(39.67,40.00){\vector(1,1){0.2}}
\emline{39.00}{39.33}{19}{39.67}{40.00}{20}
\put(42.67,48.33){\makebox(0,0)[cc]{$m$}}
\put(47.67,26.33){\makebox(0,0)[cc]{$\bar m$}}
\put(35.33,5.67){\makebox(0,0)[cc]{$l$}}
\put(35.33,64.00){\makebox(0,0)[cc]{$\bar l$}}
\put(5.33,35.67){\makebox(0,0)[cc]{$u$}}
\put(65.33,35.67){\makebox(0,0)[cc]{$\bar u$}}
\put(5.33,5.67){\makebox(0,0)[cc]{$\lambda _0$}}
\put(65.33,5.67){\makebox(0,0)[cc]{$\lambda _3$}}
\put(5.33,64.33){\makebox(0,0)[cc]{$\lambda _1$}}
\put(65.67,64.33){\makebox(0,0)[cc]{$\lambda _2$}}
\put(10.86,35.09){\vector(0,1){0.2}}
\emline{10.86}{35.02}{21}{10.86}{35.09}{22}
\put(34.32,9.86){\vector(1,0){0.2}}
\emline{34.10}{9.86}{23}{34.32}{9.86}{24}
\put(41.81,28.31){\vector(1,-1){0.2}}
\emline{41.63}{28.50}{25}{41.81}{28.31}{26}
\end{picture}
\end{center}

\vspace{0,2cm}

\noindent
The simplified notation for the flows is clear from the
picture.
Like in (\ref{lambda}), we set
\beq
\l _{\gamma }^{\a \b}=\frac{1}{\l _{\gamma } -\l _{\a }}
-\frac{1}{\l _{\b } -\l _{\a }}, \;\;\;\;\;\;\;\;\;
\l _{\b }^{\a \gamma }=-
\l _{\gamma }^{\a \b }
\label{lambda1}
\eeq
for all possible values of the pairwise distinct indices.

Let $\overrightarrow{\l _{0}\l _{1}}$ be the reference flow,
as before. The left (resp., right) adjacent flows are
$\overrightarrow{\l _{0}\l _{2}}$,
$\overrightarrow{\l _{0}\l _{3}}$
(resp.,
$\overrightarrow{\l _{1}\l _{2}}$,
$\overrightarrow{\l _{1}\l _{3}}$).
Each of them has its own $M$-operator of the form
(\ref{Z0}) (resp., (\ref{Z0a})).

The key point is to extend the trivial commutation
(\ref{comm1}) to {\it all} the flows in the graph adjacent to $u$:
\beq [ {\cal M}_{u}^{l},\,
{\cal M}_{u}^{m} ]=[ {\cal M}_{u}^{l},\,
\bar {\cal M}_{u}^{\bar l} ]=[ \bar {\cal M}_{u}^{\bar l},\,
\bar {\cal M}_{u}^{\bar m} ]=0\, .
\label{comm2}
\eeq
In contrast to eq.\,(\ref{comm1}), these are non-trivial
requirements which give bilinear equations for $\tau$.
Written in terms of $M$-operators, commutation relations
(\ref{comm2}) are discrete Zakharov-Shabat equations.

In the following proposition we use the notation like
$M_{u}^{l}=M_{u}^{l}(u,l, \bar l , \ldots )$ to indicate
the dependence of $M$-operators on the variables.
\begin{prop}
The discrete Zakharov-Shabat equations
\beq
M^{m}_{u}(m,l+1)
M^{l}_{u}(m,l)=
M^{l}_{u}(m+1,l)
M^{m}_{u}(m,l)\,,
\label{Z3}
\eeq
\beq
\bar M^{\bar m}_{u}(\bar m,\bar l+1)
\bar M^{\bar l}_{u}(\bar m,\bar l)=
\bar M^{\bar l}_{u}(\bar m+1,\bar l)
\bar M^{\bar m}_{u}(\bar m,\bar l)\,,
\label{Z4}
\eeq
\beq
\bar M^{\bar l}_{u}(l+1,\bar l)
M^{l}_{u}(l,\bar l)=
M^{l}_{u}(l,\bar l +1)
\bar M^{\bar l}_{u}(l,\bar l)
\label{Z5}
\eeq
are equivalent to the following bilinear relations for $\tau$:
\beq
\lambda _{2}^{01} \tau ^{l+1, m}_{u}
\tau ^{l, m+1}_{u+1}-
\lambda _{3}^{01} \tau ^{l, m+1}_{u}
\tau ^{l+1, m}_{u+1}+
H_{1}(l,m;u)\tau ^{l+1, m+1}_{u}
\tau ^{l, m}_{u+1}=0\,,
\label{Z8}
\eeq
\beq
\lambda _{3}^{01} \tau ^{\bar l+1, \bar m}_{u}
\tau ^{\bar l, \bar m+1}_{u+1}-
\lambda _{2}^{01} \tau ^{\bar l, \bar m+1}_{u}
\tau ^{\bar l+1, \bar m}_{u+1}+
H_{2}(\bar l, \bar m ;u)\tau ^{\bar l, \bar m}_{u}
\tau ^{\bar l +1, \bar m +1}_{u+1}=0\,,
\label{Z8a}
\eeq
\beq
\lambda _{3}^{01} \tau ^{l, \bar l +1}_{u}
\tau ^{l+1, \bar l }_{u}-
\lambda _{2}^{01} \tau ^{l, \bar l +1}_{u+1}
\tau ^{l+1, \bar l}_{u-1}+
H_{3}(l, \bar l ;u)\tau ^{l, \bar l}_{u}
\tau ^{l +1, \bar l +1}_{u}=0\,,
\label{Z8b}
\eeq
respectively, where $H_i$ are
arbitrary functions such that
$H_i (l,m; u+1)=H_i (l,m ; u)$.
\end{prop}
{\it Proof.} The proof consists in the straightforward commutation
of $M$-operators.
The $M$-operators read:
\beq
M^{l}_{u}=e^{\p _{u}}- \l _{3}^{01}
\frac{ \tau ^{l}_{u}\tau ^{l+1}_{u+1}}
{ \tau ^{l+1}_{u}\tau ^{l}_{u+1}},
\label{Z1}
\eeq
\beq
\bar M^{\bar m}_{u}=1-\l _{3}^{01}
\frac{ \tau ^{\bar m}_{u-1}\tau ^{\bar m+1}_{u+1}}
{ \tau ^{\bar m+1}_{u}\tau ^{\bar m}_{u}}
e^{-\p _{u}}\,,
\label{Z2}
\eeq
$M_{u}^{m}$ is given by (\ref{Z1}) with the changes
$l\rightarrow m$ and $3\rightarrow 2$,
$\bar M_{u}^{\bar l}$ is given by (\ref{Z2}) with the changes
$\bar m \rightarrow \bar l$ and $3\rightarrow 2$.
The details for eq.\,(\ref{Z3}) are given below.

Eq.\,(\ref{Z3})
reads:
\begin{eqnarray}
&&\left (e^{\p _{u}}-\l _{2}^{01}
\frac{ \tau ^{m,l+1}_{u}\tau ^{m+1, l+1}_{u+1} }
{ \tau ^{m+1, l+1}_{u}\tau ^{m, l+1}_{u+1} } \right )
\left (e^{\p _{u}}-\lambda _{3}^{01}
\frac{ \tau ^{m,l}_{u}\tau ^{m, l+1}_{u+1} }
{ \tau ^{m, l+1}_{u}\tau ^{m, l}_{u+1} } \right )
\nonumber \\
&=&\left (e^{\p _{u}}-\l _{3}^{01}
\frac{ \tau ^{m+1,l}_{u}\tau ^{m+1, l+1}_{u+1} }
{ \tau ^{m+1, l+1}_{u}\tau ^{m+1, l}_{u+1} } \right )
\left (e^{\p _{u}}-\lambda _{2}^{01}
\frac{ \tau ^{m,l}_{u}\tau ^{m+1, l}_{u+1} }
{ \tau ^{m+1, l}_{u}\tau ^{m, l}_{u+1} } \right ).
\label{Z6}
\end{eqnarray}
The terms $e^{2\p _{u}}$ and those which do not contain shift
operators cancel automatically. The comparison of the coefficients in
front of $e^{\p _{u}}$ yields
\begin{eqnarray}
&&\frac{ \tau ^{m,l}_{u+1}}{\tau ^{m,l}_{u+2}}
\left ( \lambda _{3}^{01}
\frac{ \tau ^{m,l+1}_{u+2}}{\tau ^{m,l+1}_{u+1}}-
\lambda _{2}^{01}
\frac{ \tau ^{m+1,l}_{u+2}}{\tau ^{m+1,l}_{u+1}}
\right )= \nonumber \\
&=&\frac{ \tau ^{m+1,l+1}_{u+1}}{\tau ^{m+1,l+1}_{u}}
\left ( \lambda _{3}^{01}
\frac{ \tau ^{m+1,l}_{u}}{\tau ^{m+1,l}_{u+1}}-
\lambda _{2}^{01}
\frac{ \tau ^{m,l+1}_{u}}{\tau ^{m,l+1}_{u+1}}
\right ),
\label{a}
\end{eqnarray}
or
\beq
\frac{
\lambda _{3}^{01}\tau ^{m+1, l}_{u}\tau ^{m, l+1}_{u+1}-
\lambda _{2}^{01}\tau ^{m, l+1}_{u}\tau ^{m+1, l}_{u+1}}
{\lambda _{3}^{01}\tau ^{m+1, l}_{u+1}\tau ^{m, l+1}_{u+2}-
\lambda _{2}^{01}\tau ^{m, l+1}_{u+1}\tau ^{m+1, l}_{u+2}}=
\frac{
\tau ^{m,l}_{u+1}\tau ^{m+1, l+1}_{u}}
{\tau ^{m,l}_{u+2}\tau ^{m+1, l+1}_{u+1}}.
\label{Z7}
\eeq
The denominators in both sides differ from the numerators by the
shift $u\rightarrow u+1$.
Therefore, their ratio is a
"quasiconstant" in $u$, so the equation is equivalent to
eq.\,(\ref{Z8}). This completes the proof.

Now we restore the
"equality of rights" of elementary flows by imposing
the requirement that Zakharov-Shabat equations should hold for
{\it any} choice of the reference flow. For example, let $l$
be the reference flow. Construct ${\cal M}$-operators for the
flows $u, \bar u$ adjacent to $l$ (see the figure). Then we
require that the operators
${\cal M}_{l}^{u}$,
${\cal M}_{l}^{m}$
$\bar {\cal M}_{l}^{\bar u}$,
$\bar {\cal M}_{l}^{-\bar m}$ commute with each other
(of course some of them commute automatically due to (\ref{comm1})).
Note, however, that ${\cal M}$-operators constructed with respect
to different reference flows are not required to be commuting,
e.g.
$[ \bar {\cal M}_{u}^{\bar m},\,
{\cal M}_{\bar l}^{\bar m} ] \neq 0$.

\begin{th}
Let $x$ be any one of the elementary flows shown in the picture and
let $v, \bar v$ be the corresponding left and right adjacent flows
such that $x, v , \bar v$ are independent. Then the commutativity
conditions
$$[{\cal M}_{x}^{v}, \, \bar {\cal M}_{x}^{\bar v}]=0$$
imposed simultaneously
for any three independent reference flows $x$ are equivalent
to the equations
\beq
\lambda _{1}^{03} \tau ^{l, m+1}_{u}
\tau ^{l+1, m}_{u+1}+
\lambda _{2}^{01} \tau ^{l+1, m}_{u}
\tau ^{l, m+1}_{u+1}+
\l _{3}^{02} \tau ^{l+1, m+1}_{u}
\tau ^{l, m}_{u+1}=0\,,
\label{hir1}
\eeq
\beq
\lambda _{1}^{02} \tau ^{\bar l, \bar m+1}_{u}
\tau ^{\bar l+1, \bar m}_{u+1}+
\l _{2}^{03} \tau ^{\bar l, \bar m}_{u}
\tau ^{\bar l +1, \bar m +1}_{u+1}+
\lambda _{3}^{01} \tau ^{\bar l+1, \bar m}_{u}
\tau ^{\bar l, \bar m+1}_{u+1}=0\,,
\label{hir2}
\eeq
\beq
\lambda _{3}^{01} \tau ^{l, \bar l +1}_{u}
\tau ^{l+1, \bar l }_{u}-
\lambda _{3}^{02} \tau ^{l, \bar l }_{u}
\tau ^{l+1, \bar l +1}_{u}=
\l _{2}^{01} \tau ^{l, \bar l +1}_{u+1}
\tau ^{l +1, \bar l }_{u-1}\, .
\label{hir3}
\eeq
\end{th}
{\it Sketch of proof.} By virtue of the previous proposition,
it is enough to show that the functions $H_i$ are constants:
$H_1 = -H_2 =- H_3 =\l _{3}^{02}$. This can be done straightforwardly
by writing down bilinear equations arising from Zakharov-Shabat
equations for $M$-operators corresponding to each choice of
the reference flow and demanding their consistency with each other.

We see that equations (\ref{hir1}), (\ref{hir3}) coincide with
the KP and Toda-like forms of HBDE (\ref{HBDE4}), (\ref{HBDE6}),
respectively. (Eq.\,(\ref{hir2}) coincides with the KP-like
form after the change $u\rightarrow -u$.)
The three equations differ by the choice of the
independent variables only that agrees with substitutions
(\ref{lin1})-(\ref{lin3a}).
The transition from one triple of independent variables to
another should be done according to the rules (i) and (ii)
(see (\ref{2-cycle}), (\ref{3-cycle})). Using these rules, it is
easy to see that the three equations (\ref{hir1})-(\ref{hir2})
are equivalent to each other.

The 4-variable MKP-like form (\ref{HBDE4a}) of the
Hirota equation follows from (\ref{hir1}) by applying
the rule (\ref{3-cycle}). Namely, fix an extra label
$\mu _{0}$ and consider the flows
$\overrightarrow{\mu _{0}\l _{\a }}$, $\a = 0, \ldots \, 3$
with time variables $p_{\a }$.
From (\ref{3-cycle}) we have, for instance,
$\tau _{u, p_0 +1}^{m+1, p_2 }=\tau _{u, p_0 }^{m, p_2 +1}$
and so on. This change of variables converts equation
(\ref{hir1}) into eq.\,(\ref{HBDE4a}).

\subsection{Matrix realization of the zero curvature condition}

We restrict ourselves by giving an example which
illustrate the general scheme outlined in the
introduction to this section.

Consider the graph of flows which is a reduced
version of the one from Sect.\,4.2:

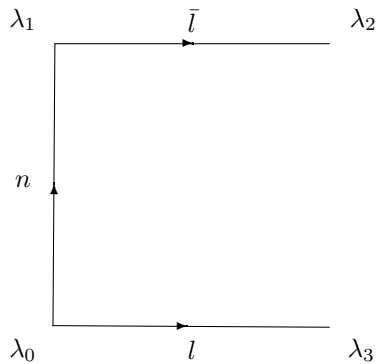
\begin{center}
\special{em:linewidth 0.4pt}
\unitlength 0.75mm
\linethickness{0.4pt}
\begin{picture}(65.67,64.33)
\emline{10.67}{10.00}{1}{59.67}{9.67}{2}
\emline{59.67}{60.00}{3}{11.00}{60.00}{4}
\emline{11.00}{60.00}{5}{10.67}{10.00}{6}
\put(35.33,60.00){\vector(1,0){0.2}}
\emline{35.00}{60.00}{7}{35.33}{60.00}{8}
\put(35.33,5.67){\makebox(0,0)[cc]{$l$}}
\put(35.33,64.00){\makebox(0,0)[cc]{$\bar l$}}
\put(5.33,35.67){\makebox(0,0)[cc]{$n$}}
\put(5.33,5.67){\makebox(0,0)[cc]{$\lambda _0$}}
\put(65.33,5.67){\makebox(0,0)[cc]{$\lambda _3$}}
\put(5.33,64.33){\makebox(0,0)[cc]{$\lambda _1$}}
\put(65.67,64.33){\makebox(0,0)[cc]{$\lambda _2$}}
\put(10.86,35.09){\vector(0,1){0.2}}
\emline{10.86}{35.02}{9}{10.86}{35.09}{10}
\put(34.32,9.86){\vector(1,0){0.2}}
\emline{34.10}{9.86}{11}{34.32}{9.86}{12}
\end{picture}
\end{center}

The variables which we do not need here are swithed off.
The simplified ad hoc notation is clear from the figure.
This choice of independent variables corresponds to
the discretized 2DTL equation.

Our goal here is to
write the zero curvature condition
with another choice of the reference flow.
Specifically, let it be the composite flow labeled by the
pair of vectors
$\overrightarrow{\l _{0}\l _{3}},\,
\overrightarrow{\l _{1}\l _{2}}$. Let $y$ be the corresponding
"composite" time variable.
According to the definition given in Sect.\,3.1,
the $\tau$-function depends on $y$ as follows:
$$
\tau ^{l, \bar l }_{u}[y]:=\tau ^{l+y, \bar l +y }_{u}.
$$
In other words, we set, by definition,
$\p _{y}:=\p _{l}+\p _{\bar l}$, so
the shift operator $e^{\p _{y}}$ acts to
the $\tau$-function by shifting $l, \bar l$ simultaneously:
$$
e^{\p _{y}}\tau _{u}^{l,\bar l }=
\tau _{u}^{l+1, \bar l +1}
e^{\p _{y}}\,.
$$

Introduce the following difference operators with $2\times2$-matrix
coefficients:
\begin{eqnarray}
&&L_{n}(l, \bar l)=\left ( \begin{array}{ccc}
e^{\p _{y}}+
\nu \displaystyle{\frac{
\tau _{n}^{l,\bar l }\tau _{n+1}^{l+1, \bar l } }
{ \tau _{n}^{l+1 ,\bar l }\tau _{n+1}^{l, \bar l }}}+
\mu
\displaystyle{\frac{ \tau _{n+1}^{l+1,\bar l +1}
\tau _{n-1}^{l+1, \bar l } }
{ \tau _{n}^{l+1 ,\bar l +1}\tau _{n}^{l+1, \bar l }}}&&
-\mu \nu \displaystyle{\frac{\tau _{n+1}^{l+1, \bar l +1}}
{\tau _{n}^{l+1, \bar l +1}}}\\ && \\
\displaystyle{\frac{\tau _{n}^{l, \bar l }}
{\tau _{n+1}^{l, \bar l }}}&&0 \end{array} \right ),
\nonumber \\
&&M_{n}(l, \bar l)=\left ( \begin{array}{ccc}
1&&
-\mu \nu \displaystyle{\frac{\tau _{n+1}^{l, \bar l +1}}
{\tau _{u}^{l, \bar l +1}}}\\ && \\
\displaystyle{\frac{\tau _{n-1}^{l+1 , \bar l +1}}
{\tau _{n}^{l+1, \bar l +1}}}&&
-e^{\p _{y}}-
\mu
\displaystyle{\frac{ \tau _{n+1}^{l,\bar l +1}
\tau _{n-1}^{l+1, \bar l +1} }
{ \tau _{n}^{l ,\bar l +1}\tau _{n}^{l+1, \bar l +1}}}
\end{array} \right ),
\label{M4}
\end{eqnarray}
where we put
$\mu \equiv \l _{2}^{01}$ ,
$\nu \equiv \l _{3}^{01}$
for the sake of brevity.
\begin{prop}
The matrix discrete Zakharov-Shabat equation
\beq
L_n (l, \bar l +1 )M_n (l,\bar l )=M_{n+1} (l, \bar l )
L_n (l,\bar l )
\label{M5}
\eeq
is equivalent to the bilinear relation
\beq
\tau _{n}^{l,\bar l +1}\tau _{n}^{l+1, \bar l}-H_n (l, \bar l )
\tau _{n}^{l,\bar l }\tau _{n}^{l+1, \bar l +1}=
( \mu /\nu )
\tau _{n+1}^{l,\bar l +1}\tau _{n-1}^{l+1, \bar l}\,,
\label{HBDE10}
\eeq
where $H_n (l,\bar l )$ is periodic in $n$ with period 1:
$H_{n+1}(l,\bar l )=H_n (l, \bar l )$.
\end{prop}
This bilinear equation coincides with eq.\,(\ref{Z8b}).
We omit the proof since it is absolutely straightforward after the
$L$-$M$ pair is given. A way to derive
matrix $M$-operators from the scalar ones is discussed in
Sect.\,5.

Like in Theorem 4.1, the validity of the zero curvature condition
for $M$-operators
constructed with respect to all possible
independent reference flows implies the bilinear equation
with a fixed constant function $H$. It has the form (\ref{HBDE6}).

{\bf Remark 4.1}\,In the 2DTL interpretation,
the operator $M_n$ generates evolution in the
chiral discrete "space-time" variable
$\bar l$
whereas $L_n$ generates shifts along the $n$-lattice.
In our scheme, both $M_n$ and $L_n$ are "$M$-operators"
rather than "$L$-operators". We write $L_n$ according to the
tradition which is justified by the case when an additional
reduction of the 2DTL is implied.

It is instructive to look at the continuous version of this zero
curvature condition. It provides the zero curvature representation
of the 2DTL with the
composite continuous reference
flow defined by the vector field
$\p _{y}:=\p _{t_1} +\p _{\bar t _1}$ (see Sect.\,3).
This representation naturally arises when one embeds the 2DTL into the
2-component KP hierarchy. The Zakharov-Shabat equation
\beq
\p _{\bar t_1 }L_n =M_{n+1}L_n -L_n M_n \,,
\label{M6}
\eeq
where
\begin{eqnarray}
&&L_n =\left ( \begin{array}{ccc}
\p _{y} -\p _{t_1 }(\log \displaystyle{\frac{\tau _{n+1}}{\tau _{n}})}&&
-\displaystyle{\frac{\tau _{n+1}}{\tau _{n}}} \\ && \\
\displaystyle{\frac{\tau _{n}}{\tau _{n+1}}} && 0 \end{array} \right ),
\nonumber \\
&&M_n =\left ( \begin{array}{ccc}
0&&
\displaystyle{\frac{\tau _{n+1}}{\tau _{n}}} \\ && \\
-\displaystyle{\frac{\tau _{n-1}}
{\tau _{n}}} && \p _{y} \end{array} \right )
\label{M7}
\end{eqnarray}
is equivalent to
\beq
\p _{t_1 }\p _{\bar t_{1}}\log \frac{ \tau _{n+1}}{\tau _{n}}=
\frac{\tau _{n+1}\tau _{n-1}}
{\tau _{n}^{2}}-
\frac{\tau _{n+2}\tau _{n}}
{\tau _{n+1}^{2}}\,,
\label{M9}
\eeq
which is the 2DTL equation in the bilinear form.

{\bf Remark 4.2}\,\,The 1D Toda chain (1DTC) is a
reduction of the 2DTL such that
$\tau _{n}$ does not depend on $t_1 +\bar t_1 $, i.e., $\p _{y}$
commutes with $\tau _{n}$. Therefore, in this case $\p _{y}$ can be
considered as a $c$-number. Identifying it with the spectral parameter,
one recognizes eqs.\,(\ref{M7}) as the standard
$L$-$M$-pair for the
1DTC realized by $2\times 2$-matrices depending
on the spectral parameter (see e.g. \cite{FadTakh}).

\section{Linearization of the Hirota equations}

Zero curvature conditions studied in the previous section are
equivalent to compatibility of an overdetermined system of linear
difference equations
for a "wave function" $\psi$.
These linear equations are called {\it auxiliary
linear problems} (ALP). They play a very important role in the theory.
Common solutions to ALP carry the complete
information about solutions to the nonlinear equations. All the
properties of the latter can be translated into the language of the
ALP. This is what we mean by the linearization of HBDE.

In accoradance with the diversity of zero curvature representations
there are many types of the ALP. This section deals with the most
important examples.

We begin with the scalar linear problems associated with
the $M$-operators (\ref{Z1}), (\ref{Z2})
for elementary discrete flows adjacent to
the reference one. They are simple first order linear difference
equations with coefficients expressed through the $\tau$-function.
The formal solution of a special form is called (formal)
Baker-Akhiezer function. It depends on a spectral parameter.
Baker-Akhiezer functions are formal analogues of Bloch
solutions.
The formula for Baker-Akhiezer functions in terms of the
$\tau$-function was suggested for the first time in \cite{Date1}.
General solutions to the ALP are linear combinations of
Baker-Akhiezer functions with different spectral parameters.
In a similar way, one may define dual Baker-Akhiezer functions
as formal solutions to the linear problems for adjoint operators.

Given a solution to the ALP, one may consider
B\"acklund transformations.
Furthermore, a "duality" between
coefficient functions and solutions of the ALP allows one to
define a chain of successive B\"acklund transformations described
by the B\"acklund flow.
We consider two types of the
B\"acklund flows.
It is shown that in the particular case
when the solutions to the ALP are
Baker-Akhiezer functions the B\"acklund flows can be identified
with elementary discrete flows adjacent to the reference flow.

There is a "gauge freedom" in the ALP which can be fixed by
certain normalization of $\psi$.
Mostly we use the gauge which leads to the simplest
possible form of the linear equations.
Another choice -- $z_0$-gauge -- is briefly
discussed in Sect.\,5.4. This gauge makes equations more
symmetric for the price of introducing an auxiliary point
$z_0 \in {\bf C}$ and complication of coefficient functions.

The ALP associated with the matrix $M$-operators
are also discussed. In fact matrix $M$-operators can be
most conveniently derived using the scalar ALP. The matrix
linear problems are obtained by rearranging the scalar ones.
More precisely, in order to rearrange
the scalar ALP in such a way that the reference flow
is taken to be composite, one has to pass to
difference operators with matrix coefficients.

\subsection{Scalar linear problems}

The commutativity of ${\cal M}$-operators (\ref{comm2})
implies that they have a common set of eigenfunctions.
Equivalently,
the discrete Zakharov-Shabat equations (\ref{Z3})-(\ref{Z5})
for $M$-operators imply
compatibility of the linear problems
\beq
M^{l}_{u}\psi ^{l, \bar l}(u)=
\psi ^{l+1, \bar l}(u)\,,
\label{S1}
\eeq
\beq
\bar M^{\bar l}\psi ^{l, \bar l}(u)=
\psi ^{l, \bar l +1}(u)
\label{S2}
\eeq
for any elementary discrete
flows $l, \bar l$ adjacent to $u$.
Note that the "eigenvalues" put equal to 1 in the r.h.s.
can be
made arbitrary by changing normalization of
$\psi $.
Our choice in (\ref{S1}), (\ref{S2}) is most convenient
in the purely discrete case though it does not lead to
a smooth continuum limit.

More explicitly, eqs.\,(\ref{S1}), (\ref{S2}) read
(see (\ref{Z1}), (\ref{Z2})):
\beq
\psi ^{l,\bar l}(u+1)
-\l _{3}^{01}V^{l,\bar l}(u)\psi ^{l,\bar l}(u)
=\psi ^{l+1, \bar l}(u)\,,
\label{S3}
\eeq
\beq
\psi ^{l,\bar l}(u)
-\l _{2}^{01}C^{l,\bar l}(u)\psi ^{l,\bar l}(u-1)
=\psi ^{l, \bar l +1}(u)\,,
\label{S4}
\eeq
where
\beq
V^{l,\bar l}(u):=
\frac{ \tau ^{l,\bar l}_{u}\tau ^{l+1, \bar l}_{u+1}}
{\tau ^{l+1,\bar l}_{u}\tau ^{l, \bar l}_{u+1}}\,,
\label{S5}
\eeq
\beq
C^{l,\bar l}(u):=
\frac{ \tau ^{l,\bar l +1}_{u+1}\tau ^{l, \bar l}_{u-1}}
{\tau ^{l,\bar l +1}_{u}\tau ^{l, \bar l}_{u}}\,,
\label{S6}
\eeq

These formulas become more symmetric in terms of the "unnormalized"
wave function
\beq
\rho ^{l, \bar l}_{u}=\psi ^{l, \bar l}(u)
\tau ^{l, \bar l }_{u}\,.
\label{S7}
\eeq
Plugging this into (\ref{S3}), (\ref{S4}), we get:
\beq
\tau ^{l+1, \bar l}_{u}
\rho ^{l,\bar l}_{u+1}-\l _{3}^{01}
\tau ^{l+1, \bar l}_{u+1}
\rho ^{l,\bar l}_{u}=
\tau ^{l, \bar l}_{u+1}
\rho ^{l+1,\bar l}_{u}\,,
\label{S8}
\eeq
\beq
\tau ^{l, \bar l +1}_{u}
\rho ^{l,\bar l}_{u}-\l _{2}^{01}
\tau ^{l, \bar l +1}_{u+1}
\rho ^{l,\bar l}_{u-1}=
\tau ^{l, \bar l}_{u}
\rho ^{l,\bar l +1}_{u}\,,
\label{S9}
\eeq

Let us show that a carefully performed continuum limit of
these equations yields the familiar linear problems for the
2DTL. Let
$$
\l _{3}= \l _{0} +\epsilon ,\;\;\;\;\;\;\;
\l _{2}= \l _{1} +\bar \epsilon , \;\;\;\;\;\;
\epsilon , \bar \epsilon \rightarrow 0\,,
$$
so
$$
e^{\p _{l}}\rightarrow 1-\epsilon \p _{t_1 },
\;\;\;\;\;\;
e^{\p _{\bar l}}\rightarrow 1-\bar \epsilon \p _{\bar t_1 }
$$
by virtue of Miwa's rule (\ref{miwa1}) applied to the
discrete flows
$\overrightarrow{\l _{0}\l _{3}}$,
$\overrightarrow{\l _{1}\l _{2}}$, respectively.
Let us change the normalization of the $\psi$-function
introducing the $\varphi$-function as follows:
\beq
\varphi ^{l,\bar l }(u)=
(\l _{0}-\l _{3})^{l}
\psi ^{l,\bar l }(u)
\label{chi}
\eeq
so the linear problems read
\begin{eqnarray}
&&(\l _{0}-\l _{3}) \varphi ^{l,\bar l}(u+1)
-\l (\l _{3}-\l _{1})
V^{l,\bar l}(u)\varphi ^{l,\bar l}(u)
=\varphi ^{l+1, \bar l}(u)\,,
\nonumber \\
&&(\l _{0}-\l _{2}) \varphi ^{l,\bar l}(u)
-\l (\l _{2}-\l _{1})
C^{l,\bar l}(u)\varphi ^{l,\bar l}(u-1)
=(\l _{0}-\l _{2}) \varphi ^{l, \bar l +1}(u)\,,
\label{S34a}
\end{eqnarray}
where $\l \equiv (\l _{1}-\l _{0})^{-1}$. We get as
$\epsilon , \bar \epsilon \rightarrow 0$:
\beq
\left \{ \begin{array}{l}
\p _{t_1}\varphi (u)=\varphi (u+1)+
\left (\l + \p _{t_1}\log
\displaystyle{\frac{\tau _{u+1}}{\tau _{u}}}
\right ) \varphi (u) \\ \\
\p _{\bar t_1}\varphi (u)=\l ^{2}
\displaystyle{\frac{\tau _{u+1}\tau _{u-1}}
{\tau _{u}^{2}}}\varphi (u-1)\,.
\end{array} \right.
\label{S10}
\eeq
The transformation
$$
\tau _{u}\rightarrow \l ^{-u^2}
e^{-\l ut_1 }\tau _{u}
$$
eliminates the constant $\l$ turning the linear problems
into the familiar form
\beq
\left \{ \begin{array}{l}
\p _{t_1}\varphi (u)=\varphi (u+1)+
v(u) \varphi (u)\\ \\
\p _{\bar t_1}\varphi (u)=c(u)\varphi (u-1)\,.\end{array}\right.
\label{S10a}
\eeq
Here
$$
v(u)=\p _{t_1}\log \frac{\tau _{u+1}}
{\tau _{u}}\,,
\;\;\;\;\;\;\;\;
c(u)=\frac{\tau _{u+1}\tau _{u-1}}
{\tau _{u}^{2}}
$$
(for the sake of notational simplicity we use here the same letter
for the transformed function).
The Zakharov-Shabat equation
$$
\left [
\p _{t_1}-e^{\p _{u}}-v(u),\,\,
\p _{\bar t_1}- c(u)e^{-\p _{u}} \right ]=0
$$
yields the 2DTL equation in the form
\beq
\p _{t_1 }\p _{\bar t_{1}}\log \frac{ \tau _{u+1}}{\tau _{u}}=
\frac{\tau _{u+1}\tau _{u-1}}
{\tau _{u}^{2}}-\frac{\tau _{u+2}\tau _{u}}{\tau _{u+1}^{2}}\,.
\label{toda}
\eeq
These are continuous analogues of eqs.\,(\ref{Z5}), (\ref{Z7}),
respectively.

\subsection{B\"acklund transformations}

The ALP in the form
(\ref{S8}), (\ref{S9}) have a remarkable property emphasized
in ref.\,\cite{SS}: they are symmetric under interchanging
$\tau$ and $\rho$. Furthermore, one may treat them as linear problems
for the function $\tau$, the compatibility condition being a
bilinear equation for $\rho$. This equation is again HBDE
of the same form. In ref.\,\cite{SS},
this fact was refered
to as the {\it duality} between "potentials" $\tau$
and "wave functions" $\rho$. This "duality" shows up most
transparently in the fully discretized case.

More precisely, rewriting eqs.\,(\ref{S3}), (\ref{S4}) as linear
equations for
$$
\tilde \psi ^{l,\bar l}(u)=
\frac{ \tau ^{l+1, \bar l+1}_{u+1}}
{ \rho ^{l+1, \bar l+1}_{u+1}} =
\left ( \psi ^{l+1, \bar l+1}(u+1)\right ) ^{-1}
$$
(see (\ref{S7})), we get
\beq
\left ( e^{-\p _{u}}-\l _{3}^{01}
\tilde V^{l,\bar l}(u)\right )\tilde \psi ^{l,\bar l}(u)=
\tilde \psi ^{l-1, \bar l}(u)\,,
\label{D1}
\eeq
\beq
\left (1-\l _{2}^{01}
\tilde C^{l,\bar l}(u+1)e^{\p _{u}} \right )
\tilde \psi ^{l,\bar l}(u)=
\tilde \psi ^{l, \bar l -1}(u)\,,
\label{D2}
\eeq
where $\tilde V$ and $\tilde C$ are given by the same formulas
(\ref{S5}), (\ref{S6}) with $\rho$ in place of $\tau$.
The difference operators in the
l.h.s. are adjoint to the operators (\ref{Z1}), (\ref{Z2})
with $\tau \rightarrow \rho$. The formal adjoint operator is
defined by the rule $(f(u)e^{k\p _{u}})^{\dag}=e^{-k\p _{u}}f(u)$.
It then follows that the compatibility conditions are described by
Theorem 4.1 with $\tau$ replaced by $\rho$.

Therefore, passing from a given solution $\tau$ to $\rho$ we have got
a new solution to HBDE. This is a B\"acklund-type transformation
also known under the names
"Darboux" or "B\"acklund-Darboux" transformation. For a
comprehensive discussion of transformations of this kind see
\cite{Matveev}.
The bilinear form of B\"acklund transformations was suggested by
R.Hirota \cite{Hirota4}.

One may repeat the procedure once again starting from $\rho$ and,
moreover, consider a chain of successive transformations of this
type. It is natural to introduce an additional
discrete variable $b$ to mark
steps of the "flow" along this chain
and let $\tau ^{l,\bar l}_{u,b}$,
$\rho ^{l,\bar l}_{u,b}$ be $\tau$ and $\rho$ at $b$-th step,
respectively.
The first B\"acklund flow is
defined by
\beq
\tau ^{l,\bar l}_{u, b+1}=
\rho ^{l,\bar l}_{u,b}\,.
\label{D3a}
\eeq
This means that $\tau$ at the next step of the "B\"acklund time" $b$
is put equal to a solution $\rho$ of the linear equations (\ref{S8}),
(\ref{S9}). Then these linear
problems become {\it bilinear equations} for $\tau _{b}$:
\beq
\tau _{u,b}^{l+1}\tau _{u+1, b+1}^{l}-\l _{3}^{01}
\tau _{u+1, b}^{l+1}\tau _{u, b+1}^{l}-
\tau _{u+1, b}^{l}\tau _{u, b+1}^{l+1}=0\,,
\label{D4}
\eeq
\beq
\tau _{u,b}^{\bar l+1}\tau _{u,b+1}^{\bar l}(u)-
\tau _{u,b}^{\bar l}(u)\tau _{u,b+1}^{\bar l+1}=
\l _{2}^{01}
\tau _{u+1,b}^{\bar l+1}\tau _{u-1, b+1}^{\bar l}\,,
\label{D5}
\eeq
where $\bar l$ (resp., $l$) in eq.\,(\ref{D4}) (resp., (\ref{D5}))
is skipped.

Similarly, defining the second B\"acklund flow (the B\"acklund "time"
is now denoted by $\bar b$),
\beq
\tau ^{l,\bar l}_{u, \bar b+1}=
\rho ^{l,\bar l}_{u-1, \bar b},
\label{D3b}
\eeq
we get from (\ref{S8}), (\ref{S9}):
\beq
\l _{3}^{01}
\tau _{u, \bar b}^{l+1}\tau _{u,\bar b+1}^{l}+
\tau _{u, \bar b}^{l}\tau _{u, \bar b+1}^{l+1}=
\tau _{u+1, \bar b+1}^{l}\tau _{u-1, \bar b}^{l+1}\,,
\label{D6}
\eeq
\beq
\tau _{u, \bar b}^{\bar l}\tau _{u+1, \bar b+1}^{\bar l+1}-
\tau _{u, \bar b}^{\bar l+1}\tau _{u+1, \bar b+1}^{\bar l}
+\l _{2}^{01}
\tau _{u+1, \bar b}^{\bar l+1}\tau _{u, \bar b+1}^{\bar l}=0\,.
\label{D7}
\eeq
In these equations one immediately recognizes different forms
of HBDE. A time discretization of the Toda chain by means
of Darboux transformations was considered in \cite{SZ}.

The B\"acklund flows can be defined by a zero curvature
condition. Given any solution $\psi$ to the ALP (\ref{S3}),
(\ref{S4}), introduce the operator
\beq
{\cal B}_{u}^{b}=e^{-\p _{b}}\left (
e^{\p _{u}}-\frac{\psi (u+1)}{\psi (u)}\right ).
\label{D8}
\eeq
Then eq.\,(\ref{D4}) is represented as the commutativity
condition $[{\cal B}_{u}^{b},\,\,{\cal M}_{u}^{l}]=0$.
A similar ${\cal B}$-operator exists for the second B\"acklund flow.

\subsection{Baker-Akhiezer functions}

Each of the ALP
(\ref{S3}), (\ref{S4}) is a first order linear difference
equation in two variables. Assuming HBDE (\ref{hir1})-(\ref{hir3})
hold, we are going to construct a 1-parametric family of
their common solutions of a special form.
These solutions $\psi (u) =\psi (u;z)$
are called
{\it Baker-Akhiezer functions}. They depend on
the {\it spectral parameter}
$z \in {\bf C}$.

Let us swith on an extra elementary flow shown by the dotted line:

\vspace{0.3cm}

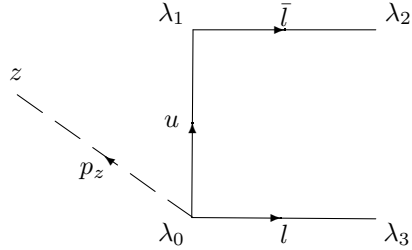
\begin{center}
\special{em:linewidth 0.4pt}
\unitlength 0.50mm
\linethickness{0.4pt}
\begin{picture}(114.00,64.33)
\emline{59.00}{10.00}{1}{108.00}{9.67}{2}
\emline{108.00}{60.00}{3}{59.33}{60.00}{4}
\emline{59.33}{60.00}{5}{59.00}{10.00}{6}
\put(83.66,60.00){\vector(1,0){0.2}}
\emline{83.33}{60.00}{7}{83.66}{60.00}{8}
\put(83.66,5.67){\makebox(0,0)[cc]{$l$}}
\put(83.66,64.00){\makebox(0,0)[cc]{$\bar l$}}
\put(53.66,35.67){\makebox(0,0)[cc]{$u$}}
\put(53.66,5.67){\makebox(0,0)[cc]{$\lambda _0$}}
\put(113.66,5.67){\makebox(0,0)[cc]{$\lambda _3$}}
\put(53.66,64.33){\makebox(0,0)[cc]{$\lambda _1$}}
\put(114.00,64.33){\makebox(0,0)[cc]{$\lambda _2$}}
\put(59.19,35.09){\vector(0,1){0.2}}
\emline{59.19}{35.02}{9}{59.19}{35.09}{10}
\put(82.65,9.86){\vector(1,0){0.2}}
\emline{82.43}{9.86}{11}{82.65}{9.86}{12}
\emline{59.33}{10.00}{13}{52.67}{14.67}{14}
\emline{49.33}{17.00}{15}{42.67}{21.67}{16}
\emline{39.33}{24.00}{17}{32.67}{28.67}{18}
\emline{29.33}{31.00}{19}{22.67}{35.67}{20}
\emline{19.33}{38.00}{21}{12.67}{42.67}{22}
\put(36.07,26.29){\vector(-4,3){0.2}}
\emline{36.52}{25.96}{23}{36.07}{26.29}{24}
\put(12.33,48.00){\makebox(0,0)[cc]{$z$}}
\put(32.67,22.33){\makebox(0,0)[cc]{$p_z$}}
\end{picture}
\end{center}

\vspace{0.3cm}

\noindent
The corresponding time variable is $p_z$.
Let
\beq
\l _{z}^{\a \b}=\frac{1}{z -\l _{\a }}
-\frac{1}{\l _{\b } -\l _{\a }}.
\label{lambda2}
\eeq
Note the identity
\beq
\l _{z}^{\a \b }\l _{z}^{\b \gamma }=
\l _{\a }^{\b \gamma }\l _{z}^{\a \gamma }\,.
\label{lambda3}
\eeq
Then, assuming the 3-term Hirota equations hold for
for the triples $(u,l, p_z)$ and $(u, \bar l , p_z)$ of
independent variables,
\beq
\psi ^{l,\bar l}(u;z)=\left.
(\l _{z}^{01})^{u}(\l _{z}^{03})^{l}
\left (\frac{
\l _{z}^{02}}{\l _{z}^{01}}\right )^{\bar l }
\frac{ \tau _{u, p_z +1}^{l, \bar l }}
{ \tau _{u, p_z }^{l, \bar l }}\right |_{p_z =0}
\label{psi}
\eeq
is a formal common solution to eqs.\,(\ref{S3}), (\ref{S4})
for any $z$. Indeed, under the
substitution (\ref{psi}) eq.\,(\ref{S3}) turns into eq.\,(\ref{hir1})
for the triple
$(u,l, p_z)$ while eq.\,(\ref{S4})
turns into eq.\,(\ref{hir3}) for
the triple $(u, \bar l , p_z)$. Therefore,
the new label $z$ is identified with the spectral parameter.
Formula (\ref{psi}) for the $\psi$-function coincides with the
Japanese formula
\cite{JimboMiwa},\,\cite{Date1} because due to (\ref{miwa1})
we have
$$
\left. \frac{ \tau _{u, p_z +1}^{l, \bar l }}
{ \tau _{u, p_z }^{l, \bar l }}\right |_{p_z =0}=
\frac{\tau _{u}^{l, \bar l}(-[z-\l _{0}])}
{\tau _{u}^{l, \bar l}(0)}.
$$
The general solution to the ALP can be represented in the
form
\beq
\psi (u)=\int d^2 z \mu (z)\psi (u;z)
\label{gen}
\eeq
with arbitrary "measure" $\mu (z)$ on the complex plane.
In other words, this is a linear combination of Baker-Akhiezer
functions with different spectral parameters.

Note that the "${\cal B}$-operator" (\ref{D8})
in which $\psi$ is taken to be
the Baker-Akhiezer function coincides with an ${\cal M}$-operator.
Indeed, we have the following formula for
the $M$-operator (\ref{Z1}) in terms of the
Baker-Akhiezer function:
\beq
M^{l}_{u}=\lim _{z\rightarrow \l _{3}}
\left ( e^{\p _{u}}-\frac{\psi (u+1; z)}{\psi (u; z)}\right ).
\label{S14}
\eeq

The {\it dual Baker-Akhiezer function}
$\psi ^{*}$ is defined by
\beq
\psi ^{*l,\bar l}(u;z)=\left.
(\l _{z}^{01})^{-u}(\l _{z}^{03})^{-l}
\left (\frac{
\l _{z}^{02}}{\l _{z}^{01}}\right )^{-\bar l }
\frac{ \tau _{u, p_z -1}^{l, \bar l }}
{ \tau _{u, p_z }^{l, \bar l }}\right |_{p_z =0}.
\label{dualpsi}
\eeq
It satisfies the equations
\begin{eqnarray}
&&\left ( M_{u}^{l}(u-1, l-1)\right )^{\dag}
\psi ^{*l}(u;z)=\psi ^{*l-1}(u;z)\,,
\nonumber \\
&&\left ( \bar M_{u}^{\bar l}(u-1, \bar l-1)\right )^{\dag}
\psi ^{*\bar l}(u;z)=\psi ^{*\bar l-1}(u;z)\,,
\label{S34d}
\end{eqnarray}
where
the difference operators in the
l.h.s. are formally adjoint to the operators (\ref{Z1}), (\ref{Z2}).

\subsection{The $z_0$-gauge}

Equations (\ref{S3}), (\ref{S4}) imply a specific choice
of normalization of the $\psi$-function. Indeed, multiplying
$\psi$ by any function, one can change the form of the equations.
This is a kind of a "gauge freedom". There is no canonical way
to fix the gauge. The gauge that we
systematically use throughout the paper is the most economic one
in the sense that
the ALP for discrete flows have the simplest
possible form. Here we
are going to discuss another choice which has its own
advantages.

This more general gauge requires to fix an additional
point
$z_0 \in {\bf C}$ different from the
vertices of the graph of flows.
The gauge is defined by the following normalization condition
for the Baker-Akhiezer function $\Psi (u;z)$:
\beq
\Psi (u;z_0 )=1\,.
\label{norm}
\eeq
We call it {\it $z_0$-gauge}. Given this condition,
it is natural to represent $\Psi$ in the form
$$
\Psi (u;z)=\frac{\psi (u;z)}{\psi (u; z_0 )}
$$
and rewrite the ALP (\ref{S3}), (\ref{S4}) for $\psi$
in terms of $\Psi$.
In this way, we get
\beq
\left (\l _{z_0 }^{01}U(u,l)e^{\p _{u}}
-\l _{z_0 }^{03}W(u,l)e^{\p _{l}} \right )\Psi (u;z)
=\l _{3}^{01}\Psi (u;z)\,,
\label{N1}
\eeq
where
\beq
U(u,l)=\frac{\tau _{u,p_0 }^{l+1}\tau _{u+1, p_0 +1}^{l}}
{\tau _{u,p_0 +1}^{l}\tau _{u+1, p_0}^{l+1}},
\;\;\;\;\;\;\;\;
W(u,l)=\frac{\tau _{u,p_0 +1}^{l+1}\tau _{u+1, p_0 }^{l}}
{\tau _{u,p_0 +1}^{l}\tau _{u+1, p_0}^{l+1}}.
\label{N2}
\eeq
A general prescription for
writing down the equation should be clear from
comparison with the picture.

\vspace{0.3cm}

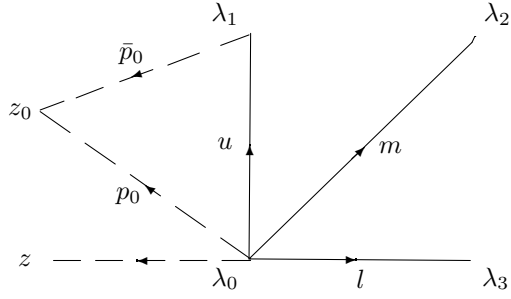
\begin{center}
\special{em:linewidth 0.4pt}
\unitlength 0.60mm
\linethickness{0.4pt}
\begin{picture}(114.00,64.33)
\emline{59.00}{10.00}{1}{108.00}{9.67}{2}
\emline{59.33}{60.00}{3}{59.00}{10.00}{4}
\put(83.66,5.67){\makebox(0,0)[cc]{$l$}}
\put(53.66,35.67){\makebox(0,0)[cc]{$u$}}
\put(53.66,5.67){\makebox(0,0)[cc]{$\lambda _0$}}
\put(113.66,5.67){\makebox(0,0)[cc]{$\lambda _3$}}
\put(53.66,64.33){\makebox(0,0)[cc]{$\lambda _1$}}
\put(114.00,64.33){\makebox(0,0)[cc]{$\lambda _2$}}
\put(59.19,35.09){\vector(0,1){0.2}}
\emline{59.19}{35.02}{5}{59.19}{35.09}{6}
\put(82.65,9.86){\vector(1,0){0.2}}
\emline{82.43}{9.86}{7}{82.65}{9.86}{8}
\emline{59.33}{10.00}{9}{52.67}{14.67}{10}
\emline{49.33}{17.00}{11}{42.67}{21.67}{12}
\emline{39.33}{24.00}{13}{32.67}{28.67}{14}
\emline{29.33}{31.00}{15}{22.67}{35.67}{16}
\emline{19.33}{38.00}{17}{12.67}{42.67}{18}
\put(36.07,26.29){\vector(-4,3){0.2}}
\emline{36.52}{25.96}{19}{36.07}{26.29}{20}
\emline{59.00}{10.00}{21}{108.33}{58.00}{22}
\emline{108.33}{58.00}{23}{109.00}{59.33}{24}
\emline{109.00}{59.33}{25}{109.00}{59.33}{26}
\put(84.53,34.71){\vector(1,1){0.2}}
\emline{81.76}{32.23}{27}{84.53}{34.71}{28}
\emline{59.33}{9.67}{29}{53.00}{9.67}{30}
\emline{49.67}{9.67}{31}{43.67}{9.67}{32}
\emline{40.33}{9.67}{33}{34.33}{9.67}{34}
\emline{31.00}{9.67}{35}{25.00}{9.67}{36}
\emline{21.67}{9.67}{37}{15.67}{9.67}{38}
\put(34.33,9.66){\vector(-1,0){0.2}}
\emline{35.11}{9.66}{39}{34.33}{9.66}{40}
\emline{59.00}{59.67}{41}{52.67}{57.33}{42}
\emline{48.00}{55.67}{43}{41.00}{53.00}{44}
\emline{37.67}{52.00}{45}{30.33}{49.33}{46}
\emline{26.67}{48.00}{47}{20.67}{45.67}{48}
\emline{18.00}{45.00}{49}{12.67}{43.00}{50}
\put(33.05,50.34){\vector(-3,-1){0.2}}
\emline{34.79}{50.98}{51}{33.05}{50.34}{52}
\put(90.33,34.67){\makebox(0,0)[cc]{$m$}}
\put(8.33,42.67){\makebox(0,0)[cc]{$z_0$}}
\put(9.33,9.67){\makebox(0,0)[cc]{$z$}}
\put(32.33,24.00){\makebox(0,0)[cc]{$p_0$}}
\put(33.00,55.33){\makebox(0,0)[cc]{$\bar p_0$}}
\end{picture}
\end{center}

\vspace{0.3cm}

\noindent
With this prescription at hand,
a similar equation can be written for
any pair of flows such that one of them is reference and
another one is left adjacent.

A nice feature of the $z_0$-gauge is that there is no need to
care about
equations for right adjacent flows. They are automatically
produced by the same prescription if one changes the "orientation"
of the reference flow
(i.e., consider $\overrightarrow{\l _{1}\l _{0}}$
as the reference flow). In
order to express everything in the same variables, one should
apply the rules (\ref{2-cycle}) ($u \rightarrow -u$) and
(\ref{3-cycle}) ($\bar p_{0} \rightarrow p_0$).
We stress that eq.\,(\ref{S4})
can not be obtained from eq.\,(\ref{S3}) in this way.
In that gauge we need two types of equations separately
for left and right adjacent flows.

The Baker-Akhiezer function that solves all these equations
in the $z_0$-gauge has the form
\beq
\Psi (u;z)=\prod _{\a \b} \left (
\frac{ \l _{z}^{\a \b }}
{\l _{z_0 }^{\a \b }}\right )^{l_{\a \b }} \left.
\frac{\tau _{p_0 , p_z +1}}{\tau _{p_0 +1, p_z }}
\right |_{p_0 =p_z =0}.
\label{N3}
\eeq
As usual, the $\tau$-functions depend on all the skipped
variables as parameters. Due to (\ref{lambda3}) the form
of the prefactor is consistent with
(\ref{2-cycle}), (\ref{3-cycle}).

Our previous gauge is a limiting case of the $z_0$-gauge
as $z_0 \rightarrow \l _{0}$. However, the limit is singular
and, therefore, it needs a regularization. As a result, the
symmetry of the $z_0$-gauge becomes broken.

\subsection{Two-component formalism}

Linear equations (\ref{S3}), (\ref{S4}) can be brought into another
form in which they become first order partial difference equations
for a 2-component vector function. Their compatibility yields
matrix Zakharov-Shabat equations presented in Sect.\,4.3.

Let us use the notation of
Sect.\,4.3 and, accordingly, denote $\psi _n \equiv \psi (n)$,
$V_{n}^{l,\bar l}\equiv V^{l,\bar l}(n)$,
$C_{n}^{l,\bar l}\equiv C^{l,\bar l}(n)$.
Then eqs.\,(\ref{S3}), (\ref{S4}) read:
\beq
\psi _{n}^{l+1, \bar l}=
\psi _{n+1}^{l,\bar l}-\nu
V_{n}^{l,\bar l}\psi _{n}^{l,\bar l}\,,
\label{2.1}
\eeq
\beq
\psi _{n}^{l, \bar l +1}=\psi _{n}^{l,\bar l}-
\mu C_{n}^{l,\bar l}\psi _{n-1}^{l,\bar l}\,.
\label{2.2}
\eeq
These equations allow us to find out how the vector
\beq
\left ( \begin{array}{c} \psi _{n}^{l,\bar l} \\ \\
\psi _{n-1}^{l,\bar l} \end{array} \right )
\label{vector}
\eeq
transforms under shifts in $n$ and $\bar l$.

Combining eqs.\,(\ref{2.1}), (\ref{2.2}) we have, for instance
(here $\p _{y}\equiv \p _{l} +\p _{\bar l}$):
\begin{eqnarray}
&&\psi _{n+1}^{l,\bar l}= \psi _{n}^{l+1,\bar l}+\nu
V_{n}^{l,\bar l}\psi _{n}^{l,\bar l} \nonumber \\
&=&\psi _{n}^{l +1,\bar l +1}+\mu
C_{n}^{l+1, \bar l}\psi _{n-1}^{l+1,\bar l}+\nu
V_{n}^{l,\bar l}\psi _{n}^{l,\bar l} \nonumber \\
&=&(e^{\p _{y}}+\nu
V_{n}^{l,\bar l})\psi _{n}^{l,\bar l}+\mu
C_{n}^{l+1,\bar l}
(\psi _{n}^{l,\bar l}
-\nu V_{n-1}^{l,\bar l}\psi _{n-1}^{l,\bar l})
\nonumber \\
&=&(e^{\p _{y}}+\nu V_{n}^{l,\bar l}+\mu
C_{n}^{l+1,\bar l})\psi _{n}^{l,\bar l}-\mu \nu
C_{n}^{l+1,\bar l}V_{n-1}^{l,\bar l}
\psi _{n-1}^{l,\bar l}\,.
\label{2.3}
\end{eqnarray}
Proceeding in the same way, we get:
\begin{eqnarray}
&&\left (\begin{array}{c}
\psi _{n+1}^{l,\bar l} \\ \\
\psi _{n}^{l,\bar l} \end{array} \right )=
\left ( \begin{array}{ccc}
e^{\p _{y}}+\nu V_{n}^{l,\bar l}+
\mu C_{n}^{l+1,\bar l} &&
-\mu \nu C_{n}^{l+1,\bar l}V_{n-1}^{l,\bar l}\\ && \\
1 && 0 \end{array} \right )
\left (\begin{array}{c}
\psi _{n}^{l,\bar l} \\ \\
\psi _{n-1}^{l,\bar l} \end{array} \right ),
\nonumber \\
&&\left (\begin{array}{c}
\psi _{n}^{l,\bar l +1} \\ \\
\psi _{n-1}^{l,\bar l +1} \end{array} \right )=
\left ( \begin{array}{ccc}
1&&
-\mu C_{n}^{l,\bar l}\\ && \\
(\nu V_{n-1}^{l,\bar l+1})^{-1} &&
-(\nu V_{n-1}^{l,\bar l+1})^{-1}
\big (e^{\p _{y}}+\mu
C_{n}^{l,\bar l} \big )
\end{array} \right )
\left (\begin{array}{c}
\psi _{n}^{l,\bar l} \\ \\
\psi _{n-1}^{l,\bar l} \end{array} \right ).
\label{2.5}
\end{eqnarray}
The operators in the r.h.s.
provide a matrix $L$-$M$-pair for HBDE which differs from
(\ref{M4})
by a "gauge" transformation.
Recall that the Baker-Akhiezer function
(\ref{psi}) has $\tau _{n}$ in the denominator,
so the two components
of the vector (\ref{vector}) have different denominators.
In the 2-component formalism, it is natural to
require the denominators to be the same. This condition
partially fixes the gauge.

Thus, introducing the vector $(\psi _{n},\, \chi _{n})$ with the
second component
\beq
\chi _{n}^{l,\bar l}=
\frac{\tau _{n-1}^{l,\bar l}}{\tau _{n}^{l,\bar l}}
\psi _{n-1}^{l,\bar l}\,,
\label{2.6}
\eeq
we rewrite eqs.\,(\ref{2.5}) in the form
\beq
L_{n}(l,\bar l)
\left ( \begin{array}{c}
\psi _{n}^{l,\bar l} \\
\chi _{n}^{l,\bar l}\end{array}\right )=
\left ( \begin{array}{c}
\psi _{n+1}^{l,\bar l} \\
\chi _{n+1}^{l,\bar l}\end{array}\right ),
\label{2.7}
\eeq
\beq
M_{n}(l,\bar l)
\left ( \begin{array}{c}
\psi _{n}^{l,\bar l} \\
\chi _{n}^{l,\bar l}\end{array}\right )=
\nu \left ( \begin{array}{c}
\psi _{n}^{l,\bar l +1} \\
\chi _{n}^{l,\bar l +1}\end{array}\right )
\label{2.8}
\eeq
with the $L$ and $M$-operators given by eqs.\,(\ref{M4}).

Equations (\ref{2.7}), (\ref{2.8}) imply
some useful difference equations for
$\psi _{n}^{l,\bar l}$. Excluding
$\chi _{n}$, we get from (\ref{2.7}):
\beq
\psi _{n}^{l+1, \bar l +1}=\psi _{n+1}^{l, \bar l}-
\left ( \nu
\frac{\tau _{n}^{l,\bar l}\tau _{n+1}^{l+1, \bar l}}
{\tau _{n}^{l+1,\bar l}\tau _{n+1}^{l, \bar l}}+
\mu
\frac{\tau _{n-1}^{l+1,\bar l}\tau _{n+1}^{l+1, \bar l +1}}
{\tau _{n}^{l+1,\bar l +1}\tau _{n}^{l+1, \bar l}}\right )
\psi _{n}^{l, \bar l}+
\mu \nu
\frac{\tau _{n-1}^{l,\bar l}\tau _{n+1}^{l+1, \bar l +1}}
{\tau _{n}^{l,\bar l }\tau _{n}^{l+1, \bar l +1}}
\psi _{n-1}^{l, \bar l}\,.
\label{2.9}
\eeq

Similarly, excluding $\chi _n$ from eq.\,(\ref{2.8}) and using
HBDE (\ref{HBDE6}), we get a 4-term equation for $\psi _n$:
\beq
\psi _{n}^{l+1, \bar l +1}-\psi _{n}^{l+1, \bar l }=
-\nu \frac{\tau _{n}^{l,\bar l +1}\tau _{n+1}^{l+1, \bar l +1}}
{\tau _{n}^{l+1,\bar l +1}
\tau _{n+1}^{l, \bar l +1}}\psi _{n}^{l, \bar l +1}
+(\nu - \mu )
\frac{\tau _{n}^{l,\bar l }\tau _{n+1}^{l+1, \bar l +1}}
{\tau _{n}^{l+1,\bar l }\tau _{n+1}^{l, \bar l +1}}
\psi _{n}^{l,\bar l }\,.
\label{2.11}
\eeq
Here it is
implied that $\tau _{n}^{l, \bar l}$ satisfies HBDE (\ref{hir3})
in the corresponding notation. For more information on
eq.\,(\ref{2.11}) see the next section.

The continuous analogue of eq.\,(\ref{2.9}) is
\beq
(\p _{t_1}+\p _{\bar t_{1}})\varphi _{n}=\varphi _{n+1}+
\left ( \p _{t_1 }\log \frac{\tau _{n+1}}{\tau _{n}}\right )
\varphi _{n}
+\frac{\tau _{n-1}\tau _{n+1}}{\tau _{n}^{2}}
\varphi _{n-1}\,.
\label{2.10}
\eeq
This is obvious from (\ref{S10})
(we write $\varphi _{n}$ instead of $\varphi (u)$ according
to the present notation).
It is a discrete non-stationary
Schr\"odinger equation. Eq.\,(\ref{2.11}) in the continuum limit
turns into
\beq
\p _{t_1}\p _{\bar t_{1}}\varphi _{n}-
\left (\p _{t_1 }\log \frac{\tau _{n+1}}{\tau _{n}}\right )
\p _{\bar t_1}\varphi _{n}-
\frac{\tau _{n-1}\tau _{n+1}}{\tau _{n}^{2}}\varphi _{n}=0\,.
\label{2.12}
\eeq
It is a continuous two-dimensional Schr\"odinger equation in
magnetic field. Its quasiperiodic solutions were studied in the paper
\cite{DKN} by means of the algebro-geometric
approach. The corresponding
theory for discrete two-dimensional equations similar to (\ref{2.11})
was proposed in \cite{Kr1}.

\section{Pseudo-difference $M$-operators}

In this section we study the general form of $M$-operators
which satisfy conditions of the zero curvature form with
$M$-operators for elementary discrete flows adjacent to
the reference flow.

Starting from the scalar ALP for a pair
of left and right adjacent flows it is not difficult to find the
$M$-operators for {\it non-adjacent flows}. Indeed,
it is possible to exclude the reference flow from the pair
of linear equations. Then
the right adjacent flow turns out to be written in terms of the
left adjacent one. Considering the latter as a new reference flow,
one obtains a general $M$-operator for {\it any} elementary
discrete flow in terms of {\it any} (elementary discrete)
reference flow. In general these are pseudo-difference operators,
i.e., they contain negative powers of first order
difference operators.

This construction can be extended to more general operators
which generate some new flows commuting with the elementary
discrete flows. We call them adjoint flows. The corresponding
pseudo-difference operators are constructed in Sect.\,6.2 with
the help of two arbitrary independent solutions
$\psi$, $\psi ^{*}$ to the ALP and the adjoint ALP.

\subsection{$M$-operators for arbitrary elementary discrete
flows and the corresponding linear problems}

Rearranging eqs.\,(\ref{S3}), (\ref{S4}), it is possible to
find out $M$-operators for arbitrary elementary discrete flows,
not only for adjacent to the reference flow. The idea is to
exclude shifts in $u$ and then consider $l$ as a new reference flow.

From (\ref{S3}) we have
$$
\psi ^{l, \bar l}(u+1)=\psi ^{l+1, \bar l}(u)
+\l _{3}^{01}V^{l, \bar l}(u)
\psi ^{l, \bar l}(u)\,.
$$
Plugging this into (\ref{S4}) written in the form
$$
\psi ^{l, \bar l}(u+1)-\psi ^{l, \bar l +1}(u+1)
=\l _{2}^{01}C^{l, \bar l}(u+1)
\psi ^{l, \bar l}(u)\,,
$$
in place of $\psi ^{l, \bar l}(u+1)$
and $\psi ^{l, \bar l +1}(u+1)$, we get
$$
\psi ^{l+1, \bar l +1}(u)
+\l _{3}^{01}V^{l, \bar l +1}(u)
\psi ^{l, \bar l +1}(u)=
\psi ^{l+1, \bar l}(u)+
\big ( \l _{3}^{01}V^{l, \bar l}(u)
-\l _{2}^{01}C^{l, \bar l}(u+1)\big )
\psi ^{l, \bar l}(u)\,.
$$
Using (\ref{hir3}), we find
\begin{eqnarray}
&& \l _{3}^{01}V^{l, \bar l}(u)
-\l _{2}^{01}C^{l, \bar l}(u+1)
\nonumber \\
&=&
\l _{3}^{01}V^{l, \bar l +1}(u)
-\l _{2}^{01}C^{l+1, \bar l}(u)
=\l _{3}^{01}
\frac{\tau _{u}^{l, \bar l}\tau _{u+1}^{l+1, \bar l +1}}
{\tau _{u}^{l+1, \bar l}\tau _{u+1}^{l, \bar l +1}}\,.
\label{N4}
\end{eqnarray}
Note that the first equality
follows already from eq.\,(\ref{Z5}) which is a weaker condition
than (\ref{hir3}).

Therefore, $\psi$ obeys the following 4-term linear equation:
\beq
\psi ^{l+1, \bar l +1}(u)
-\psi ^{l+1, \bar l}(u)
=\l _{1}^{03}
\frac{\tau _{u}^{l, \bar l +1}\tau _{u+1}^{l+1, \bar l +1}}
{\tau _{u}^{l+1, \bar l +1}\tau _{u+1}^{l, \bar l +1}}
\psi ^{l, \bar l +1}(u)
+\l _{3}^{02}
\frac{\tau _{u}^{l, \bar l }\tau _{u+1}^{l+1, \bar l +1}}
{\tau _{u}^{l+1, \bar l }\tau _{u+1}^{l, \bar l +1}}
\psi ^{l, \bar l }(u)
\label{4-term}
\eeq
in which we recognize eq.\,(\ref{2.9}) from Sect.\,5.5.

The relation (\ref{N4}) allows us to rewrite it in the form
\beq
\frac{\tau _{u}^{l+1, \bar l}}
{\tau _{u-1}^{l+1, \bar l}}\,
\tilde \Delta _{l} \,
\frac{\tau _{u}^{l, \bar l +1}}
{\tau _{u+1}^{l, \bar l +1}}\,
\bar \Delta _{\bar l} \,
\, \psi ^{l, \bar l }(u)
+\l _{2}^{01}
\psi ^{l, \bar l }(u)=0\,,
\label{4-term1}
\eeq
where
\beq
\tilde \Delta _{l} \equiv  e^{\p _{l}}+\l _{3}^{01}\,,
\;\;\;\;\;\;\;\;
\bar \Delta _{\bar l} \equiv  e^{\p _{\bar l}}- 1\,.
\label{Delt}
\eeq
This equation looks like a discrete 2-dimensional Laplace
equation in a curved space.
It can be formally rewritten as
\beq
\left (
\bar \Delta _{\bar l}+\l _{2}^{01}
\frac{\tau _{u+1}^{l, \bar l +1}}
{\tau _{u}^{l, \bar l +1}}\, \tilde \Delta _{l}^{-1}
\frac{\tau _{u-1}^{l+1, \bar l}}{\tau _{u}^{l+1, \bar l}}
\right )
\psi ^{l, \bar l }(u)=0\,,
\label{4-term2}
\eeq
or, finally,
\beq
\psi ^{l, \bar l +1}(u)=
\left ( 1-
\frac{\tau _{u+1}^{l, \bar l +1}}
{\tau _{u}^{l, \bar l +1}}
\, \frac{
\l _{2}^{01}}{e^{\p _{l}}+
\l _{3}^{01}}\,
\frac{\tau _{u-1}^{l+1, \bar l}}
{\tau _{u}^{l+1, \bar l}}
\right )
\psi ^{l, \bar l }(u)\,.
\label{4-term3}
\eeq
To avoid a confusion, we stress that
the operator inside the brackets acts to the variable $l$
whereas $u$ enters as a parameter. This operator
should be identified with
the $M$-operator generating the flow $\bar l$ realized as a
{\it pseudo-difference operator in $l$}.

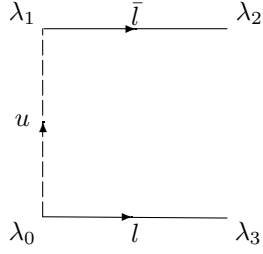
\begin{center}
\special{em:linewidth 0.4pt}
\unitlength 0.50mm
\linethickness{0.4pt}
\begin{picture}(65.67,64.33)
\emline{10.67}{10.00}{1}{59.67}{9.67}{2}
\emline{59.67}{60.00}{3}{11.00}{60.00}{4}
\put(35.33,60.00){\vector(1,0){0.2}}
\emline{35.00}{60.00}{5}{35.33}{60.00}{6}
\put(35.33,5.67){\makebox(0,0)[cc]{$l$}}
\put(35.33,64.00){\makebox(0,0)[cc]{$\bar l$}}
\put(5.33,35.67){\makebox(0,0)[cc]{$u$}}
\put(5.33,5.67){\makebox(0,0)[cc]{$\lambda _0$}}
\put(65.33,5.67){\makebox(0,0)[cc]{$\lambda _3$}}
\put(5.33,64.33){\makebox(0,0)[cc]{$\lambda _1$}}
\put(65.67,64.33){\makebox(0,0)[cc]{$\lambda _2$}}
\put(10.86,35.09){\vector(0,1){0.2}}
\emline{10.86}{35.02}{7}{10.86}{35.09}{8}
\put(34.32,9.86){\vector(1,0){0.2}}
\emline{34.10}{9.86}{9}{34.32}{9.86}{10}
\emline{10.67}{10.00}{11}{10.67}{14.00}{12}
\emline{10.67}{15.33}{13}{10.67}{19.33}{14}
\emline{10.67}{20.67}{15}{10.67}{24.67}{16}
\emline{10.67}{26.00}{17}{10.67}{30.00}{18}
\emline{10.67}{31.33}{19}{10.67}{35.33}{20}
\emline{10.67}{37.00}{21}{10.67}{41.00}{22}
\emline{10.67}{42.33}{23}{10.67}{46.33}{24}
\emline{10.67}{47.67}{25}{10.67}{51.67}{26}
\emline{10.67}{53.00}{27}{10.67}{57.00}{28}
\emline{10.67}{58.33}{29}{10.67}{60.00}{30}
\end{picture}
\end{center}

\noindent
In other words, letting $l$ to be the reference flow, we,
therefore, have got an $M$-operator for the flow $\bar l$ which
is not adjacent to $l$.

In the limit when the points $\l _{0}$, $\l _{1}$ merge,
the flow $\bar l$ becomes left adjacent to $l$. Let us
demonstrate how the corresponding difference $M$-operator
is reproduced from the r.h.s. of eq.\,(\ref{4-term3}).
Let $\l _{1}-\l _{0}=\epsilon$, $\epsilon \rightarrow 0$.
We have
$$
\begin{array}{l}
\displaystyle{
\frac{\l _{2}^{01}}{e^{\p _{l}}+\l _{3}^{01}}}
\rightarrow 1+\epsilon
\left (e^{\p _{l}}+\l _{3}^{02} \right )+O(\epsilon ^{2})\,, \\ \\
\displaystyle{\frac{\tau _{u\pm 1}^{l, \bar l}}
{\tau _{u}^{l, \bar l}}}
\rightarrow 1+O(\epsilon )\,, \\ \\
\displaystyle{\frac{\tau _{u+1}^{l, \bar l +1}
\tau _{u-1}^{l+1, \bar l }}
{\tau _{u}^{l+1, \bar l}
\tau _{u}^{l, \bar l +1}}}
\rightarrow 1-\epsilon \l _{3}^{02}\left (1-
\displaystyle{\frac{\tau ^{l, \bar l}
\tau ^{l+1, \bar l +1}}
{\tau ^{l+1, \bar l}
\tau ^{l, \bar l +1}}} \right )
+O(\epsilon ^{2})
\end{array}
$$
as $\epsilon \rightarrow 0$ (in the last line eq.\,(\ref{hir3})
is used), so the naive limit of the r.h.s. (\ref{4-term3}) is zero.
However, we should take into account the
change of normalization which is implied when the former flow
$\bar l$ turns into a left adjacent flow to $l$. This is
achieved by
$\psi ^{\bar l}\rightarrow
(-\epsilon )^{\bar l} \psi ^{\bar l}$,
so in the limit the right $M$-operator
$$
M_{l}^{\bar l}=e^{\p _{l}}-
\l _{2}^{03}
\frac{\tau ^{l, \bar l }
\tau ^{l+1, \bar l +1}}
{\tau ^{l+1, \bar l}
\tau ^{l, \bar l +1}}
$$
is reproduced.

For illustrative purposes, let us give continuous analogues
of the above formulas. Rather than to perform the limit directly,
it is much easier to use the continuous version
(\ref{S10a}) of the linear
problems from the very beginning. Making use of eq.\,(\ref{toda}),
we find the analogue of eq.\,(\ref{4-term}):
\beq
\p _{t_1 }\p _{\bar t_1}\varphi (u)-v(u)
\p _{\bar t_1}\varphi (u)-c(u)\varphi (u)=0\,.
\label{4cont}
\eeq
The analogues of eqs.\,(\ref{4-term1}), (\ref{4-term3}) read
\beq
\frac{\tau _{u}}{\tau _{u-1}}\p _{t_1 }
\frac{\tau _{u}}{\tau _{u+1}}
\p _{\bar t_1 }\varphi (u)=\varphi (u)\,,
\label{4cont1}
\eeq
\beq
\left (\p _{\bar t_1 }-
\frac{\tau _{u+1}}{\tau _{u}}\p _{t_1 }^{-1}
\frac{\tau _{u-1}}{\tau _{u}}\right )\varphi (u)=0\,,
\label{4cont2}
\eeq
respectively.

\subsection{Adjoint flows}

Finally, we extend the above scheme to incorporate more
general flows which we call {\it adjoint}.
Let
\beq
A_{l}^{a}=1+w\Delta _{l}^{-1}w^{*}\,,
\;\;\;\;\;\;\;\;
\Delta _{l}=e^{\p _{l}}-1
\label{a1}
\eeq
be a pseudo-difference operator, where $w$, $w^{*}$ are
as yet arbitrary functions of all the time variables.
By $a$ we denote the time variable corresponding to the
adjoint flow that we are going to define.
In this section, the reference flow is $l$.
The $M$-operator for an elementary discrete flow $p$
(see the figure) is
\beq
M_{l}^{p}=e^{\p _{l}}- \l_ {p}
\frac{\tau _{l}^{p}\tau _{l+1}^{p+1}}
{\tau _{l+1}^{p}\tau _{l}^{p+1}}\,,
\;\;\;\;\;\;\;\;
\l _{p}\equiv \l _{4}^{03}\,.
\label{aM}
\eeq

\vspace{0.2cm}

\begin{center}
\special{em:linewidth 0.4pt}
\unitlength 0.50mm
\linethickness{0.4pt}
\begin{picture}(114.67,64.00)
\emline{59.67}{9.67}{1}{108.67}{9.34}{2}
\emline{108.67}{59.67}{3}{60.00}{59.67}{4}
\put(84.33,59.67){\vector(1,0){0.2}}
\emline{84.00}{59.67}{5}{84.33}{59.67}{6}
\put(84.33,5.34){\makebox(0,0)[cc]{$l$}}
\put(84.33,63.67){\makebox(0,0)[cc]{$\bar l$}}
\put(54.33,35.34){\makebox(0,0)[cc]{$n$}}
\put(54.33,5.34){\makebox(0,0)[cc]{$\lambda _0$}}
\put(114.33,5.34){\makebox(0,0)[cc]{$\lambda _3$}}
\put(54.33,64.00){\makebox(0,0)[cc]{$\lambda _1$}}
\put(114.67,64.00){\makebox(0,0)[cc]{$\lambda _2$}}
\put(59.86,34.76){\vector(0,1){0.2}}
\emline{59.86}{34.69}{7}{59.86}{34.76}{8}
\put(83.32,9.53){\vector(1,0){0.2}}
\emline{83.10}{9.53}{9}{83.32}{9.53}{10}
\emline{59.67}{9.67}{11}{59.67}{13.67}{12}
\emline{59.67}{15.00}{13}{59.67}{19.00}{14}
\emline{59.67}{20.34}{15}{59.67}{24.34}{16}
\emline{59.67}{25.67}{17}{59.67}{29.67}{18}
\emline{59.67}{31.00}{19}{59.67}{35.00}{20}
\emline{59.67}{36.67}{21}{59.67}{40.67}{22}
\emline{59.67}{42.00}{23}{59.67}{46.00}{24}
\emline{59.67}{47.34}{25}{59.67}{51.34}{26}
\emline{59.67}{52.67}{27}{59.67}{56.67}{28}
\emline{59.67}{58.00}{29}{59.67}{59.67}{30}
\emline{59.33}{9.67}{31}{18.33}{46.67}{32}
\put(14.33,49.67){\makebox(0,0)[cc]{$\lambda _4$}}
\put(32.76,33.63){\vector(-1,1){0.2}}
\emline{33.45}{33.00}{33}{32.76}{33.63}{34}
\put(32.23,27.29){\makebox(0,0)[cc]{$p$}}
\end{picture}
\end{center}

\begin{prop}
The commutativity condition
\beq [e^{-\p _{a}} A_{l}^{a},\,
e^{-\p _{p}} M_{l}^{p}]=0
\label{a2}
\eeq
holds only if $w$ and $w^{*}$ satisfy the linear equations
\beq
\left \{ \begin{array}{l}
\big (e^{\p _{l}}+\l _{p}V_{l}^{p, a+1}\big )w_{l}^{p}
=\omega w_{l}^{p+1} \\ \\
\big (e^{-\p _{l}}+\l _{p}V_{l}^{p, a}\big )w^{* p+1}_{l}
=\omega w^{* p}_{l}\,,
\end{array} \right.
\label{a3}
\eeq
where
$$
V_{l}^{p,a}\equiv \frac{\tau _{l}^{p,a}\tau _{l+1}^{p+1, a}}
{\tau _{l+1}^{p,a}\tau _{l}^{p+1, a}}
$$
and $\omega$ is an arbitrary constant.
\end{prop}
The proof is by straightforward computation.
Equations (\ref{a3}) are necessary conditions for vanishing
of the pseudo-difference part of the commutator. Here are
main steps of the proof. Eq.\,(\ref{a2}),
$$
\left ( e^{\p _{l}}-\l _{p} V_{l}^{p, a+1}\right )
\left (1+w_{l}^{p}\Delta _{l}^{-1}w_{l}^{* p}\right )
=\left (1+w_{l}^{p+1}\Delta _{l}^{-1}w_{l}^{* p+1}\right )
\left ( e^{\p _{l}}-\l _{p} V_{l}^{p, a}\right ),
$$
can be rewritten as
\begin{eqnarray}
&& \l _{p} \left ( V_{l}^{p,a+1}-V_{l}^{p,a}\right )
+w_{l}^{p+1}w^{* p+1}_{l-1}-w_{l+1}^{p}w^{* p}_{l}
\nonumber \\
&=& \left ( w_{l+1}^{p}-\l _{p}V_{l}^{p,a+1}w_{l}^{p}-
\omega w_{l}^{p+1}\right )
\Delta _{l}^{-1}w^{* p}_{l}
\nonumber \\
&-& w_{l}^{p+1}\Delta _{l}^{-1}
\left ( w_{l-1}^{* p+1}-\l _{p}V_{l}^{p,a}w_{l}^{* p+1}-
\omega w_{l}^{* p}\right ).
\label{a4}
\end{eqnarray}
Since the l.h.s. does not contain negative powers of
$\Delta _{l}$, the r.h.s. should be zero. This condition
implies eqs.\,(\ref{a3}).

The ALP for the left adjacent $p$ to the reference flow $l$
and its adjoint read (cf. (\ref{S3}))
\beq
\left \{ \begin{array}{l}
\psi ^{p}(l+1)-\l _{p}V_{l}^{p}\psi ^{p}(l)=\psi ^{p+1}(l)\\ \\
\psi ^{* p}(l-1)-\l _{p}V_{l-1}^{p-1}\psi ^{* p}(l)=
\psi ^{* p-1}(l)\end{array}\right.\,.
\label{a5}
\eeq
Whence we identify
$$
w_{l}^{p,a}=\psi ^{p, a+1}(l)\,,
\;\;\;\;\;\;\;\;
w_{l}^{* p,a}=\psi ^{* p, a}(l +1)\,,
$$
where $\psi$, $\psi ^{*}$ are arbitrary solutions to the linear
problems (\ref{a5}).

The operator (\ref{a1}) acquires the form
\beq
A_{l}^{a}=1+\psi ^{a+1}(l)\Delta _{l}^{-1}\psi ^{* a}(l+1)\,.
\label{a6}
\eeq
The commutativity condition (\ref{a2}) is equivalent to
a nonlinear equation for $\tau _{l}^{p, a}$. The adjoint flow
$a$ is defined by two {\it arbitrary} solutions
$\psi$, $\psi ^{*}$ to the linear problems (\ref{a5}).
For continuous hierarchies, pseudo-differential analogues
of the operators (\ref{a6}) and corresponding adjoint flows
were studied in \cite{Orl}.

As an example, let us show that taking
$\psi$, $\psi ^{*}$ to be the Baker-Akhiezer function
$\psi (l;z)$ and its dual, one reproduces eq.\,(\ref{4-term3}).
According to the prescription of Sect.\,5.3, the Baker-Akhiezer
function and its dual read
\beq
\psi ^{a}(l;z)=\left.
\zeta ^{a}(\l _{z}^{03})^{l}
\frac{ \tau _{p_z +1}^{l,a}}
{ \tau _{p_z }^{l,a}}\right |_{p_z =0},
\;\;\;\;\;\;
\psi ^{* a}(l;z)=\left.
\zeta ^{-a}(\l _{z}^{03})^{-l}
\frac{ \tau _{p_z -1}^{l,a}}
{ \tau _{p_z }^{l,a}}\right |_{p_z =0},
\label{psia}
\eeq
where $\zeta ^{a}$ is a normalization factor specified below.
Here $p_z$ is the time variable corresponding to the
flow $\overrightarrow{\l _{0} z}$ left adjacent to $l$.
Let $z\rightarrow \l _{1}$. In the limit $p_z$ coincides with $u$.
Plugging (\ref{psia}) with $\zeta =\l _{z}^{02}$ into (\ref{a6}),
we do reproduce the operator in the r.h.s. of eq.\,(\ref{4-term3}).

\section{On the hierarchy of bilinear difference equations}

Integrable partial differential equations always can be included in
an {\it infinite hierarchy}. Infinite number of commuting flows
generates infinite number of evolution equations.

The hierarchies of discrete integrable equations are less studied.
First of all, it is not quite clear what are "higher discrete flows"
on the space of pseudo-difference operators. An understanding of this
matter is necessary if one is going to extend the Zakharov-Shabat
formalism to the higher
Hirota equations known in the literature \cite{Miwa2},\,\cite{OHTI}.

There are two "complimentary" points of view on this matter.
First, one might consider 3-term HBDE
(\ref{hir1}) as a counterpart of the whole infinite hierarchy itself.
In this case it should be understood as an infinite set of equations
(continuously numbered by labels $\l _{\a}$'s) for a function
of infinitely many variables $l_{\a \b }$ associated with
$\overrightarrow{\l _{\a} \l _{\b }}$.
Second, one might expect that
composite discrete flows are good candidates
for true analogues of the higher continuous flows. This
is justified by analysing the continuum limit. Indeed,
to get a higher
continuous flow as a limiting case, one should start from
a composite discrete flow with specially adjusted labels.

Our goal in this section is
to show how these two approaches could be
consistent with each other. A natural conjecture is that
the $N$-term "higher" Hirota
equations for a function of $N$ variables are consequences of the
basic 3-term equation (\ref{hir1}) treated {\it
as a hierarchy}. This means that the 3-term equation is supposed
to hold for each
triple of the $N$ variables with
the corresponding $\lambda _{\alpha}$'s. To
support this conjecture, the case of 4-term HBDE is considered in detail.
Besides, an extension of the Zakharov-Shabat scheme to this case is
suggested.

\subsection{Higher equations of the hierarchy}

Higher Hirota difference equations known in the
literature \cite{OHTI} are written for a function
$\tau (l_1 , \ldots \,, l_N )$ of $N$ variables.
They have the form
\beq
\left | \begin{array}{llllllll}
1& z_1 & z_{1}^{2}&&\ldots &&z_{1}^{N-2}& \tau _1 \hat \tau _{1}\\
&&&&&&&\\
1& z_2 & z_{2}^{2}&&\ldots &&z_{2}^{N-2}& \tau _2 \hat \tau _{2}\\
&&&&&&&\\
\ldots &\ldots &\ldots &&\ldots && \ldots &\ldots \\
&&&&&&&\\
1& z_N & z_{N}^{2}&&\ldots &&z_{N}^{N-2}& \tau _N \hat \tau _{N}
\end{array}\right | =0\,,
\label{h1}
\eeq
where $z_i$ are arbitrary constants and
\begin{eqnarray}
&&\tau _{i}\equiv \tau (l_1 , l_2 ,\ldots ,l_{i-1}, l_{i}+1, l_{i+1},
\ldots , l_N )\,, \nonumber\\
&&\hat \tau _{i}\equiv
\tau (l_1 +1, l_2 +1,\ldots ,l_{i-1}+1, l_{i}, l_{i+1}+1,
\ldots , l_N +1)\,.
\label{h1a}
\end{eqnarray}
In the more compact form they read
\beq
\sum _{j=1}^{N} \Lambda _{j}\tau _{j}\hat \tau _{j}=0\,.
\label{h2}
\eeq
The constants $\Lambda _j$ are subject to the only condition
$\sum _{j=1}^{N}\Lambda _{j}=0.$ For $N=3$ one gets usual
Hirota's bilinear difference equation, where $z_i$ are certain
rational functions of $\l _{\a}$'s. Like in the 3-term case,
the variables $l_i$ will be identified with elementary discrete
flows.

The transformation
\beq
\tau (l_1 ,\ldots , l_N )\rightarrow
\exp \left [ \frac{1}{2N-4}\sum _{k=1}^{N}\log \Lambda _{k}
\left (\sum _{j=1, \ne k}^{N}l_{j}\right )^{2}
\right ]
\tau (l_1 ,\ldots , x_N )
\label{h3}
\eeq
sends eq.\,(\ref{H2}) to the canonical form,
\beq
\sum _{j=1}^{N} \tau _{j}\hat \tau _{j}=0\,,
\label{h4}
\eeq
which does not contain any free parameters.

An analogue of Hirota's original form,
\beq
\left (\sum _{j=1}^{N}
\Lambda _{j}\exp (D_{x_{j}})\right )
\tau \cdot \tau =0\,,
\label{h5}
\eeq
is obtained from eq.\,(\ref{H2}) by the linear change of
variables
\beq
x_j = -l_j +\frac{1}{N-2}\sum _{i=1}^{N}l_i
\label{h6}
\eeq
which generalizes (\ref{lin1a}).

\subsection{Zero curvature conditions for composite flows}

We are going to show that the zero curvature
condition written for the composite discrete flows
introduced in Sect.\,3.1 lead to the "higher" bilinear
equations of the form (\ref{h1}). The
"higher" $M$-operators are obtained as products of the
elementary ones.

In this section we deal with the following graph of flows:

\vspace{0.5cm}

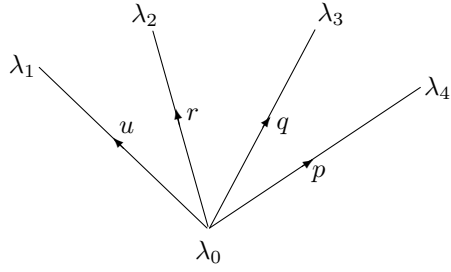
\begin{center}
\special{em:linewidth 0.4pt}
\unitlength 1mm
\linethickness{0.4pt}
\begin{picture}(59.99,38.80)
\emline{29.33}{10.33}{1}{7.00}{31.67}{2}
\emline{29.67}{10.33}{3}{57.67}{29.00}{4}
\emline{29.67}{10.33}{5}{43.67}{36.67}{6}
\emline{29.44}{10.52}{7}{22.10}{36.52}{8}
\put(17.01,22.09){\vector(-1,1){0.2}}
\emline{18.21}{20.89}{9}{17.01}{22.09}{10}
\put(25.37,25.37){\vector(-1,4){0.2}}
\emline{25.67}{23.88}{11}{25.37}{25.37}{12}
\put(37.61,24.92){\vector(3,4){0.2}}
\emline{36.71}{23.73}{13}{37.61}{24.92}{14}
\put(43.28,19.40){\vector(3,2){0.2}}
\emline{41.93}{18.50}{15}{43.28}{19.40}{16}
\put(29.55,7.16){\makebox(0,0)[cc]{$\lambda _0$}}
\put(4.78,31.94){\makebox(0,0)[cc]{$\lambda _1$}}
\put(21.04,38.65){\makebox(0,0)[cc]{$\lambda _2$}}
\put(45.81,38.80){\makebox(0,0)[cc]{$\lambda _3$}}
\put(59.99,28.95){\makebox(0,0)[cc]{$\lambda _4$}}
\put(44.17,17.46){\makebox(0,0)[cc]{$p$}}
\put(39.40,23.88){\makebox(0,0)[cc]{$q$}}
\put(27.46,25.52){\makebox(0,0)[cc]{$r$}}
\put(18.65,23.73){\makebox(0,0)[cc]{$u$}}
\end{picture}
\end{center}

\noindent
The reference flow is $u$. Other notation is clear from the picture.
For simplicity, we consider
left adjacent flows only.
All what follows can be easily
reformulated in terms of the right adjacent flows.

According to the definition of composite flows
(Sect.\,3.1), we introduce
a "higher" $M$-operator $M_{u}^{qr}$ generating evolution in
the composite flow labeled by
the pair of vectors
$\overrightarrow{\l _{0}\l _{2}}$,
$\overrightarrow{\l _{0}\l _{3}}$
as the product of elementary $M$-operators of the form (\ref{Z1}):
\beq
M^{qr}_{u} = e^{\p _r }M^{q}_{u}
e^{-\p _r }M^{r}_{u}\,,
\label{H1}
\eeq
or, in more details,
\beq
M^{qr}_{u}(q,r) = M^{q}_{u}(q,r+1)
M^{r}_{u}(q,r)\,.
\label{H2}
\eeq
Due to the zero curvature condition for elementary flows
$M_{u}^{qr}=M_{u}^{rq}$.
The compatibility
of this composite flow with an elementary flow
$\overrightarrow{\l _{0}\l _{4}}$ reads
\beq
M^{p}_{u}(p, q+1, r+1)
M^{qr}_{u}(p,q,r)=
M^{qr}_{u}(p+1,q,r)
M^{p}_{u}(p,q,r)\,.
\label{H3}
\eeq
Clearly, this zero curvature condition is an immediate consequence
of eq.\,(\ref{Z3}) and definition (\ref{H1}) provided eq.\,(\ref{Z3})
holds for any pair of the flows from the triple
$p,q,r$.

A few words about the notation.
For the notational simplicity,
we put
$$
\l _{2}^{01}=\l _{r}\,, \;\;\;\;\;\;
\l _{3}^{01}=\l _{q}\,, \;\;\;\;\;\;
\l _{4}^{01}=\l _{p}\,.
$$
Since in this section we deal with too many variables,
it is also convenient to change slightly the notation
for the $\tau$-function writing the argument $u$ in brackets:
$\tau _{u}\rightarrow \tau (u)$. Other variables are indices,
as before. This will remind us that $u$ is the reference
variable. Needless to say that unshifted variables
will be skipped when possible. Throughout this section,
the notation
\beq
V_{u}^{(p)} (p,q,r)=
\frac{ \tau ^{p,q,r}(u)\tau ^{p+1, q,r}(u+1)}
{ \tau ^{p+1,q,r}(u)\tau ^{p,q,r}(u+1)}
\label{H4}
\eeq
is used.

Now we are ready to elaborate eq.\,(\ref{H3}) explicitly.
Eq.\,(\ref{H3}) reads
\begin{eqnarray}
&&\left ( e^{\p _{u}}-\lambda _{p}V_{u}^{(p)}(p,q+1,r+1)\right )
\left ( e^{\p _{u}}-\lambda _{q}V_{u}^{(q)}(p,q,r+1)\right )
\left ( e^{\p _{u}}-\lambda _{r}V_{u}^{(r)}(p,q,r)\right )
\nonumber \\
&=&\left ( e^{\p _{u}}-\lambda _{q}V_{u}^{(q)}(p+1,q,r+1)\right )
\left ( e^{\p _{u}}-\lambda _{r}V_{u}^{(r)}(p+1,q,r)\right )
\left ( e^{\p _{u}}-\lambda _{p}V_{u}^{(p)}(p,q,r)\right ).
\label{H5}
\end{eqnarray}
The coefficients in front of $e^{2\p _{u}}$ give
\begin{eqnarray}
&&\lambda _{p}V_{u}^{(p)}(p,q+1,r+1)+
\lambda _{q}V_{u+1}^{(q)}(p,q,r+1)+
\lambda _{r}V_{u+2}^{(r)}(p,q,r)
\nonumber \\
&=&\lambda _{p}V_{u+2}^{(p)}(p,q,r)+
\lambda _{q}V_{u}^{(q)}(p+1,q,r+1)+
\lambda _{r}V_{u+1}^{(r)}(p+1,q,r)\,.
\label{H6}
\end{eqnarray}
This is a direct corollary of
3-term HBDE. To see this, recall eq.\,(\ref{a}) from the
proof of Theorem 4.1. In the present notation it reads
\beq
\lambda _{p}V_{u}^{(p)}(p,q+1,r+1)+
\lambda _{q}V_{u+1}^{(q)}(p,q,r+1)=
\lambda _{p}V_{u+1}^{(p)}(p,q,r+1)+
\lambda _{q}V_{u}^{(q)}(p+1,q,r+1)\,.
\label{H7}
\eeq
Here $r+1$ enters as a parameter. A similar equation can be written
down for the pair $p,r$, where now $q$ enters as a parameter:
\beq
\lambda _{p}V_{u+1}^{(p)}(p,q,r+1)+
\lambda _{r}V_{u+2}^{(r)}(p,q,r)=
\lambda _{p}V_{u+2}^{(p)}(p,q,r)+
\lambda _{r}V_{u+1}^{(r)}(p+1,q,r)\,.
\label{H8}
\eeq
Taking their sum, we obtain eq.\,(\ref{H6}).

Comparing coefficients in front of $e^{\p _{u}}$, we get:
\begin{eqnarray}
&&\lambda _{p}\lambda _{q}V_{u}^{(q)}(p,q,r+1)V_{u}^{(p)}(p,q+1,r+1)+
\lambda _{q}\lambda _{r}V_{u+1}^{(r)}(p,q,r)V_{u+1}^{(q)}(p,q,r+1)
\nonumber \\
&+&\lambda _{p}\lambda _{r}V_{u}^{(p)}(p,q+1,r+1)V_{u+1}^{(r)}(p,q,r)
\nonumber \\
&=&\lambda _{p}\lambda _{q}V_{u}^{(q)}(p+1,q,r+1)V_{u+1}^{(p)}(p,q,r)+
\lambda _{q}\lambda _{r}V_{u}^{(r)}(p+1,q,r)V_{u}^{(q)}(p+1,q,r+1)
\nonumber \\
&+&\lambda _{p}\lambda _{r}V_{u+1}^{(p)}(p,q,r+1)V_{u+1}^{(r)}(p,q,r)\,.
\label{H9}
\end{eqnarray}
In a similar way,
it is easy to show that this equality also follows from HBDE.
Of course this is a trivial result provided the zero curvature
conditions (\ref{Z3}) hold for both elementary $M$-operators
in the product (\ref{H2}).

The fact which is {\it not} obvious from the very beginning is that
eqs.\,(\ref{H7}), (\ref{H8}), (\ref{H9}) imply one of the {\it higher
Hirota equations}
(namely, (\ref{h1}) at $N=4$). Proceeding like in the
proof of Proposition 4.1, we rewrite (\ref{H9}) in the form
\begin{eqnarray}
&&\frac{\tau (u+1)}{\tau (u+2)}\left (
\lambda _{q}\lambda _{r}
\frac{\tau ^{q+1, r+1}(u+2)}{\tau ^{q+1, r+1}(u+1)}-
\lambda _{p}\lambda _{r}
\frac{\tau ^{p+1, r+1}(u+2)}{\tau ^{p+1, r+1}(u+1)} \right.
\nonumber \\
&-&\left. \lambda _{p}\lambda _{q}
\frac{\tau ^{p+1, r+1}(u)\tau ^{p+1, q+1, r+1}(u+1)\tau ^{p+1}(u+2)}
{\tau ^{p+1, q+1, r+1}(u)\tau ^{p+1, r+1}(u+1)\tau ^{p+1}(u+1)}
\right )
\nonumber \\
&=&\frac{\tau ^{p+1, q+1, r+1}(u+1)}{\tau ^{p+1, q+1, r+1}(u)}\left (
\lambda _{q}\lambda _{r}
\frac{\tau ^{p+1}(u)}{\tau ^{p+1}(u+1)}-
\lambda _{p}\lambda _{q}
\frac{\tau ^{r+1}(u)}{\tau ^{r+1}(u+1)} \right.
\nonumber \\
&-&\left. \lambda _{p}\lambda _{r}
\frac{\tau ^{q+1, r+1}(u)\tau (u+1)\tau ^{r+1}(u+2)}
{\tau ^{r+1}(u+1)\tau ^{q+1, r+1}(u+1)\tau (u+2)}
\right )
\label{H11}
\end{eqnarray}
(cf.\,(\ref{a})). Multiplying both sides by
$(\lambda _{q}-\lambda _{r})\tau ^{q+1, r+1}(u+1)
\tau ^{p+1}(u+1)$ and using HBDE in the form
\beq
(\lambda _{q}-\lambda _{r})\tau ^{q+1, r+1}(u)\tau (u+1)=
\lambda _{q}\tau ^{r+1}(u)\tau ^{q+1}(u+1)-
\lambda _{r}\tau ^{q+1}(u)\tau ^{r+1}(u+1)\,,
\label{H12}
\eeq
we get
\beq
\frac{\tau (u+1)\tau ^{p+1, q+1, r+1}(u)}
{\tau (u+2)
\tau ^{p+1, q+1, r+1}(u+1)}A^{p,q,r}(u)=B^{p,q,r}(u)\,,
\label{H13}
\eeq
where
\begin{eqnarray*}
&&A^{p,q,r}(u)=
(\lambda _{q}-\lambda _{r})\lambda _{q}\lambda _{r}
\tau ^{q+1, r+1}(u+2)\tau ^{p+1}(u+1) \nonumber \\
&-&
\lambda _{p}\lambda _{q}
\frac{\tau ^{p+1, q+1}(u+2)}{\tau ^{p+1, q+1, r+1}(u)}
(\lambda _{q}\tau ^{p+1, r+1}(u)\tau ^{q+1, r+1}(u+1))\nonumber \\
&+&
\lambda _{p}\lambda _{r}
\frac{\tau ^{p+1, r+1}(u+2)}{\tau ^{p+1, q+1, r+1}(u)}
(\lambda _{r}\tau ^{p+1, q+1}(u)\tau ^{q+1, r+1}(u+1))\,,
\nonumber \\
&&B^{p,q,r}(u)=
(\lambda _{q}-\lambda _{r})\lambda _{q}\lambda _{r}
\tau ^{p+1}(u)\tau ^{q+1, r+1}(u+1) \nonumber \\
&-&
\lambda _{p}\lambda _{q}
\frac{\tau ^{r+1}(u)}{\tau (u+2)}
(\lambda _{q}\tau ^{p+1}(u+1)\tau ^{q+1}(u+2))\nonumber \\
&+&
\lambda _{p}\lambda _{r}
\frac{\tau ^{q+1}(u)}{\tau (u+2)}
(\lambda _{r}\tau ^{p+1}(u+1)\tau ^{r+1}(u+2))\,.
\end{eqnarray*}
The last two terms can be transformed further using
eq.\,(\ref{H12}). The result is
\begin{eqnarray*}
&&A^{p,q,r}(u)=
h^{p,q,r}(u+1)-
\lambda _{p}^{2}(\lambda _{q}-\lambda _{r})
\frac{\tau ^{q+1, r+1}(u)\tau ^{p+1, q+1, r+1}(u+1)\tau ^{p+1}(u+2)}
{\tau ^{p+1, q+1, r+1}(u)}\,,
\nonumber \\
&&B^{p,q,r}(u)=
h^{p,q,r}(u)-
\lambda _{p}^{2}(\lambda _{q}-\lambda _{r})
\frac{\tau ^{q+1, r+1}(u)\tau (u+1)\tau ^{p+1}(u+2)}{\tau (u+2)}\,,
\end{eqnarray*}
where
\begin{eqnarray}
&&h^{p,q,r}(u) \!=\!
\lambda _{q}\lambda _{r}(\lambda _{q}-\lambda _{r})
\tau ^{p+1}(u)
\tau ^{q+1, r+1}(u+1)\!+\!
\lambda _{r}\lambda _{p}(\lambda _{r}-\lambda _{p})
\tau ^{q+1}(u)
\tau ^{p+1, r+1}(u+1)\nonumber \\
&+\!&\lambda _{p}\lambda _{q}(\lambda _{p}-\lambda _{q})
\tau ^{r+1}(u)\tau ^{p+1, q+1}(u+1)\,.
\label{H18}
\end{eqnarray}
Finally, eq.\,(\ref{H13}) turns into
\beq
\frac{\tau (u+1)\tau ^{p+1, q+1, r+1}(u)}
{\tau (u+2)
\tau ^{p+1, q+1, r+1}(u+1)}=
\frac{h^{p,q,r}(u)}
{h^{p,q,r}(u+1)}
\,,
\label{H19}
\eeq
which leads to
\beq
h^{p,q,r}(u)+(\lambda _{p}-\lambda _{q})
(\lambda _{q}-\lambda _{r})
(\lambda _{r}-\lambda _{p}) \tau (u+1)
\tau ^{p+1, q+1, r+1}(u)=0\,.
\label{H20}
\eeq
Eq.\,(\ref{H20}) has the form (\ref{h1}) at $N=4$. This completes
the calculation.

$M$-operators for arbitrary composite flows
can be defined as a straightforward generalization
of (\ref{H1}):
\beq
M^{p_N \ldots p_{2}p_{1}}_{u}=
\exp \left ( \sum _{j=1}^{N}\p _{p_j}\right )
\prod _{i=1}^{N} \left ( e^{-\p _{p_i}}M^{p_i}_{u}\right ).
\label{H21}
\eeq
Note that the order of the operators in the product is not essential
since the operators
$e^{-\p _{p_i}}M^{p_i}_{u}$ commute due to the zero curvature
condition (\ref{Z3}). For simplicity it is assumed that
all the flows $p_i$ are left adjacent to
the reference flow $u$. Operators (\ref{H21}) generates discrete
analogues of higher flows of the KP hierarchy (see the corresponding
graph of flows in Sect.\,3.2).
There is now straightforward to write down
similar operators for right adjacent flows which would generate
higher flows of the discrete 2DTL hierarchy.

{\bf Conjecture.} All the higher HBDE (\ref{h1}) follow from the
compatibility of the composite flows generated by the $M$-operators
(\ref{H21}) and elementary discrete flows.

The calculation given above shows that the conjecture is true for
the 4-term bilinear equation.
Unfortunately, we are not aware of any other proof
than this sophisticated calculation that is hard to perform in the
general case. The conjecture claims that all the higher
bilinear equations are
corollaries of eq.\,(\ref{hir1}) considered as a hierarchy, i.e., valid
for all triples of adjacent flows.

\section{Reductions of Hirota's equations}

The hierarchy of discrete Hirota's equations admits several important
reductions. {\it Reduction} means imposing a constraint compatible
with the hierarchy, so that the number of independent variables
gets reduced. In this way one is able to construct discrete analogues
of KdV, SG and other interesting equations.

The simplest way to impose a constraint
is to require the $\tau$-function
to be {\it stationary} with respect to a
particular flow $s$ (possibly, up
to a "gauge" transformation (\ref{inv1})).
Non-trivial examples emerge when the stationary
flow is composite. As for the
commutation representation, there are two options.

First, the stationary
flow can be reference. Then $M$-operators become free of
differentiation because the symbol $\p _{s}$ commutes with all the
coefficients. In other words, $\p _{s}$ can be considered as a
$c$-number and identified with a {\it spectral parameter}. This is
a natural origin of $M$-operators depending on
(rational) spectral parameter.

Alternatively,
one may take as the reference flow any flow $y$ other than $s$. Then
any $M$-operator $M^{(f)}$ generating a flow $f$ contains operators
$\p _{y}$. Since coefficients of the operator $M^{(f)}$ do not depend
on $s$, the compatibility condition for the flows $s$ and $f$ acquires
the Lax-type form
$$
M^{(s)}(f+1)M^{(f)}=
M^{(f)}M^{(s)}(f)\,,
$$
where $M^{(f)}$ plays the role of {\it Lax operator}. Unlike the
Zakharov-Shabat scheme, where each zero curvature condition
involves two different time flows (apart from the reference flow),
Lax equations are written for each flow separately. The Lax
equation represents the time flow as a similarity (i.e.,
isospectral) transformation of the Lax operator.
It is natural to call $M^{(s)}$ {\it the $L$-operator of the reduced
hierarchy}. This is a
natural origin of $L$-operators which are difference
(or differential) rather than pseudo-difference (or pseudo-differential)
operators\footnote{The Lax operator for the hierarchy without
any reduction is in general an infinite series in negative powers
of first order difference operators. The theory based on the
Lax representation with Lax operators of this kind is
not considered here.}.
In the light of this general scheme, we
pass to considering examples.

\subsection{KdV-like reductions}

{\it 0). Discrete d'Alembert equation
(a trivial example)}. Let $u$ be an elementary
discrete flow dealt with in Sects.\,3,4. The stationarity condition with
respect to this flow, $\tau (u+1)=
\tau (u)$, immediately leads (see (\ref{hir1})) to the relation
\beq
\tau ^{l+1, m}\tau ^{l, m+1}=
\tau ^{l,m}\tau ^{l+1, m+1}\,,
\label{R1}
\eeq
where $l,m$ are any other two elementary flows. This is the discrete
2D d'Alembert equation written in "light cone" coordinates. The general
solution is $\tau ^{l,m}=\chi _{+}(l)\chi _{-}(m)$ with arbitrary
$\chi _{\pm }$; this is just the allowed "gauge" freedom (\ref{inv1a}) of
the $\tau$-function. So we see that this reduction is too strong --
it respects trivial solutions only. In this sense it is incompatible
with the hierarchy. To get nontrivial examples, one should either impose
stationarity conditions with respect to the composite ("higher")
flows or impose periodic conditions in $u$ with period $N>1$
(e.g. $\tau (u+2)=\tau (u)$).

{\it 1). Discrete KdV equation}.
Consider the graph of flows

\vspace{0.2cm}

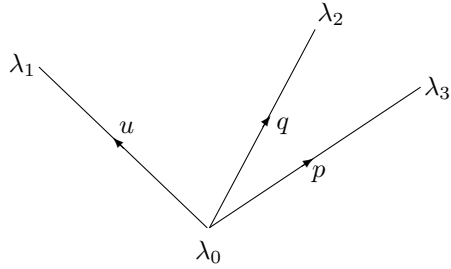
\begin{center}
\special{em:linewidth 0.4pt}
\unitlength 1.00mm
\linethickness{0.4pt}
\begin{picture}(59.99,38.80)
\emline{29.33}{10.33}{1}{7.00}{31.67}{2}
\emline{29.67}{10.33}{3}{57.67}{29.00}{4}
\emline{29.67}{10.33}{5}{43.67}{36.67}{6}
\put(17.01,22.09){\vector(-1,1){0.2}}
\emline{18.21}{20.89}{7}{17.01}{22.09}{8}
\put(37.61,24.92){\vector(3,4){0.2}}
\emline{36.71}{23.73}{9}{37.61}{24.92}{10}
\put(43.28,19.40){\vector(3,2){0.2}}
\emline{41.93}{18.50}{11}{43.28}{19.40}{12}
\put(29.55,7.16){\makebox(0,0)[cc]{$\lambda _0$}}
\put(4.78,31.94){\makebox(0,0)[cc]{$\lambda _1$}}
\put(45.81,38.80){\makebox(0,0)[cc]{$\lambda _2$}}
\put(59.99,28.95){\makebox(0,0)[cc]{$\lambda _3$}}
\put(44.17,17.46){\makebox(0,0)[cc]{$p$}}
\put(39.40,23.88){\makebox(0,0)[cc]{$q$}}
\put(18.65,23.73){\makebox(0,0)[cc]{$u$}}
\end{picture}
\end{center}

\noindent
and put $\l _{2}^{01}=\l _{q}$,
$\l _{3}^{01}=\l _{p}$ for brevity. In this notation,
eq.\,(\ref{hir1}) takes the form
\beq
\lambda _{q}
\tau _{u}^{p+1, q}\tau _{u+1}^{p, q+1}
-\lambda _{p}
\tau _{u}^{p, q+1}\tau _{u+1}^{p+1, q}+
(\lambda _{p}-\lambda _{q})
\tau _{u}^{p+1, q+1}\tau _{u+1}^{p,q} =0\,.
\label{hir1a}
\eeq
To get the discrete KdV reduction, we impose the constraint
\beq
\tau ^{p+1,q+1}_{u}=\tau ^{p,q}_{u}\,,
\label{stat1}
\eeq
i.e., the $\tau$-function is stationary with respect to the
composite flow labeled by the pair of vectors
$\overrightarrow{\l _{0}\l _{2}}$, $\overrightarrow{\l _{0}\l _{3}}$.
This condition converts the 3-dimensional equation (\ref{hir1a})
into the following 2-dimensional one:
\beq
\lambda _{q} \tau _{u}^{p+1}\tau _{u+1}^{p-1}-\lambda _{p}
\tau _{u}^{p-1}\tau _{u+1}^{p+1}+
(\lambda _{p}-\lambda _{q})
\tau _{u}^{p}\tau _{u+1}^{p} =0\,.
\label{R7}
\eeq
This is the discrete KdV equation in the bilinear form
\cite{HirotaKdV},\,\cite{OHTI}.
The discrete KdV equation is also known in the form \cite{OHTI}
\beq
V(u,p)-V(u-1, p-1)=\kappa (V^{-1}(u, p-1)-V^{-1}(u-1, p))\,.
\label{DKdV}
\eeq
The equivalence to eq.\,(\ref{R7}) follows from the identification
$$
V(u,p)=\frac{\tau _{u}^{p}\tau _{u+1}^{p+1}}
{\tau _{u}^{p+1}\tau _{u+1}^{p}}\,,\;\;\;\;\;\;\;\;
\kappa = \frac{\lambda _{q}}{\lambda _{p}}\,.
$$

Let us turn to $M$ and $L$ operators.
Following the history of the KdV equation, we begin with
difference operators with scalar coefficients.
Let $u$ be the reference variable. Then the (composite) stationary
flow is generated by the $M$-operator of the type (\ref{H2}):
\beq
M^{pq}_{u}=M_{u}^{p}(p,q+1)M_{u}^{q}(p,q)=L^{(KdV)}=
e^{2\p_{u}}
-\left ( \lambda _{p}
\frac{\tau _{u}^{p-1}\tau _{u+1}^{p}}
{\tau _{u}^{p}\tau _{u+1}^{p-1}}+\lambda _{q}
\frac{\tau _{u+1}^{p}\tau _{u+2}^{p-1}}
{\tau _{u+1}^{p-1}\tau _{u+2}^{p}}\right )e^{\p _{u}}+
\lambda _{p}\lambda _{q}\,.
\label{R8}
\eeq
We call this 2-nd order difference
operator {\it the Lax operator of the discrete KdV equation}.
The spectral
problem $L^{(KdV)}\psi =E\psi$ is a discrete analogue of the
stationary Schr\"odinger equation that is
an auxiliary linear problem for KdV. The $p$-evolution generated by the
$M$-operator of the type (\ref{Z1}) is isospectral:
$L^{(KdV)}(p+1)=M_{u}^{p}L^{(KdV)}(p)(M_{u}^{p})^{-1}.$

If the reference flow coincides with the stationary one, we get
2$\times$2 matrix $M$-operators depending on a spectral parameter
$z$. The spectral parameter is an eigenvalue of the shift operator
along the stationary flow acting to the $\psi$-function:
$\exp (\p _{p}+\p _{q})\psi = z^2 \psi$.
Consider the vector
$$
\left (\begin{array}{c}
\psi _{u}^{p} \\  \\  \\
\chi _{u}^{p} \end{array} \right )=
\left (\begin{array}{c}
\psi ^{p}(u) \\  \\
\displaystyle{\frac{\tau _{u}^{p+1}}{\tau _{u}^{p}}}
\psi _{u}^{p+1} \end{array} \right ).
$$
Repeating the argument of
Sect.\,5.5, we obtain the following linear problems for
shifts in $u$ and $p$ respectively:
\beq
\left (\begin{array}{c}
\psi _{u+1}^{p} \\ \\  \\  \\
\chi _{u+1}^{p} \end{array} \right )=
\left ( \begin{array}{ccc}
\l _{p}\displaystyle{\frac{
\tau _{u}^{p}\tau _{u+1}^{p+1}}
{\tau _{u}^{p+1}\tau _{u+1}^{p}}} &&
\displaystyle{\frac{\tau _{u}^{p}}{\tau _{u}^{p+1}}}\\ && \\
z^2 \displaystyle{\frac{\tau _{u+1}^{p+1}}{\tau _{u+1}^{p}}}
&& \l _{q} \end{array} \right )
\left (\begin{array}{c}
\psi _{u}^{p} \\ \\ \\  \\
\chi _{u}^{p} \end{array} \right ),
\label{R9}
\eeq
\beq
\left (\begin{array}{c}
\psi _{u}^{p+1} \\ \\ \\ \\
\chi _{u}^{p+1} \end{array} \right )=
\left ( \begin{array}{ccc}
0 &&
\displaystyle{\frac{\tau _{u}^{p}}{\tau _{u}^{p+1}}}\\ && \\
z^2 \displaystyle{\frac{\tau _{u}^{p+2}}{\tau _{u}^{p+1}}}
&& \l _{q}-\l _{p} \end{array} \right )
\left (\begin{array}{c}
\psi _{u}^{p} \\ \\  \\  \\
\chi _{u}^{p} \end{array} \right ).
\label{R10}
\eeq
Compatibility of these linear problems yields the discrete
matrix Zakharov-Shabat equations with the spectral parameter $z$.

Equations for the first component $\psi _{u}^{p}$ read:
\beq
\psi _{u+1}^{p}+
\frac{1}{\l _{p}-\l _{q}}
\left ( \lambda _{q}^{2}
\frac{\tau _{u-1}^{p+1}\tau _{u+1}^{p-1}}
{\tau _{u-1}^{p}\tau _{u+1}^{p}}-
\lambda _{p}^{2}
\frac{\tau _{u-1}^{p-1}\tau _{u+1}^{p+1}}
{\tau _{u-1}^{p}\tau _{u+1}^{p}} \right )\psi _{u}^{p}=
(z^2 -\l _{p}\l _{q})\psi _{u-1}^{p}\,,
\label{R11}
\eeq
\beq
\psi _{u}^{p+1}+(\l _{p}-\l _{q})
\frac{ (\tau _{u}^{p})^{2} }{\tau _{u}^{p-1}\tau _{u}^{p+1}}
\psi _{u}^{p}=z^2 \psi ^{p-1}_{u}\,.
\label{R12}
\eeq
Eq.\,(\ref{R11}) coincides with the spectral problem for the Lax
operator (\ref{R8}) provided the $\tau$-function obeys the
bilinear relation (\ref{R7}).

{\it 2). Discrete Boussinesq equation \cite{Miwa3}.}
In this example the
$\tau$-function $\tau ^{p,q}_{u}$ satisfies
$\tau ^{p+1, q+1}_{u+1}=\tau ^{p,q}_{u}$.
Eq.\,(\ref{hir1a}) reduces
to
\beq
\lambda _{q}\tau ^{p+1, q}\tau ^{p-1,q}-
\lambda _{p}\tau ^{p, q+1}\tau ^{p,q-1}+
(\lambda _{p}-\lambda _{q})\tau ^{p+1, q+1}\tau ^{p-1,q-1}=0\,,
\label{R13}
\eeq
where $u$ enters implicitly as a parameter. Different kinds of
Lax and Zakharov-Shabat
representations for this equation can be written down
straightforwardly.

In a similar way, it is possible to define $A_{n}$-type reductions.
In this case the $\tau$-function obeys the condition
$\exp \big (\sum _{\alpha =1}^{n+1}\p _{p_{\alpha }} \big )
\tau =\tau$.
This becomes an actual reduction
for higher Hirota equations
with the number of variables greater than $n$.

\subsection{Discrete time 1D Toda chain and its relatives}

This group of examples includes the discrete time 1D Toda chain
(1DTC), discrete AKNS system (in particular, discrete nonlinear
Schr\"odinger equation), discrete time relativistic Toda chain
and discrete Heisenberg ferromagnet (HF).
These models differ by choosing dependent and independent variables
while the type of reduction is essentially the same.

Let the graph of flows and the notation be the same as in Sect.\,4.3.
Now the $\tau$-function is required
to be stationary with respect to the composite flow
labeled by the pair of vectors
$\overrightarrow{\l _{0}\l _{3}}$,
$\overrightarrow{\l _{1}\l _{2}}$, i.e.,
\beq
\tau _{n}^{l+1, \bar l}=\tau _{n}^{l, \bar l}\,.
\label{stat2}
\eeq
The stationary flow is generated by the
"composite" $M$-operator $e^{\p _{l}}
\bar M^{\bar l}_{n}e^{-\p _{l}}M^{l}_{n}$ with the reference flow
$n$.
This operator should be identified with the Lax operator of the
discrete time 1DTC:
\beq
L^{(TC)}=e^{\p _{n}}-
\left ( \nu
\frac{\tau _{n}^{l}\tau _{n+1}^{l+1}}
{\tau _{n}^{l+1}\tau _{n+1}^{l}}+ \mu
\frac{\tau _{n-1}^{l+1}\tau _{n+1}^{l}}
{\tau _{n}^{l}\tau _{n}^{l+1}}\right )
+\nu \mu \frac{\tau _{n-1}^{l}\tau _{n+1}^{l}}
{(\tau _{n}^{l})^{2}}
e^{-\p _{n}}\,.
\label{R2}
\eeq
This is a 2-nd order difference operator
in $n$ with scalar coefficients. The spectral problem
$L^{(TC)}\psi =E\psi $ is a 1D discrete {\it stationary}
Schr\"odinger equation (cf.\,(\ref{2.9})). The $l$-dynamics does not
change the spectrum of $L^{(TC)}$.

Replacing
$e^{\p _{y}}=e^{\p _{l}+\p _{\bar l}}
\rightarrow z^2$, we get from (\ref{M4}) an
$L$-$M$ pair for the discrete time 1DTC realized by 2$\times$2
matrices depending on the spectral parameter $z$:
\beq
L_{n}^{(TC)}(z)=\left ( \begin{array}{ccc}
z^2 + \nu
\displaystyle{\frac{ \tau _{n}^{l}\tau _{n+1}^{l+1} }
{ \tau _{n}^{l+1}\tau _{n+1}^{l}}}+
\mu
\displaystyle{\frac{ \tau _{n+1}^{l}
\tau _{n-1}^{l+1}}
{ \tau _{n}^{l}\tau _{n}^{l+1}}}&&
-\mu \nu \displaystyle{\frac{\tau _{n+1}^{l}}
{\tau _{n}^{l}}}\\ && \\
\displaystyle{\frac{\tau _{n}^{l }}
{\tau _{n+1}^{l}}}&&0 \end{array} \right ),
\label{R3}
\eeq
\beq
M_{n}^{(TC)}(z)=\left ( \begin{array}{ccc}
1&&
-\mu \nu \displaystyle{\frac{\tau _{n+1}^{l-1}}
{\tau _{n}^{l-1}}}\\ && \\
\displaystyle{\frac{\tau _{n-1}^{l}}
{\tau _{n}^{l}}}&&
-z^2 -\mu
\displaystyle{\frac{ \tau _{n+1}^{l-1}
\tau _{n-1}^{l} }
{ \tau _{n}^{l-1}\tau _{n}^{l}}}
\end{array} \right ).
\label{R4}
\eeq
Correspondingly, the 2D discrete Schr\"odinger equation (\ref{2.11})
becomes a 1D spectral problem in the variable $l$,
\beq
\psi _{n}^{l+1}+(\nu -\mu )
\frac{\tau _{n}^{l}\tau _{n+1}^{l-1}}
{\tau _{n}^{l+1}\tau _{n+1}^{l-1}}\psi _{n}^{l}=
z^2 \left ( \psi _{n}^{l} +\nu
\frac{\tau _{n}^{l-1}\tau _{n+1}^{l}}
{\tau _{n}^{l}\tau _{n+1}^{l-1}}\psi _{n}^{l-1}\right ),
\label{R5a}
\eeq
with the non-standard dependence on the spectral parameter.
Another useful form of this equation is
\beq
\bar \psi _{n}^{l+1}-\nu
\frac{\tau _{n}^{l-1}\tau _{n+1}^{l}}
{\tau _{n}^{l}\tau _{n+1}^{l-1}}\bar \psi _{n}^{l-1}=
\left ( z+(\mu -\nu )z^{-1} \frac{\tau _{n}^{l}\tau _{n+1}^{l}}
{\tau _{n}^{l+1}\tau _{n+1}^{l-1}}\right )
\bar \psi _{n}^{l}\,,
\label{R5}
\eeq
where $\bar \psi _{n}^{l}=z^{-l}\psi _{n}^{l}$.

The equation of motion in the bilinear form is
the 2-dimensional reduction of eq.\,(\ref{HBDE6}):
\beq
\nu \tau _{n}^{l+1}\tau _{n}^{l-1}-
\mu \tau _{n+1}^{l-1}\tau _{n-1}^{l+1}=
(\nu -\mu )(\tau _{n}^{l})^{2}\,,
\label{R6}
\eeq
which after the linear change $l\rightarrow m=l+n$,
$\tau _{n}^{l}\rightarrow \tau _{n}(m)$ coincides
with Hirota's original form of this equation \cite{Hirota2}:
\beq
(1+g^{-2})\tau _{n}(m+1)\tau _{n}(m-1)-
\tau _{n+1}(m)\tau _{n-1}(m)=g^{-2}(\tau _{n}(m))^{2}\,,
\label{TC1}
\eeq
where
$$
g^{-2}=\frac{\nu }{\mu }-1\,.
$$
In terms of the new dependent variable
\beq
\phi _{n}(m)=\log \frac{\tau _{n+1}(m)}{\tau _{n}(m)}
\label{TC2}
\eeq
eq.\,(\ref{TC1}) acquires the form
\beq
\exp \big ( \phi _{n}(m+1)+\phi _{n}(m-1)-2\phi _{n}(m)\big )=
\frac{ 1+g^{2}\exp (\phi _{n+1}(m)-\phi _{n}(m))}
{ 1+g^{2}\exp (\phi _{n}(m)-\phi _{n-1}(m))}
\label{TC3}
\eeq
that is the discrete time 1DTC equation studied by Yu.Suris
\cite{Suris1}.

The continuum limit in $m$ is straightforward: set $m \rightarrow
m/\epsilon $, $g^2 =-\epsilon ^{2}$, then
$\phi _{n}(m\pm 1)\rightarrow \phi _{n}\pm \epsilon \phi _{n}'+
\frac{1}{2}\epsilon ^{2}\phi _{n}''$ as $\epsilon \rightarrow 0$.
Expanding eq.\,(\ref{TC3}) in $\epsilon$, we get the familiar
1DTC equation:
\beq
\phi _{n}''=e^{\phi _{n}-\phi _{n-1}}-
e^{\phi _{n+1}-\phi _{n}}\,.
\label{TC4}
\eeq

It is interesting to note that eq.\,(\ref{TC3}) has another
continuum limit yielding the sine-Gordon equation. To see this, let us
redefine the field $\phi$ before taking the limit:
\beq
\phi _{n}(m)=i(-1)^{n-m}\varphi _{n}(m)\,,
\label{redef}
\eeq
so the equation reads
\beq
\exp \big ( i\varphi _{n+1}(m)+i\varphi _{n-1}(m)-i\varphi _{n}(m+1)
-i\varphi _{n}(m-1)\big )=
\frac{ 1+g^{-2}\exp (i\varphi _{n+1}(m)+i\varphi _{n}(m))}
{ 1+g^{-2}\exp (-i\varphi _{n-1}(m)-i\varphi _{n}(m))}.
\label{TC5}
\eeq
It is the field $\varphi _{n}(m)$ that is now assumed to have a
smooth continuum limit in $n,m$. Setting
$m \rightarrow
m/\epsilon $,
$n \rightarrow
n/\epsilon $,
$g^{-2}=-i\epsilon ^{2}$, $\epsilon \rightarrow 0$ and expanding
in $\epsilon$ as before, we get the SG equation
\beq
(\p _{n}^{2}-\p _{m}^{2})\varphi =2\sin (2\varphi )\,.
\label{TC6}
\eeq
(The same limit for the field $\phi$ in eq.\,(\ref{TC3}) would give
the d'Alembert equation
$(\p _{n}^{2}-\p _{m}^{2})\varphi =0$.) A discrete analogue of the
SG equation of a different kind is given below in Sect.\,8.3.

{\bf Remark 8.1}\, The discrete KdV reduction discussed
in Sect.\,8.1 is formally a particular case of the present
reduction when $\l _{0}$ and $\l _{1}$ merge. However, it is
more convenient to consider them separately.
Note also that there exists a continuum limit
leading to the KdV equation
from eq.\,(\ref{R6}), too.
In this sense the discrete time 1DTC is
sometimes considered as a discretization of the KdV equation.

The discrete version of the decoupled nonlinear Schr\"odinger equation
(called also the AKNS system) is essentially the same reduction, i.e.,
the stationary flow is the same. The difference is in picking out other
independent flows to be involved in the equations.
Specifically, instead of the
$n$-flow one may consider any other elementary discrete flow
$p$ left adjacent to $l$.

We would like to show how to derive the discrete AKNS system
directly from the bilinear Hirota equations. The basic bilinear
equations in question are
\beq
\lambda _{p}\tau _{n}^{p, \bar l +1}\tau _{n}^{p+1, \bar l}+
(\mu -\lambda _{p})
\tau _{n}^{p, \bar l }\tau _{n}^{p+1, \bar l +1}=
\mu \tau _{n+1}^{p, \bar l +1}\tau _{n-1}^{p+1, \bar l}\,,
\label{R17}
\eeq
\beq
\nu \tau _{n}^{p+1, \bar l +1}\tau _{n+1}^{p, \bar l}-
\lambda _{p}\tau _{n}^{p, \bar l }\tau _{n+1}^{p+1, \bar l +1}=
(\nu-\lambda _{p})
\tau _{n}^{p+1, \bar l }\tau _{n+1}^{p, \bar l +1}\,.
\label{R18}
\eeq
The first one is eq.\,(\ref{hir3}) written for the triple
$(p,n,\bar l)$. The second one is obtained from eq.\,(\ref{hir1}) written
for the triple $(l,p,n)$ by taking into account the invariance
$\tau ^{l+1, \bar l +1}=\tau ^{l,\bar l}$. It is easy to see that
in terms of the quantities
\beq
Q^{p,\bar l}=\frac{ \tau _{n+1}^{p, \bar l}}
{ \tau _{n}^{p, \bar l}}\,, \;\;\;\;\;\;
R^{p,\bar l}=\frac{ \tau _{n-1}^{p, \bar l}}
{ \tau _{n}^{p, \bar l}}
\label{R19}
\eeq
($n$ is fixed) these equations can be rewritten as follows:
\beq
(\nu -\lambda _{p})
(\lambda _{p}-\mu )
Q^{p, \bar l +1}=
(\lambda _{p}-\mu \, Q^{p, \bar l +1}R^{p+1, \bar l})
(\nu \, Q^{p, \bar l}-\lambda _{p}Q^{p+1, \bar l +1})\,,
\label{R20}
\eeq
\beq
(\nu -\lambda _{p})
(\lambda _{p}-\mu )
R^{p+1, \bar l }=
(\lambda _{p}-\mu \, Q^{p, \bar l +1}R^{p+1, \bar l})
(\nu \, R^{p+1, \bar l +1}-\lambda _{p}R^{p, \bar l })\,.
\label{R21}
\eeq
This system is equivalent to the discrete AKNS
system given in ref.\,\cite{Miwa3}.

There exists another choice of dependent variables which converts
eqs.\,(\ref{R6}), (\ref{R17}), (\ref{R18}) into a discrete analogue
of the relativistic Toda chain (RTC) \cite{Ruij}. Passing again to the
variable $m=l+n$, like in eq.\,(\ref{TC2}), we put
\beq
x_{m}(p)=-\log \frac{\tau _{n}^{p}(m+1)}{\tau _{n}^{p}(m)}\,.
\label{RTC1}
\eeq
The claim is that the equation of motion for $x_{m}(p)$ has the form
\beq
\!\! \frac{
\big (1\! -\! \alpha \exp (x_{m+1}(p)\!-\! x_{m}(p))\big )\!
\big (1\!-\!\beta \exp (x_{m}(p)\! -\! x_{m-1}(p-1))\big ) \!
\big (1\!-\!\gamma \exp (x_{m}(p)\!-\!x_{m}(p+1)\big ) }
{\big (1\!-\!\alpha \exp (x_{m}(p)\!-\!x_{m-1}(p))\big )  \!
\big (1\!-\!\beta \exp (x_{m+1}(p+1)\!-\!x_{m}(p))\big )  \!
\big (1\!-\!\gamma \exp (x_{m}(p-1)\!-\!x_{m}(p))\big ) }\!=\! 1\,,
\label{RTC2}
\eeq
where
\beq
\alpha = \frac{\nu }{\mu -\mu }\,,
\;\;\;\;\;
\beta = \frac{\nu (\mu -\lambda _{p})}
{\lambda _{p}(\mu -\nu )}\,,
\;\;\;\;\;
\gamma = \frac{\lambda _{p}}{\lambda _{p}-\mu }\,,
\label{RTC3}
\eeq
which is the discrete time RTC equation
suggested in ref.\,\cite{Suris2}.

Let us outline the way to derive eq.\,(\ref{RTC2})\footnote{The idea
of this derivation is due to S.Kharchev (unpublished).}. The basic
bilinear relations
(\ref{R6}), (\ref{R17}), (\ref{R18}) read
\beq
(\nu -\mu )(\tau _{n}^{p}(m))^2 -
\nu \tau _{n}^{p}(m+1)\tau _{n}^{p}(m-1)=
-\mu \tau _{n+1}^{p}(m)\tau _{n-1}^{p}(m)\,,
\label{RTC4}
\eeq
\beq
(\mu -\lambda _{p})\tau _{n}^{p}(m+1)\tau _{n}^{p+1}(m)+
\lambda _{p}\tau _{n}^{p+1}(m+1)\tau _{n}^{p}(m)=
\mu \tau _{n+1}^{p}(m+1)\tau _{n-1}^{p+1}(m)\,,
\label{RTC5}
\eeq
\beq
\nu \tau _{n+1}^{p}(m+1)\tau _{n}^{p+1}(m-1)-
\lambda _{p}\tau _{n}^{p}(m)\tau _{n+1}^{p+1}(m)=
(\nu -\lambda _{p})\tau _{n}^{p+1}(m)\tau _{n+1}^{p}(m)\,,
\label{RTC6}
\eeq
respectively. It is straightforward to show that the following two
bilinear relations are direct corollaries of the basic ones:
\beq
(\nu -\lambda _{p})\tau _{n+1}^{p}(m)\tau _{n}^{p+1}(m+1)-
(\mu -\lambda _{p})\tau _{n+1}^{p+1}(m)\tau _{n}^{p}(m+1)=
(\nu -\mu )\tau _{n+1}^{p}(m+1)\tau _{n}^{p+1}(m)\,,
\label{RTC7}
\eeq
\beq
\lambda _{p}(\nu -\mu )
\tau _{n}^{p}(m)\tau _{n}^{p+1}(m)+
\nu (\mu -\lambda _{p})
\tau _{n}^{p}(m+1)\tau _{n}^{p+1}(m-1)=
\mu (\nu -\lambda _{p})
\tau _{n+1}^{p}(m)\tau _{n-1}^{p+1}(m)\,.
\label{RTC8}
\eeq
Eq.\,(\ref{RTC7}) is obtained by eliminating $\tau _{n}^{p}(m)$ from
(\ref{RTC5}), (\ref{RTC6}) (i.e. by dividing them by
$\tau _{n}^{p+1}(m+1)$,
$\tau _{n+1}^{p+1}(m)$ respectively and taking their sum) and making
use of eq.\,(\ref{RTC4}).
Eq.\,(\ref{RTC8}) is obtained
in a similar way by eliminating $\tau _{n}^{p+1}(m)$ from
(\ref{RTC5}), (\ref{RTC6}) and making
use of eq.\,(\ref{RTC7}). Now eq.\,(\ref{RTC2}) easily follows from
(\ref{RTC4}), (\ref{RTC5}) and (\ref{RTC8}).

{\bf Remark 8.2}\, The bilinearization of the usual (continuous time)
RTC was suggested in ref.\,\cite{OKS}. The equivalence of the RTC and the
"semi-discretized" AKNS system (with discrete "space" and continuous
time variables) was recently proved in ref.\,\cite{KMZh}.

We conclude this subsection with a few words about the
discrete Heisenberg ferromagnet
\cite{Miwa2}. This equation fits the scheme in the following way.
The reduction is the same.
In addition to the flow $p$
from the previous example, one should introduce
yet another elementary discrete flow $q$ left adjacent to
$l$. The $\tau$-function now depends on
four variables: $\tau = \tau _{n}^{\bar l}(p,q)$. Fix $n$, $\bar l$  and
consider the following four functions of $p, q$:
$\tau _{n}^{\bar l + 1}(p,q)$,
$\tau _{n}^{\bar l - 1}(p,q)$,
$\tau _{n + 1}^{\bar l + 1}(p,q)$,
$\tau _{n - 1}^{\bar l - 1}(p,q)$.
It can be shown (using,
in particular, one of the higher
Hirota equations in 4 variables) that certain combinations
of these functions satisfy a system of nonlinear
difference equations in the
variables $p,q$. This system is equivalent to the discrete HF model
discussed in detail in ref.\,\cite{Miwa2}, where it was treated in a
slightly different way as a part of the reduced 2-component 2DTL
hierarchy. As in the case of the discrete AKNS system, the
aforementioned embedding into
the 1-component 2DTL hierarchy leads to equivalent equations of motion.
We omit the details.

\subsection{Periodic reductions}

Periodic reductions of the continuous 2DTL hierarchy give rise to a
number of very important equations. For example, the 2-periodic
reduction $\tau _{n+2}=\tau _{n}$ contains the sine-Gordon (SG)
equation. The same periodic constraint can be imposed in the
discretized set-up, thus providing us with a discrete analogue of
the SG equation.

We remark that periodic reductions can be treated on equal footing
with stationary reductions. Indeed, the flow $p\rightarrow p+2$ is,
formally, a degenerate case of a composite flow
when the corresponding labels pairwise merge on the
complex plane. The periodicity
$\tau ^{p+2}=\tau ^{p}$ means stationarity with respect to this
degenerate "composite" flow. However, this point of view does not
seem to be useful in practice. Usually it is more convenient to treat
periodic reductions separately.

Let us consider the 2DTL-like form of HBDE (\ref{HBDE6})
with the constraint
\beq
\tau _{n+2}^{l,\bar l}=\tau _{n}^{l,\bar l}\,.
\label{stat3}
\eeq
It becomes the following system of two-dimensional equations:
\begin{eqnarray}
&&\nu \tau _{0}^{l,\bar l+1}\tau _{0}^{l+1, \bar l}-
(\nu -\mu )\tau _{0}^{l,\bar l}\tau _{0}^{l+1, \bar l +1}=
\mu \tau _{1}^{l,\bar l+1}\tau _{1}^{l+1, \bar l}\,,
\nonumber \\
&&\nu \tau _{1}^{l,\bar l+1}\tau _{1}^{l+1, \bar l}-
(\nu -\mu )\tau _{1}^{l,\bar l}\tau _{1}^{l+1, \bar l +1}=
\mu \tau _{0}^{l,\bar l+1}\tau _{0}^{l+1, \bar l}\,.
\label{R22}
\end{eqnarray}
The SG field $\Phi ^{l, \bar l}$ on the square lattice $(l, \bar l)$
is given by
\beq
\Phi ^{l, \bar l}=\frac{1}{2}\log \frac{\tau _{0}^{l, \bar l}}
{\tau _{1}^{l, \bar l}}.
\label{R23}
\eeq
Rearranging eqs.\,(\ref{R22}), one gets a closed equation for
$\Phi ^{l, \bar l}$,
\beq
\nu \, \mbox{sinh} \big (\Phi ^{l, \bar l}+\Phi ^{l+1, \bar l +1}-
\Phi ^{l, \bar l +1}- \Phi ^{l+1, \bar l }\big )=
\mu \, \mbox{sinh} \big (\Phi ^{l, \bar l}+\Phi ^{l+1, \bar l +1}+
\Phi ^{l, \bar l +1}+ \Phi ^{l+1, \bar l }\big )\,,
\label{R24}
\eeq
which is known as discrete SG equation \cite{Hirota3} written in the
light cone coordinates.

Let us mention another useful form of the discrete SG
equation \cite{FV},\,\cite{pendulum}. Set
\beq
S^{l, \bar l}=\exp \big (-2\Phi ^{l+1, \bar l}-2\Phi ^{l, \bar l +1}
\big )=
\frac {\tau _{1}^{l+1, \bar l}\tau _{1}^{l, \bar l +1}}
{\tau _{0}^{l+1, \bar l}\tau _{0}^{l, \bar l +1}}\,,
\label{R25}
\eeq
\beq
\tilde S^{l, \bar l}=\exp
\big (-2\Phi ^{l, \bar l}-2\Phi ^{l+1, \bar l +1}
\big )=
\frac {\tau _{1}^{l, \bar l}\tau _{1}^{l+1, \bar l +1}}
{\tau _{0}^{l, \bar l}\tau _{0}^{l+1, \bar l +1}}\,,
\label{R25a}
\eeq
then, evidently,
$$
\tilde S^{l, \bar l +1}
\tilde S^{l+1, \bar l }=
S^{l, \bar l }
S^{l+1, \bar l +1}\,.
$$
On the other hand, the discrete SG equation implies that
\beq
\tilde S^{l, \bar l } =
\frac{\mu -\nu S^{l, \bar l}}
{\mu S^{l, \bar l}-\nu }\,,
\label{R24a}
\eeq
so eq.\,(\ref{R24}) converts into
\beq
S^{l, \bar l}S^{l+1, \bar l +1}=
\frac{ (\mu - \nu S^{l, \bar l +1})
(\mu - \nu S^{l+1, \bar l })}
{ (\mu S^{l, \bar l +1} -\nu )
(\mu S^{l+1, \bar l } - \nu )}
\label{R26}
\eeq
(cf.\,(\ref{Ysys})).

We now turn to the zero curvature representation. Let $l$ be the
reference flow. The shift $n\rightarrow n+2$ is generated by the
scalar $L$-operator
\beq
L^{(SG)}=
e^{2\p _{l}}+
\nu \left (
\frac{\tau _{1}^{l, \bar l}\tau _{0}^{l+1, \bar l}}
{\tau _{0}^{l, \bar l}\tau _{1}^{l+1, \bar l}}+
\frac{\tau _{0}^{l+1, \bar l}\tau _{1}^{l+2, \bar l}}
{\tau _{1}^{l+1, \bar l}\tau _{0}^{l+2, \bar l}}\right )
e^{\p _{l}}+\nu ^{2}\,,
\label{SG1}
\eeq
However, this representation
is not convenient for describing evolution in $\bar l$.

The matrix $L$-$M$ pair with spectral parameter is more appropriate
here. To derive it, one should take the "stationary" flow $2n$ to be
the reference one and repeat the arguments given in Sect.\,5.5 with
necessary modifications. The operator $e^{2\p _{n}}$ should be
substituted by a spectral parameter $z^2$. Omitting the details,
we present the result.

The auxiliary linear problems read
\beq
-\nu ^{-1}
\left ( \begin{array}{c} \psi ^{l+1, \bar l} \\ \\ \\ \\
\chi ^{l+1, \bar l} \end{array} \right )=
\left ( \begin{array}{ccc}
\displaystyle{\frac{\tau _{0}^{l, \bar l}\tau _{1}^{l+1, \bar l}}
{\tau _{0}^{l+1, \bar l}\tau _{1}^{l, \bar l}}} &&
-\displaystyle{\frac{z}{\nu }
\frac{\tau _{0}^{l, \bar l}}{\tau _{1}^{l, \bar l}}}\\ && \\
-\displaystyle{\frac{z}{\nu }
\frac{\tau _{1}^{l+1, \bar l}}{\tau _{0}^{l+1, \bar l}}} && 1
\end{array} \right )
\left ( \begin{array}{c} \psi ^{l, \bar l} \\ \\  \\ \\
\chi ^{l, \bar l} \end{array} \right ),
\label{SG2}
\eeq
\beq
\mu
\left ( \begin{array}{c} \psi ^{l, \bar l +1} \\ \\  \\ \\
\chi ^{l, \bar l +1} \end{array} \right )=
\left ( \begin{array}{ccc}
1 &&
-\displaystyle{\frac{\mu }{z}
\frac{\tau _{1}^{l, \bar l+1 }}{\tau _{0}^{l, \bar l +1}}}\\ && \\
-\displaystyle{\frac{\mu }{z}
\frac{\tau _{0}^{l, \bar l}}{\tau _{1}^{l, \bar l}}} &&
\displaystyle{\frac{\tau _{0}^{l, \bar l}\tau _{1}^{l, \bar l +1}}
{\tau _{0}^{l, \bar l +1}\tau _{1}^{l, \bar l}}}
\end{array} \right )
\left ( \begin{array}{c} \psi ^{l, \bar l} \\ \\  \\ \\
\chi ^{l, \bar l} \end{array} \right )
\label{SG3}
\eeq
which is akin to (\ref{R9}).
Denoting the matrices in the r.h.s. of eqs.\,(\ref{SG2}), (\ref{SG3})
by $M^{(+)}$, $M^{(-)}$ respectively, we can write the compatibility
condition
\beq
M^{(+)}(l, \bar l +1)
M^{(-)}(l, \bar l )=
M^{(-)}(l+1, \bar l )
M^{(+)}(l, \bar l )
\label{SG4}
\eeq
which yields the discrete SG equation.

$N$-periodic reductions ($\tau _{n+N}=\tau _{n}$) can be treated
in a similar way. They correspond to $N$-periodic Toda lattices in
discrete time. It is also possible to impose periodicity
with respect to any one of the composite
flows. In the remaining part of this subsection we briefly comment
on an important class of such reductions which are discrete analogues
of Intermediate Long Wave (ILW) equations.

The universal form of reductions from the 2DTL to the family of
continuous ILW equations is most transparently written in terms
of the $\tau$-function of the 2DTL hierarchy. The reduction to the
$\mbox{ILW}_{k}$ equation reads \cite{LOPZ}
\beq
\tau _{n+k}(t_{1}+h, t_{2}, \ldots ;\, \bar t_{1}, \bar t_{2},
\ldots )=
\tau _{n}(t_{1}, t_{2}, \ldots ;\, \bar t_{1}, \bar t_{2},
\ldots )\,,
\label{ILW1}
\eeq
where $h$ is a fixed parameter. This parameter interpolates between the
$k$-periodic reduction ($h=0$) and the Benjamin-Ono equation
($h \rightarrow \infty $). In words, the $\tau$-function should not
depend on a particular combination of $n$ and $t_1$. This suggests
a discretization of the
$\mbox{ILW}_{k}$ equation. According to our general rules
of discretization, one
should substitute $t_1$ by an elementary discrete flow $p$.
Then it is natural
to substitute eq.\,(\ref{ILW1}) by
\beq
\tau _{n+k}^{p+l}=\tau _{n}^{p}\,,
\label{ILW2}
\eeq
where $l$ and $k$ are integer parameters. The particular cases are
the discrete KdV equation ($l=k=1$) and the $k$-periodic reduction
($l=0$). In the continuum limit we get the
continuous
$\mbox{ILW}_{k}$ equation.

\subsection{Discrete Liouville equation}

The discrete Liouville equation (DLE) and its $A_n$-generalizations
\cite{Hirota6}
(discrete time 2DTL
with open boundaries) form a very important special class
of discrete integrable systems which in general does not fit the
reduction scheme discussed in this section. We include it here for
the only reason that the DLE is, formally, a degenerate case of the
discrete SG equation. The relationship between these two integrable
systems deserves further study.

The DLE can be obtained from the discrete SG equation as a
result of a certain scaling limit. Let us rescale
$S^{l, \bar l}\rightarrow \mu S^{l, \bar l}$
in eq.\,(\ref{R26}). Clearly, this
rescaling means a constant shift of the field:
$\Phi ^{l, \bar l}\rightarrow
\Phi ^{l, \bar l} -\frac{1}{4}\log \mu$. Then, taking
the limit
$\mu \rightarrow 0$ in
eq.\,(\ref{R26}) (keeping shifts in $\bar l$ alive!), one arrives at
the DLE
\beq
S_{L}^{l, \bar l}S_{L}^{l+1, \bar l +1}=
(\nu ^{-1}- S_{L}^{l, \bar l +1})
(\nu ^{-1}- S_{L}^{l+1, \bar l })\,,
\label{R27}
\eeq
where, formally,
\beq
S_{L}^{l, \bar l}=\lim _{\mu
\rightarrow 0} \big ( \mu ^{-1}S^{l, \bar l }\big )\,.
\label{SL}
\eeq
Setting
\beq
S_{L}^{l, \bar l}=\exp
\big (-2\Phi _{L}^{l+1, \bar l}-2\Phi _{L}^{l, \bar l +1}
\big )\,,
\label{R28}
\eeq
we get, in place of eq.\,(\ref{R24}), the DLE written in terms of
the discrete Liouville field \cite{Hirota5}:
\beq
2\nu \, \mbox{sinh}
\big (\Phi _{L}^{l, \bar l}+\Phi _{L}^{l+1, \bar l +1}-
\Phi _{L}^{l, \bar l +1}- \Phi _{L}^{l+1, \bar l }\big )=
\exp \big (\Phi _{L}^{l, \bar l}+\Phi _{L}^{l+1, \bar l +1}+
\Phi _{L}^{l, \bar l +1}+ \Phi _{L}^{l+1, \bar l }\big )\,,
\label{R29}
\eeq
or, in a simpler form,
\beq
\exp \big ( -2\Phi _{L}^{l+1, \bar l}-2\Phi _{L}^{l, \bar l +1}\big )-
\exp \big (-2\Phi _{L}^{l, \bar l}-2\Phi _{L}^{l+1, \bar l +1}\big )
= \nu ^{-1}\,.
\label{R29a}
\eeq

In the continuum limit one should put
$l\rightarrow \nu ^{1/2}x_{+}$,
$\bar l \rightarrow \nu ^{1/2}x_{-}$,
$S_{L}^{l, \bar l}\rightarrow \exp \big (-4\Phi (x_{+}, x_{-})\big )$.
Expanding in $\nu ^{-1} \rightarrow 0$, we get the continuous
Liouville equation
\beq
2\p _{x_{+}}\p _{x_{-}}\Phi (x_{+}, x_{-})=e^{4\Phi (x_{+}, x_{-})}\,.
\label{Lcontin}
\eeq

The bilinear form of eq.\,(\ref{R27}) is available via the
substitution
\beq
S_{L}^{l, \bar l}=
\frac{T^{1}(l+1, \bar l )T^{1}(l, \bar l +1)}
{T^{0}(l+1, \bar l )T^{2}(l, \bar l +1)}
\label{subst}
\eeq
after which the DLE becomes equivalent (up to a "gauge freedom",
see below) to
the bilinear relation
\beq
T^{a}(l+1, \bar l)T^{a}(l, \bar l +1)-
T^{a}(l, \bar l)T^{a}(l+1, \bar l +1)=
\nu ^{-1}T^{a-1}(l+1, \bar l)T^{a+1}(l, \bar l +1)
\label{R30}
\eeq
with the condition
\beq
T^{a}(l, \bar l)=0 \;\;\;\;\;\;\mbox{unless} \;\; a=0,\,1,\,2\,.
\label{R31}
\eeq
This condition implies the discrete d'Alembert equation (\ref{R1})
for $T^0$ and $T^2$, so they have to have a factorized form
$T^{0}(l, \bar l )=\chi ^{0}(l)\bar \chi ^{0}(\bar l)$,
$T^{2}(l, \bar l )=\chi ^{2}(l)\bar \chi ^{2}(\bar l)$
with arbitrary and independent functions $\chi ^{0,2}$,
$\bar \chi ^{0,2}$. This is just the aforementioned gauge freedom.

The striking similarity between eqs.\,(\ref{R25}) and (\ref{R29})
is transparent after the replacing $T^{a}(l, \bar l )\rightarrow
\tau _{a}^{l, \bar l }$. Furthermore, taking into account the
periodicity $\tau _{2}^{l, \bar l}=\tau _{0}^{l, \bar l}$, they become
formally identical. (Equivalently, using the gauge freedom, one can set
$T^{2}(l, \bar l )=
T^{0}(l, \bar l )$ in eq.\,(\ref{R29}).) It would be interesting to
link them directly on the level of solutions.

\section*{Acknowledgements}

I am grateful to A.Gorsky, D.Lebedev, O.Lipan, A.Marshakov,
A.Mironov, A.Orlov, P.Wiegmann
and, especially, to S.Kharchev and I.Krichever
for many interesting and helpful discussions.
It is a pleasure to thank CMAT de l'Ecole Polytechnique, where this
work was initiated, for the hospitality.
This work was supported in part by RFBR grant 97-02-19085 and
by ISTC grant 015.

\section*{Appendix: bilinear difference equations from
continuous hierarchies}
\def\theequation{A\arabic{equation}}
\setcounter{equation}{0}

In the Appendix we give an alternative point of view to
the difference Hirota equations.
It relies on the famous Miwa transformation (\ref{miwa1})
which so far was obscure in our exposition.
Given a continuous integrable hierarchy
(such as KP or 2DTL), this relation
can be used {\it as a definition} of the elementary discrete
flows. This definition leads to the same discrete flows
as in Sect.\,4.2. This approach has as many advantages as
disadvantages. The main advantage is much more direct and
instructive connection with the grassmannian
approach to continuous hierarchies and their $\tau$-functions.
The main disadvantage
is misleadingly less invariant formulation which is
inconvenient in some cases.

\subsubsection*{The Miwa transformation}

Let $\tau (t_1 , t_2 , t_3 , \ldots )
\equiv \tau (t)$ be the $\tau$-function of the
continuous KP hierarchy. It is a function of infinite number of
"times" $t_i$ and satisfies infinitely many bilinear equations.
The $\tau$-function solves all equations of the hierarchy
simultaneously.

In general the
$\tau$-function can be represented as an infinite dimensional
determinant \cite{JimboMiwa}-\cite{SW}. It turns out that there
exists a choice of independent variables such that the determinant
reduces to a finite dimensional one. This choice is provided by
the Miwa transformation \cite{Miwa1}:
\beq
t_k =t_{k}^{(0)}-\frac{1}{k}\sum _{\alpha \in I} p_{\alpha }
\mu _{\alpha}^{-k}\,, \;\;\;\;\;\;k=1,2,\ldots
\label{NE4}
\eeq
Here, the summation runs over a finite set $I$, $t_{k}^{(0)}$ are
"background values" of the times, $\mu _{\alpha}$ are arbitrary
complex numbers (called Miwa's variables) and $p_{\alpha}$ are
integers (sometimes called multiplicities of $\mu _{\alpha}$'s).

{\bf Remark}\, The Miwa transformation plays an important role in
revealing integrable structures of matrix models of 2D gravity.
In particular, the easiest proof of the
fact that the partition functions of the Kontsevich
model \cite{K} and its generalizations \cite{KMMMZ} are $\tau$-functions
of the KP hierarchy relies on Miwa's transformation.

In what follows we use the short hand notation (\ref{sh}).

\noindent
{\bf Important fact:} The $\tau$-function of the KP hierarchy obeys
the identity
\beq
\tau \left (t^{(0)}+\sum _{\alpha =1}^{N}\big ([\nu _{\alpha}^{-1}]
-[\mu _{\alpha}^{-1}]\big )\right ) =
\frac{ \tau (t^{(0)})
\prod _{\alpha, \beta}^{N}(\nu _{\alpha}-\mu _{\beta})}
{\prod _{\alpha >\beta}^{N}(\nu _{\alpha}-\nu _{\beta})
\prod _{\alpha <\beta}^{N}(\mu _{\alpha}-\mu _{\beta})}
\det _{1\leq \alpha , \beta \leq N}K(\nu _{\alpha}, \mu _{\beta}),
\label{NE5}
\eeq
where
\beq
K(\nu , \mu )=\frac{\tau (t^{(0)}+[\nu ^{-1}]-[\mu ^{-1}])}
{(\nu - \mu )\tau (t^{(0)})}.
\label{NE6}
\eeq
Here $N\geq 1$ and $\mu _{\alpha}, \nu _{\alpha}$ are arbitrary
complex numbers. A useful particular case of this formula is
\beq
\tau \left (t^{(0)}-\sum _{\alpha =1}^{N}[\mu _{\alpha}^{-1}]\right )=
\frac{\det _{1\leq \alpha , \beta \leq N}
\big (\varphi _{\alpha}(\mu _{\beta})\big )}
{\prod _{\alpha <\beta}^{N}(\mu _{\alpha}-\mu _{\beta})},
\label{NE7}
\eeq
where
\beq
\varphi _{m}(\mu )=\frac{1}{(m-1)!}
\lim _{\nu \rightarrow \infty}\nu ^{2m-1}\frac{\p ^{m-1}}
{\p \nu ^{m-1}}K(\nu , \mu )\,.
\label{NE8}
\eeq
When one translates the KP theory into the language of
free fermions \cite{JimboMiwa} formula (\ref{NE5}) becomes nothing
else than Wick's theorem, $K(\nu , \mu )$ being the fermionic
propagator on a Riemann surface.

Instead of treating eq.\,(\ref{NE5}) as an identity, one may
follow another way. Given a function $K(\nu ,\mu )$ with a simple
pole at $\nu = \mu$, this equation can be used as a
{\it definition} of the l.h.s. This simply means that we disregard the
dependence on background times $t_{k}^{(0)}$ assuming they are fixed.
The $\tau$-function
in Miwa's variables satisfy certain bilinear relations to which
formula (\ref{NE5}) gives a solution in the form of a
finite dimensional determinant.

In the case of the 2DTL hierarchy the Miwa transformation goes in
a similar way. The $\tau$-function $\tau _{n}(t_1 , t_2 ,
\ldots ;\, {\bar t}_{1}, {\bar t}_{2}, \ldots )\equiv
\tau _{n}(t;{\bar t})$ depends on the discrete time $n$ and two
infinite sets of continuous times $t_i$ and ${\bar t}_{i}$. We set
\begin{eqnarray}
&&t_k =t_{k}^{(0)}-\frac{1}{k}\sum _{\alpha \in I} p_{\alpha }
\mu _{\alpha}^{-k}\,,  \nonumber \\
&&{\bar t}_k =
{\bar t}_{k}^{(0)}-\frac{1}{k}\sum _{\alpha \in \bar I}
{\bar p}_{\alpha }
{\bar \mu } _{\alpha}^{k}\,, \;\;\;\;\;\;k=1,2,\ldots
\label{miwization2}
\end{eqnarray}
where ${\bar \mu}_{\alpha}$ is an independent
set of Miwa's variables with
multiplicities ${\bar p}_{\alpha}$.

The following analogue of eq.\,(\ref{NE5}) holds:
\beq
\tau _{n-N}\left (t^{(0)}
-\sum _{\alpha =1}^{N}[\mu _{\alpha}^{-1}];\,
{\bar t}^{(0)}
+\sum _{\alpha =1}^{N}[\bar \mu _{\alpha}]
\right ) =
\frac{ \tau _{n}(t^{(0)};\, {\bar t}^{(0)})
\prod _{\alpha =1}^{N}\mu _{\alpha}^{N-1}}
{\prod _{\alpha < \beta}^{N}(\mu _{\alpha}-\mu _{\beta})
(\bar \mu _{\beta}-\bar \mu _{\alpha})}
\det _{1\leq \alpha , \beta \leq N}J_{n}
(\mu _{\alpha}, \bar \mu _{\beta}),
\label{NE5a}
\eeq
where
\beq
J_n (\mu , \bar \mu )=
\frac{\tau _{n-1}(t^{(0)}-[\mu ^{-1}];
\bar t^{(0)}+[\bar \mu ])}
{\tau _{n}(t^{(0)};\, \bar t^{(0)})}\,.
\label{NE6a}
\eeq
Note that in this case the function $J_n (\mu ,
\bar \mu )$ does not necessarily
have first order pole at $\mu =\bar \mu$.

\subsubsection*{Discrete flows}

Discrete equations for the $\tau$-function listed in
Sect.\,2 are obtained if one fixes Miwa's
variables $\mu _{\alpha}$ and consider flows in the multiplicities
$p_{\alpha}$. We give a few examples.

{\it Example 1}. Set
\beq
\tau ^{p_1 , p_2 , p_3 }=\tau \left (t^{(0)}-\sum _{\alpha =1}^{3}
p_{\alpha }[\mu _{\alpha }^{-1}]\right )
\label{NE9}
\eeq
and consider $\tilde t^{(0)}=
t^{(0)}-\sum _{\alpha =1}^{3}
p_{\alpha }[\mu _{\alpha }^{-1}] $ as a new "background". According
to eq.\,(\ref{NE7}),
\begin{eqnarray}
&&\tau ^{p_1 +1}=\tilde \varphi _{1}(\mu _{1})\,,\nonumber \\
&&\tau ^{p_1 +1, p_2 +1}=\frac{
\left |\begin{array}{cc}
\tilde \varphi _{1}(\mu _{1}) &
\tilde \varphi _{1}(\mu _{2}) \\
\tilde \varphi _{2}(\mu _{1}) &
\tilde \varphi _{2}(\mu _{2}) \end{array}\right | }
{\mu _{1}-\mu _{2} }
\label{NE10}
\end{eqnarray}
with some functions $\tilde \varphi _{1}, \tilde \varphi _{2}$.
Combining the zero determinant with two identical lines,
\beq
0=\left |\begin{array}{ccccc}
\tilde \varphi _{1}(\mu _{1}) &&
\tilde \varphi _{1}(\mu _{2}) &&
\tilde \varphi _{1}(\mu _{3}) \\ && \\
\tilde \varphi _{1}(\mu _{1}) &&
\tilde \varphi _{1}(\mu _{2}) &&
\tilde \varphi _{1}(\mu _{3}) \\ && \\
\tilde \varphi _{2}(\mu _{1}) &&
\tilde \varphi _{2}(\mu _{2}) &&
\tilde \varphi _{2}(\mu _{3})
\end{array}\right | ,
\label{NE11}
\eeq
and expanding it in the first line, we get an equation of the form
(\ref{HBDE4}). Since its coefficients do not depend on the
chosen background, the equation holds for all values of $p_1 , p_2 ,
p_3 $.

{\it Example 2}. Repeating the previous argument for
$$
\tau _{p_0}^{p_1 , p_2 , p_3 }=\tau \left ( t^{(0)}+
p_{0}([\mu _{0}^{-1}]-[\nu ^{-1}])+\sum _{\alpha =1}^{3}p_{\alpha }
([\nu ^{-1}]-[\mu _{\alpha }^{-1}])\right )
$$
and making use of eq.\,(\ref{NE5}), we get eq.\,(\ref{HBDE4a}).

{\it Example 3}. In eq.\,(\ref{NE5a}), let $N=2$, $\mu _1 = \mu$,
$\bar \mu _{1}=\bar \mu$, $\mu _{2}\rightarrow \infty$,
$\bar \mu _{2}\rightarrow 0$:
\begin{eqnarray}
&&\tau _{n-2} (t^{(0)}-[\mu ^{-1}];\, \bar t^{(0)}+[\bar \mu ])
\tau _{n}(t^{(0)}; \bar t^{(0)})= \nonumber \\
&=&\frac{\mu }{\bar \mu}
\left | \begin{array}{lll}
\tau _{n-1} (t^{(0)}-[\mu ^{-1}];\,
\bar t^{(0)}+[\bar \mu ]) &&
\tau _{n-1} (t^{(0)}-[\mu ^{-1}];\, \bar t^{(0)}) \\ && \\
\tau _{n-1} (t^{(0)};\, \bar t^{(0)}+[\bar \mu ]) &&
\tau _{n-1} (t^{(0)};\, \bar t^{(0)}) \end{array} \right |
\label{NE12}
\end{eqnarray}
Denoting
\beq
\tau _{n}^{l, \bar l} =
\tau _{n}(t^{(0)}-l[\mu ^{-1}];\,
\bar t^{(0)}-\bar l [\bar \mu ])\,,
\label{NE9a}
\eeq
we get the equation
\beq
\tau _{n}^{l,\bar l +1}\tau _{n}^{l+1, \bar l}-
\tau _{n}^{l,\bar l }\tau _{n}^{l+1, \bar l +1}=
({\bar \mu }/\mu )
\tau _{n+1}^{l,\bar l +1}\tau _{n-1}^{l+1, \bar l}\,.
\label{HBDE6a}
\eeq

{\it Example 4}. Example 1 can be generalized in the following way.
Consider a $N\times N$-matrix with the lines
$\varphi _{1}(\mu _i )$,
$\varphi _{1}(\mu _i )$,
$\varphi _{2}(\mu _i )$,
$\varphi _{3}(\mu _i )$,
\ldots ,
$\varphi _{N-1}(\mu _i )$, $i=1, 2 , \ldots , N$, so that
the first two lines coincide and determinant of this matrix
is zero. Then, expanding in the first row, like in
Example 1, we get the "higher"
bilinear difference equation of the form (\ref{h1}).

{\it Example 5}. At last, we show how to derive HBDE
in the KP-like form from eq.\,(\ref{NE7})
in a direct way\footnote{this argument is taken from
ref.\,\cite{KMMMZ}.}.
When two or more $\mu _{\alpha}$'s coincide, both the numerator
and denominator in the r.h.s. of eq.\,(\ref{NE7}) equal zero. Resolving
the indeterminacy, we have
\beq
\tau \left (t^{(0)}-\sum _{\alpha =1}^{N}
p_{\alpha}[\mu _{\alpha}^{-1}]\right )=
\frac{\det
\big ( M_{ij}^{ ({\cal N})} \big )}
{\prod _{\alpha <\beta}^{N}
(\mu _{\alpha}-\mu _{\beta})^{ p_{\alpha}p_{\beta} }}, \;\;\;\;
1\leq i,j \leq {\cal N}\,,
\label{NE14}
\eeq
and all $\mu _{\alpha}$'s are now distinct. Here ${\cal N}\equiv
\sum _{\alpha =1}^{N} p_{\alpha}$,\,
$M_{ij}^{({\cal N})}$
is the ${\cal N}\times {\cal N}$-matrix having the rows
\begin{eqnarray}
&&\varphi _{i}(\mu _{1}),\,
\varphi _{i}'(\mu _{1}),\,
\varphi _{i}''(\mu _{1}),\ldots ,
\varphi _{i}^{(p_1 -1)}(\mu _{1}),\,
\nonumber \\
&&\varphi _{i}(\mu _{2}),\,
\varphi _{i}'(\mu _{2}),\,
\varphi _{i}''(\mu _{2}),\ldots ,
\varphi _{i}^{(p_2 -1)}(\mu _{2}),\ldots ,
\nonumber \\
&&\varphi _{i}(\mu _{N}),\,
\varphi _{i}'(\mu _{N}),\,
\varphi _{i}''(\mu _{N}),\ldots ,
\varphi _{i}^{(p_N -1)}(\mu _{N})\,,\;\;\;\;\;
1\leq i \leq {\cal N}\,.
\label{rows}
\end{eqnarray}

We need the well known Jacobi identity for determinants:
\beq
D[i_1 |j_1 ] D[ i_2 | j_2 ]-
D[i_1 |j_2 ] D[ i_2 | j_1 ]=
D[i_1 , i_2 |j_1 , j_2 ] D\,, \;\;\;\;\; i_1 < i_2 ,\; j_1 < j_2\,,
\label{Jacobi}
\eeq
where $D$ is determinant of a square matrix and
$D[i_1 , i_2 |j_1 , j_2 ]$ denotes minors of this matrix with
$i_{1,2 }$-th rows and $j_{1,2 }$-th columns removed. Applying this
identity to the matrix $M_{ij}^{({\cal N})}$ in (\ref{NE14}) for
$i_1 = {\cal N}-1$,
$i_2 = {\cal N}$,
$j_1 =\sum _{\alpha =1}^{a}p_{\alpha }$,
$j_2 =j_1 +\sum _{\alpha =a+1}^{b}p_{\alpha }$,
$1\leq a<b\leq N$, we get, in the short hand notation,
\beq
(\mu _{a}-\mu _{b})\tau \tau ^{p_a -1, p_b -1 }=
\tau ^{p_b -1 }\hat \tau ^{p_a -1 }-
\tau ^{p_a -1 }\hat \tau ^{p_b -1 }\,,
\label{NE15}
\eeq
where $\hat \tau$ is defined by the same formula (\ref{NE14}) with
the matrix
${\hat M}_{ij}^{({\cal N}-1)}=
M_{ij}^{({\cal N}-1)}$ for $1\leq i \leq {\cal N}-2$,
${\hat M}_{{\cal N}-1, j}^{({\cal N}-1)}=
M_{{\cal N}, j}^{({\cal N})}$.

Let $\mu _c$ be a third Miwa variable (different from
$\mu _a$,
$\mu _b$) with the multiplicity $p_c$ not shown explicitly in
eq.\,(\ref{NE15}). Multiplying this equality by
$\tau ^{p_c -1}/\tau $ and then writing down a couple of similar
equations obtained by cyclic permutations of the indices $a,b,c$, we
see that the sum of these three equations coincides with
eq.\,(\ref{HBDE4}).

{\bf Remark}\,\,The discrete flows discussed
here coincide with those
introduced in the main body of the paper if one
fixes the following choice of the labels $\l _{0}$ and
$\l _{1}$: $\l _{0} =\infty$, $\l _{1}=0$. (To remove a
label to infinity, one should use a different normalization.)

\subsubsection*{Continuum limit}

As it is clear from eq.\,(\ref{NE4}), inverse Miwa's variables
$\mu _{\alpha}^{-1}$ play the role of lattice spacings for the
discrete flows. So, to perform the limit
to continuous equations, it is necessary for $\mu _{\alpha}$ to
tend to infinity with a simultaneous rescaling of $p_{\alpha}$'s.

Here is the typical example (the KP hierarchy).
In this example we follow
ref.\,\cite{OHTI}. Introduce three (a priori independent) lattice
spacings $\varepsilon _{i}=\mu _{i}^{-1}$, $i=1,2,3$, and rescale
$p_i \rightarrow p_i / \varepsilon _i$. It is then convenient to rewrite
the KP-like form (\ref{HBDE4}) of HBDE in terms of Hirota's
$D$-operator (\ref{D}):
\begin{eqnarray}
&&\left ( \varepsilon _1 (\varepsilon _2 -\varepsilon _3 )
e^{-(\varepsilon _1 /2)D_{p_1 }+
(\varepsilon _2 /2)D_{p_2 }+
(\varepsilon _3 /2)D_{p_3 } }+ \right . \nonumber \\
&+& \varepsilon _2 (\varepsilon _3 -\varepsilon _1 )
e^{(\varepsilon _1 /2)D_{p_1 }-
(\varepsilon _2 /2)D_{p_2 }+
(\varepsilon _3 /2)D_{p_3 } }+ \nonumber \\
&+&\left . \varepsilon _3 (\varepsilon _1 -\varepsilon _2 )
e^{(\varepsilon _1 /2)D_{p_1 }+
(\varepsilon _2 /2)D_{p_2 }-
(\varepsilon _3 /2)D_{p_3 } }\right )\tau \cdot \tau =0\,.
\label{NE16}
\end{eqnarray}
This equation serves as a "generating function" for
a part of the continuous KP hierarchy. To see
this, we express operators $D_{p_i }$ through Hirota's derivatives
with respect to the continuous flows $t_k$,
$$
D_{p_i }= -\sum _{k=1}^{\infty} \frac{1}{k}\varepsilon _{i}^{k-1}
D_{t_k} \,, \;\;\;\;\;\; i=1,2,3,
$$
in accordance with (\ref{NE4}). Substituting this into eq.\,(\ref{NE16})
and expanding it in a power series in $\varepsilon _i$, we have:
\begin{eqnarray}
&&\left ( \varepsilon _1 (\varepsilon _2 -\varepsilon _3 )
\sum _{j,k,l=0}^{\infty} \varepsilon _{1}^{j}
\varepsilon _{2}^{k}
\varepsilon _{3}^{l}
{\cal P}_{j}(\frac{1}{2}\tilde D)
{\cal P}_{k}(-\frac{1}{2}\tilde D)
{\cal P}_{l}(-\frac{1}{2}\tilde D)+
\right . \nonumber \\
&+& \varepsilon _2 (\varepsilon _3 -\varepsilon _1 )
\sum _{j,k,l=0}^{\infty} \varepsilon _{1}^{j}
\varepsilon _{2}^{k}
\varepsilon _{3}^{l}
{\cal P}_{j}(-\frac{1}{2}\tilde D)
{\cal P}_{k}(\frac{1}{2}\tilde D)
{\cal P}_{l}(-\frac{1}{2}\tilde D)+
\nonumber \\
&+&\left . \varepsilon _3 (\varepsilon _1 -\varepsilon _2 )
\sum _{j,k,l=0}^{\infty} \varepsilon _{1}^{j}
\varepsilon _{2}^{k}
\varepsilon _{3}^{l}
{\cal P}_{j}(-\frac{1}{2}\tilde D)
{\cal P}_{k}(-\frac{1}{2}\tilde D)
{\cal P}_{l}(\frac{1}{2}\tilde D)
\right )\tau \cdot \tau =0\,,
\label{NE17}
\end{eqnarray}
where $\tilde D\equiv (D_{t_1},\, D_{t_2}/2,\, \ldots ,
D_{t_k}/k , \,\ldots )$ and ${\cal P}_{j}(t)$ are Schur polynomials
defined by
\beq
\exp \left ( \sum _{k=1}^{\infty} t_k z^k \right )=
\sum _{m=0}^{\infty} {\cal P}_m (t)z^m \,.
\label{Schur}
\eeq
Extracting the coefficients in front of
$\varepsilon _{1}^{j}
\varepsilon _{2}^{k}
\varepsilon _{3}^{l}$, we obtain the infinite set of bilinear
equations,
\beq
\left | \begin{array}{lllll}
{\cal P}_{j-1}(\frac{1}{2}\tilde D) &&
{\cal P}_{j-1}(-\frac{1}{2}\tilde D) &&
{\cal P}_{j}(-\frac{1}{2}\tilde D) \\ &&&& \\
{\cal P}_{k-1}(\frac{1}{2}\tilde D) &&
{\cal P}_{k-1}(-\frac{1}{2}\tilde D) &&
{\cal P}_{k}(-\frac{1}{2}\tilde D) \\ &&&& \\
{\cal P}_{l-1}(\frac{1}{2}\tilde D) &&
{\cal P}_{l-1}(-\frac{1}{2}\tilde D) &&
{\cal P}_{l}(-\frac{1}{2}\tilde D) \end{array}\right |
\tau \cdot \tau =0\,,
\label{NE18}
\eeq
which for $1\leq j<k<l$ form a subset of the whole KP hierarchy in
the bilinear form.

The leading term as $\varepsilon _i \rightarrow 0$ in
(\ref{NE17}) corresponds to $(j,k,l)=(1,2,3)$ in (\ref{NE18}).
In this case eq.\,(\ref{NE18}) gives
(the bilinear form of) the KP equation itself:
\beq
(D_{t_1}^{4}-4D_{t_1}D_{t_3}+3D_{t_2}^{2})\tau \cdot \tau =0\,.
\label{KP}
\eeq
This example shows once again
that the discrete hierarchy has a more transparent
structure than the continuous one. The continuum limit brings
artificial complications.


\begin{thebibliography}{99}

\bibitem{Hirota1} R.Hirota, {\it Discrete analogue of a generalized
Toda equation}, Journ. Phys. Soc. Japan {\bf 50} (1981) 3785-3791.

\bibitem{HirotaKdV} R.Hirota, {\it Nonlinear partial difference
equations I},
Journ. Phys. Soc. Japan {\bf 43} (1977) 1424-1433.

\bibitem{Hirota2} R.Hirota, {\it Nonlinear partial difference
equations II; Discrete time Toda equations},
Journ. Phys. Soc. Japan {\bf 43} (1977) 2074-2078.

\bibitem{Hirota3} R.Hirota, {\it Nonlinear partial difference
equations III; Discrete sine-Gordon equation},
Journ. Phys. Soc. Japan {\bf 43} (1977) 2079-2086.

\newcounter{hirota4}
\setcounter{hirota4}{4}

\bibitem{Hirota4} R.Hirota, {\it Nonlinear partial difference
equations \Roman{hirota4};
B\"acklund transformation for the discrete Toda
equation},
Journ. Phys. Soc. Japan {\bf 45} (1978) 321-332.

\addtocounter{hirota4}{1}

\bibitem{Hirota5} R.Hirota, {\it Nonlinear partial difference
equations \Roman{hirota4}; Nonlinear equations reducible to
linear equations},
Journ. Phys. Soc. Japan {\bf 46} (1979) 312-319.

\bibitem{KLWZ} I.Krichever, O.Lipan, P.Wiegmann and A.Zabrodin,
{\it Quantum integrable models and discrete classical
Hirota equations}, preprint ESI-330 (1996), hep-th/9604080.

\bibitem{Z} A.Zabrodin, {\it Discrete Hirota's equation in
quantum integrable models}, preprint ITEP-TH-44/96 (1996),
hep-th/9610039.

\bibitem{KP} A.Kl\"umper and P.Pearce, {\it Conformal weights of
RSOS lattice models and their fusion hierarchies}, Physica
{\bf A183} (1992) 304-350.

\bibitem{Kuniba1} A.Kuniba, T.Nakanishi and J.Suzuki,
{\it Functional relations in solvable lattice models, I: Functional
relations and representation theory, II: Applications}, Int. Journ. Mod.
Phys. {\bf A9} (1994) 5215-5312.


\bibitem{Miwa1} T.Miwa, {\it On Hirota's difference equations},
Proc. Japan Acad. {\bf 58} Ser.A (1982) 9-12.

\bibitem{Miwa2}  E.Date, M.Jimbo and T.Miwa, {\it Method for
generating discrete soliton equations I, II}, Journ. Phys. Soc. Japan
{\bf 51} (1982) 4116-4131.

\newcounter{miwa}
\setcounter{miwa}{3}

\bibitem{Miwa3}  E.Date, M.Jimbo and T.Miwa, {\it Method for
generating discrete soliton equations
\Roman{miwa},
\addtocounter{miwa}{1}
\Roman{miwa}}, Journ. Phys. Soc. Japan {\bf 52} (1983) 388, 761.

\bibitem{Sato} M.Sato, {\it Soliton equations as dynamical systems
on infinite dimensional Grassmann manifolds}, RIMS Kokyuroku
{\bf 439} (1981) 30-46.

\bibitem{JimboMiwa} M.Jimbo and T.Miwa, {\it Solitons and infinite
dimensional Lie algebras}, Publ. RIMS, Kyoto Univ. {\bf 19} (1983)
943-1001.

\bibitem{Date2} E.Date, M.Jimbo, M.Kashiwara and T.Miwa,
{\it Transformation groups for soliton equations}, in Nonlinear
Integrable Systems, M.Jimbo and T.Miwa (eds.), World Scientific,
Singapore 1983.

\bibitem{SW} G.Segal and G.Wilson, {\it Loop groups and equations
of KdV type}, Publ. IHES {\bf 61} (1985) 5-65.

\bibitem{Zam} Al.B.Zamolodchikov, {\it On the thermodynamic Bethe
ansatz equations for reflectionless ADE scattering theories},
Phys. Lett. {\bf B253} (1991) 391-394.

\bibitem{Tateo} F.Ravanini, A.Valleriani and R.Tateo,
{\it Dynkin TBA's}, Int. Journ. Mod. Phys. {\bf A8} (1993) 1707-1727.

\bibitem{kr}
B. Dubrovin, V. Matveev and S. Novikov, {\it Non-linear equations of
Korteweg-de Vries type, finite zone linear operators and Abelian varieties},
Uspekhi Mat. Nauk {\bf 31:1} (1976) 55-136;
I.M. Krichever, {\it Nonlinear equations and elliptic curves},
Modern problems in mathematics, Itogi nauki i tekhniki,
VINITI AN USSR {\bf 23} (1983).

\bibitem{SS} S.Saito and N.Saitoh, {\it Linearization of bilinear
difference equations}, Phys. Lett. {\bf A120} (1987) 322-326;
{\it Gauge and dual symmetries and linearization of Hirota's
bilinear equations}, Journ. Math. Phys. {\bf 28} (1987) 1052-1055.

\bibitem{UT} K.Ueno and K.Takasaki, {\it Toda lattice hierarchy},
Adv. Studies in Pure Math. {\bf 4} (1984) 1-95.

\bibitem{FadTakh} L.Faddeev and L.Takhtajan, {\it Hamiltonian methods in
the theory of solitons}, Springer, 1987.

\bibitem{Date1} E.Date, M.Jimbo, M.Kashiwara and T.Miwa,
{\it Operator approach to the Kadomtsev-Petviashvili equation.
Transformation groups for soliton equations III}, Journ. Phys.
Soc. Japan, {\bf 50} (1981) 3806-3812.

\bibitem{Matveev} V.B.Matveev, {\it Darboux transformation
and explicit solutions of the Kadomtsev-Petviashvili equation
depending on functional parameters}, Lett. Math. Phys. {\bf 3}
(1979) 213-216; {\it Darboux transformation and the explicit
solutions of differential-difference and difference-difference
evolution equations I}, Lett. Math. Phys. {\bf 3} (1979) 217-222;
V.B.Matveev and M.A.Salle, {\it
Differential-difference evolution equations II
(Darboux transformation for the Toda lattice)},
Lett. Math. Phys. {\bf 3} (1979) 425-429.

\bibitem{SZ} V.Spiridonov and A.Zhedanov, {\it Discrete
Darboux transformations, the discrete-time Toda lattice
and the Askew-Wilson polynomials}, Methods and Applications
of Analysis {\bf 2} (1995) 369-398.

\bibitem{DKN} B.Dubrovin, I.Krichever and S.Novikov, {\it Schr\"odinger
equation in magnetic field and Riemann surfaces}, Doklady Akad. Nauk USSR
{\bf 229:1} (1976) 15-18.

\bibitem{Kr1} I.Krichever, {\it Two-dimensional periodic difference
operators and algebraic geometry}, Doklady Akad. Nauk USSR {\bf 285:1}
(1985) 31-36.

\bibitem{Orl} A.Orlov, {\it Symmetries for unifying different
soliton systems into a single integrable hierarchy},
preprint IINS/Oce-04/03 (1991);
A.Orlov and S.Rauch-Wojciechowski, {\it Dressing method,
Darboux transformation and generalized restricted flows
for the KdV hierarchy}, Physica {\bf D69} (1993) 77-84.

\bibitem{OHTI} Y.Ohta, R.Hirota, S.Tsujimoto and T.Imai,
{\it Casorati and discrete Gram type determinant representations
of solutions to the discrete KP hierarchy}, Journ. Phys. Soc. Japan
{\bf 62} (1993) 1872-1886.

\bibitem{Suris1} Yu.Suris, {\it Generalized Toda chains in discrete
time}, Leningrad Math. Journ. {\bf 2} (1991) 339-352;
{\it Discrete-time generalized Toda lattices: complete integrability
and relation with relativistic Toda lattices}, Phys. Lett. {\bf A145}
(1990) 113-119.

\bibitem{Ruij} S.N.M.Ruijsenaars, {\it Relativistic Toda systems},
Commun. Math. Phys. {\bf 133} (1990) 217-247.

\bibitem{Suris2} Yu.Suris, {\it A discrete time relativistic Toda
lattice}, J. Physics {\bf A29} (1996) 451-465;
{\it A collection of integrable systems of the Toda type in
continuous and discrete time, with 2$\times$2 Lax
representations}, solv-int/9703004.

\bibitem{OKS} Y.Ohta, K.Kajiwara,
J.Matsukidaira and J.Satsuma, {\it Casorati
determinant solution of the relativistic Toda lattice equation},
solv-int/9304002.

\bibitem{KMZh} S.Kharchev, A.Mironov and A.Zhedanov,
{\it Faces of relativistic Toda chain}, preprint ITEP/TH-1/95,
FIAN/TD-19/95, hep-th/9606144.

\bibitem{FV} L.D.Faddeev and A.Yu.Volkov, {\it Quantum inverse
scattering method on a spacetime lattice}, Teor. Mat. Fiz. {\bf 92}
(1992) 207-214 (in russian);
L.D.Faddeev, {\it Current-like variables in massive and massless
integrable models}, Lectures at E.Fermi Summer School, Varenna 1994,
hep-th/9406196.

\bibitem{pendulum} A.Bobenko, N.Kutz and V.Pinkall, {\it The discrete
quantum pendulum}, Phys. Lett. {\bf A177} (1993) 399-404.

\bibitem{LOPZ} D.Lebedev, A.Orlov, S.Pakuliak and A.Zabrodin,
{\it Nonlocal integrable equations as reductions of the Toda
hierarchy}, Phys. Lett. {\bf A160} (1991) 166-172.

\bibitem{Hirota6} R.Hirota, {\it Discrete two-dimensional Toda
molecule equation}, Journ. Phys. Soc. Japan {\bf 56} (1987) 4285-4288.

\bibitem{K} M.Kontsevich, {\it Intersection theory on the moduli
space of curves}, Funk. Anal. i ego Pril. {\bf 25:2} (1991) 50-57.

\bibitem{KMMMZ} S.Kharchev, A.Marshakov, A.Mironov, A.Morozov and
A.Zabrodin, {\it Towards unified theory of 2d gravity}, Nucl. Phys.
{\bf B380} (1992) 181-240.

\end{thebibliography}
\end{document}